\documentclass[preprint,12pt]{aastex}
\newcommand{\fracb}[2]{\left(\frac{#1}{#2}\right)}
\newcommand{\fracsb}[2]{\left[\frac{#1}{#2}\right]}
\newcommand{\mean}[1]{\langle{#1}\rangle}
\usepackage{rotating}

\begin{document}

\title{Interaction of a highly magnetized impulsive relativistic flow with an external medium}

\author{Jonathan Granot\altaffilmark{1,2,3}}

\altaffiltext{1}{Racah Institute of Physics, The Hebrew University, Jerusalem 91904, Israel}
\altaffiltext{2}{Raymond and Beverly Sackler School of Physics \& Astronomy, Tel Aviv University, Tel Aviv 69978, Israel}
\altaffiltext{3}{Centre for Astrophysics Research, University of Hertfordshire, College Lane, Hatfield, AL10 9AB, UK; j.granot@herts.ac.uk}

\begin{abstract}

Important astrophysical sources, such as gamma-ray bursts (GRBs) or
tidal disruption events, are impulsive -- strongly varying with
time. These outflows are likely highly magnetized near the central
source, but their interaction with the external medium is not yet
fully understood. Here I consider the combined impulsive magnetic
acceleration of an initially highly magnetized shell of plasma and its
deceleration by the external medium. I find four main dynamical
regimes, that (for a given outflow) depend on the external
density. (I) For small enough external densities the shell becomes
kinetically dominated before it is significantly decelerated, thus
reverting to the familiar unmagnetized ``thin shell'' case, which
produces bright reverse shock emission that peaks well after the
prompt GRB. (II) For larger external densities the shell remains
highly magnetized and the reverse shock is strongly suppressed. It
eventually transfers most of its energy through $pdV$ work to the
shocked external medium, whose afterglow emission peaks on a timescale
similar to the prompt GRB duration. (III) For even larger external
densities there is no initial impulsive acceleration phase. (IV) For
the highest external densities the flow remains Newtonian.

\end{abstract}

\keywords{gamma-rays burst: general --- magnetohydrodynamics (MHD) --- shock waves --- ISM: jets and outflows}

\section{Introduction}
\label{sec:introduction}

The composition of relativistic jets or outflows in different
astrophysical sources, and in particular their degree of
magnetization, is highly uncertain and of great interest.  Pulsar
winds are almost certainly Poynting flux dominated near the central
source, and the same most likely also holds for active galatic nuclei
(AGN) and tidal disruption events (TDEs) of a star by a super-massive
black hole. In AGN and TDEs, since the central accreting black hole is
super-massive, then even close to it the Thompson optical depth
$\tau_T$ may not be high enough for thermal acceleration by radiation
pressure -- the main competition to magnetic acceleration -- to work
efficiently \citep[e.g.,][]{Ghis11}. In GRBs or micro-quasars,
however, thermal acceleration could also work ($\tau_T\gg 1$ is
possible, or even likely), and the dominant acceleration mechanism is
less clear.

One of the most important open questions about outflows that start out
highly magnetized near the central source is how they convert most of
their initial electromagnetic energy to other forms, namely bulk
kinetic energy or the energy in the random motions of the particles,
which also produce the radiation we observe from these
sources. Observations of relevant sources, such as AGN, GRBs or pulsar
wind nebulae suggest that the outflow magnetization is rather low at
large distances from the source. This is known as the $\sigma$
problem, namely how to transform from $\sigma\gg 1$ near the source to
$\sigma\ll 1$ very far from the source, where the magnetization
parameter $\sigma$ is the Poynting-to-matter energy flux ratio.

Different approaches to this problem have been considered so far.
Outflows that are Poynting flux dominated near the source are usually
treated under ideal MHD, axi-symmetry and steady-state (minaly for
simplicity). Under these conditions, however, it is hard to achieve
$\sigma < 1$ (or $\sigma\ll 1$) far from the source that would enable
efficient energy dissipation in internal
shocks~\citep{Kom09,Lyub09,Lyub10a}. One possible solution to this
problem is that the magnetization remains high ($\sigma\gg 1$) also at
large distances from the source and the observed emission is powered
by magnetic reconnection rather than by internal
shocks~\citep{LB03,Lyut06,GS06,Gian08}. Alternatively, the
non-axi-symmetric kink instability could randomize the direction of
the magnetic field, causing it to behave more like a fluid and
enhancing magnetic reconnection, which both increase the acceleration
and help lower the magnetization~\citep{HB00,DS02,GS06}. Another
option that may be relevant for AGN and GRBs~\citep{Lyub10b}, is that
if the Poynting flux dominated outflow has alternating fields (e.g. a
striped wind) then the Kruskal-Schwarzschild instability (i.e. the
magnetic version of the Rayleigh-Taylor instability) of the current
sheets could lead to significant magnetic reconnection, which in turn
increases the initial acceleration resulting in a positive feedback
and self-sustained acceleration that leads to a low $\sigma$.

While most previous works have assumed a steady state (i.e. no time
dependence), here the focus is on the effects of strong time
dependence -- impulsive outflows that are initially highly magnetized,
under ideal MHD. \citet[][hereafter paper I]{GKS11} have recently
found a new impulsive magnetic acceleration mechanism for relativistic
outflows, which is qualitatively different from its Newtonian
analog~\citep{Cont95}, and can lead to kinetic energy dominance and
low magnetizations that allow for efficient dissipation in internal
shocks. Paper I focused mainly on the acceleration of an initially
highly magnetized shell of plasma into vacuum, and only briefly
discussed the effects of its interaction with the external medium.
Here I analyze in detail the effects of its interaction with an
unmagnetized external medium whose density varies as a power-law with
the distance from the central source.

Most astrophysical relativistic outflow sources, such as AGN,
micro-quasars or pulsar wind nebulae (PWN), operate more or less
steadily over long periods of time. Therefore, the deceleration of
their outflow due to its interaction with external medium becomes
important only at very large distances from the source (at the ``hot
spots'' near the leading edge of AGN or micro-quasar jets\footnote{In
such jets, at relatively small distances from the source the external
medium can provide lateral pressure support that helps in the
collimation of the jet and its early collimation induced quasi-steady
acceleration.} and at the wind termination shock in PWN). AGN or
micro-quasar jets occasionally produce bright flares, which likely
correspond to a sudden and short lived large increase in their jet
power (or energy output rate). If the resulting ejected shell (or
blob) of plasma is highly magnetized then it can accelerate by the
impulsive mechanism found in paper I. Since it would be propagating in
the evacuated channel cleared by the preceding long lived steady
outflow from the same source, the deceleration by the external medium
would become important only well after the acceleration is over. There
are, however, also sources that are both impulsive and short-lived,
such as GRBs, TDEs or potentially also relativistic outflows from
giant flares in soft gamma-repeaters.  In such sources the
deceleration because of the interaction with the external medium can
become important already during the acceleration stage, and this may
have important implications for our understanding of these sources and
the interpretation of their observations.

The deceleration of an unmagnetized uniform\footnote{A non-uniform
shell of ejecta or relativistic wind with a power-law profile have
also been considered in other
works~\citep[e.g.][]{BM76,SM00,NS06,Nousek06,GK06,Levinson10}, and can
result in a temporally extended phase of energy injection into the
external (afterglow) shock. For simplicity, however, this work is
restricted to the case of a uniform shell of ejecta.}  relativistic
shell through its interaction with the external medium has been
studied in the context of GRBs
\citep{SP95,Sari97,KS00,KZ03,NP04}. The main results are summarized
and extended to a general power-law (with the distance from the
central source) external density profile
in~\S~\ref{sec:low-sigma}. 
The deceleration of a uniform magnetized relativistic shell by an
unmagnetized external medium has also been studied
~\citep{ZK05,GMA08,MGA09,Mizuno09,Lyut11}. However, most of the
treatments so far have assumed arbitrary initial conditions just
before the deceleration radius where most of the energy is transfered
to the shocked external medium, which can result in some unrealistic
outcomes~\citep[notable exceptions are paper I
and][]{Levinson10}. 

This work self-consistently considers the combined impulsive magnetic
acceleration and deceleration by a unmagnetized external medium of an
initially highly magnetized shell.  The main results for the
acceleration into vacuum of such a highly magnetized shell (paper I)
are described in ~\S~\ref{sec:high-sigma-acc}. The test case that was
studied in detail in paper I features a magnetized shell initially at
rest whose back end leans against a conducting wall with vacuum in
front of it, with initial width $l_0$, magnetic field $B_0$, rest mass
density $\rho_0$ and magnetization 
\begin{equation}
\sigma_0 = \frac{B_0^2}{4\pi\rho_0c^2}\gg 1\ .
\end{equation}
The shell is crossed by a strong, self-similar rarefaction wave
essentially on its light crossing time so that at a radius $R_0\sim
l_0$ it reaches a typical magnetization
$\mean{\sigma}\sim\sigma_0^{2/3}$ and Lorentz factor
$\mean{\Gamma}\sim\sigma_0^{1/3}$. It then becomes
super-fast-magnetosonic and looses causal contact with the wall,
resulting in a much slower subsequent impulsive acceleration phase in
which $\mean{\Gamma}\propto R^{1/3}$.  Eventually it becomes
kinetically dominated at the coasting radius $R_c \sim R_0\sigma_0^2$,
and at larger radii it starts coasting at a constant Lorentz factor
($\mean{\Gamma}\sim\sigma_0$) and spreading radially while its
magnetization rapidly drops with radius ($\mean{\sigma}\sim R_c/R$).

The combined acceleration and deceleration for an expansion into an
unmagnetized external medium with a power-law density profile is
addressed in detail in ~\S~\ref{sec:high-sigma-acc-dec}. The test case
from paper I is generalized by replacing the vacuum with an
appropriate external medium. Five distinct dynamical regimes are
identified, and their main properties are derived and discussed. In
regime I the external density is sufficiently low that early on it
hardly affects the shell, which accelerates essentially as if into
vacuum (as described above) until well after its coasting radius
$R_c$. By the time the effects of the external medium become important
the magnetization is already low, so that regime I effectively reverts
to the unmagnetized thin shell case (where both the reverse shock
emission and afterglow emission peak on a timescale longer than that
of the prompt GRB emission). In regime II the external density is
sufficiently large that it starts to strongly affect the shell during
its impulsive acceleration phase, while it is still highly magnetized.
The shell then starts to decelerate or accelerate more slowly until it
transfers most of its energy to the shocked external medium.  In
regime II the shell is highly magnetized all the way to its
deceleration radius, and therefore this strongly suppresses the
reverse shock (which is either non-existent or very weak) and its
associated emission. Thus, regime II can be thought of as a highly
magnetized thick shell case, in which no bright reverse shock emission
is expected, and the afterglow emission peaks on a timescale
comparable to that of the prompt GRB. In regime III the external
density is high enough that from the very start it inhibits the
acceleration so that there is no impulsive acceleration phase, and the
dynamics become essentially independent of the flow composition
(i.e. of $\sigma_0$, while $\sigma$ scales linearly with $\sigma_0$
but affects only the small fraction of the total energy that is in
kinetic form, $(1+\sigma)^{-1}\approx\sigma^{-1}\ll 1$). The
observational signatures of regime III are very similar to those of
regime II. In regime IV the external density is so high that the flow
remains Newtonian all along. This regime might be relevant for a
highly magnetized jet trying to bore its way out of a massive star
progenitor in long duration GRBs. Finally, regime II$^*$ occurs only
for a highly stratified external medium for which it replaces regime
II, and where also regimes I and III all show interesting and
qualitatively different behavior compared to smaller stratifications.

Table~\ref{tab:notation} summarizes the main notations and definitions
that are used in this work in order to help the reader follow the
text. The new results found in this work are compared to previous
works in \S~\ref{sec:comp}, and their implications are discussed in
\S~\ref{sec:dis}.

\section{Deceleration of an unmagnetized impulsive relativistic flow}
\label{sec:low-sigma}

Before generalizing the dynamics to the case of a highly magnetized
outflow, I begin with a detailed description of the deceleration of
an unmagnetized shell (corresponding to $\sigma \ll 1$ where $\sigma$
is defined in the next section), that initially coasts and propagates
relativistically into an unmagnetized external medium with a power law
density profile.

For simplicity I assume spherical symmetry, and that the original
ejecta from the GRB form a uniform shell of initial Lorentz factor
$\Gamma_0$ and initial width $\Delta_0$, were a subscript `0' is used
to denote the initial value of a quantity. Bulk Lorentz factors
(denoted by $\Gamma$), as well as the radius $R$ and width $\Delta$ of
the shell are measured in the rest frame of the central source (which
is also the rest frame of the external medium, and thus serves as the
lab frame), while thermodynamic quantities like the rest-mass density
$\rho$, the number density $n$, the pressure $p$, and the internal
energy density $e$ are measured in the local rest frame of the
fluid. A reasonable variation in $\Gamma_0$ of $\delta\Gamma_0 \sim
\Gamma_0$ will result in a significant radial spreading of the shell 
from the spreading radius, $R_s \sim \Delta_0\Gamma_0^2$, so that its
(lab-frame) width evolves as $\Delta\sim\max(\Delta_0,R/\Gamma_0^2)
\sim \Delta_0\max(1,R/R_s)$. The ambient medium is assumed to have a
power law mass density profile, $\rho_1 = AR^{-k}$, where for
simplicity I consider only $k<3$, which is also the parameter range
of most physical interest. Of particular interest are the cases
$k=0$, which corresponds to a constant density medium like the ISM, and
$k=2$, which is expected for the stellar wind of a massive star
progenitor.

As the shell interacts with the external medium and sweeps it up, two
shocks are formed: a forward shock that propagates into the ambient
medium and a reverse shock that goes back into the shell and slows it
down. The shocked shell material and the shocked external medium are
separated by a contact discontinuity. There are thus four different
regions: (1) unperturbed external medium, (2) shocked external medium,
(3) shocked shell material, and (4) unperturbed shell
material. Quantities at each region are denoted by the appropriate
subscript $i=1,2,3,4$. We have $\Gamma_4 = \Gamma_0$, $\Gamma_1 = 1$,
and since regions 2 and 3 are separated by a contact discontinuity,
$\Gamma_2 = \Gamma_3 = \Gamma$ and $p_2 = p_3 = p$.  Together with the
shock jump conditions between regions 3 and 4 (for the reverse shock)
and between regions 2 and 1 (for the forward shock), the resulting set
of equations (together with the equations of state in the different
regions) can be solved to obtain $\Gamma$, $e$, $\rho_2$ and $\rho_3$
(as well as the Lorentz factors of the reverse and forward shock
fronts) as a function of $\Gamma_0$ and the density ratio $f =
\rho_4/\rho_1$ of the unperturbed shell material and external medium.
There are two limits for which there is a simple analytic solution
\citep{SP95}: for $f\gg\Gamma_0^2$ the reverse shock is Newtonian,
and\footnote{more accurately $\Gamma=\Gamma_0(1-\sqrt{\epsilon})$ and
$\Gamma_{43}=1+2\epsilon$, where $\epsilon=2\Gamma_0^2/7f\ll 1$}
$\Gamma\approx\Gamma_0$, while for $f\ll\Gamma_0^2$ the reverse shock
is relativistic, $\Gamma\approx 2^{-1/2}\Gamma_0^{1/2}f^{1/4}$ and the
relative Lorentz factor between the fluid in regions 4 and 3 is
$\Gamma_{43}\approx 2^{-1/2}\Gamma_0^{1/2}f^{-1/4}$, where
\begin{equation}\label{f}
f\equiv\frac{\rho_4}{\rho_1}=\frac{E}{4\pi Ac^2\Gamma_0^2R^{2-k}\Delta }
=\frac{l_{\rm S}^{3-k}}{(3-k)\Gamma_0^2 R^{2-k}\Delta}\ ,
\end{equation}
$E=10^{53}E_{53}\;$erg is the (isotropic equivalent) kinetic energy of
the ejecta shell, and
\begin{equation}\label{l}
l_{\rm S}=\left[\frac{(3-k)E}{4\pi Ac^2}\right]^{1/(3-k)} =
\left\{\matrix{2.5\times 10^{18}E_{53}^{1/3}n_0^{-1/3}\;{\rm cm} 
& (k=0)\ , \cr & \cr
1.8\times 10^{19}E_{53} A_*^{-1}\;{\rm cm}  & (k=2)\ , }\right.
\end{equation}
is the Sedov radius where the (isotropic equivalent) swept up mass
equals $E/c^2$. Numerical values are provided for the physically
interesting cases of $k=0$, which correspond to a uniform medium of
number density $n=n_0\;{\rm cm^{-3}}$ ($A=nm_p$ where $m_p$ is the
proton mass), and $k = 2$, which corresponds to the stellar wind of a
massive star progenitor, with $A_*=A/(5\times 10^{11}\;{\rm gr\;
cm^{-1}})$.  It is clear from Eq.~(\ref{f}) that $k=2$ is a critical
value below which $f$ decreases with radius and above which $f$
increases with radius, before the shell starts spreading (i.e.  while
$\Delta\approx\Delta_0$ and is independent of radius). Since $k=2$ is
also a physically interesting value, it will be discussed separately
below. The case $2<k<3$ will also be briefly mentioned. We shall,
however, first concentrate on $k<2$.

For $k<2$, $f$ decreases with radius. Thus the reverse shock is
initially Newtonian, and becomes relativistic at a radius $R_N$ given
by $f(R_N)=\Gamma_0^2$, or $R_N\sim\min(R_\Gamma,R_{N,0})$ with
\begin{equation}\label{R_N}
R_{N,0}=\left(\frac{E}{4\pi A c^2\Gamma_0^4\Delta_0}\right)^{1/(2-k)}
=\left[\frac{l_{\rm S}^{3-k}}{(3-k)\Gamma_0^4\Delta_0}\right]^{1/(2-k)}
=4.2\times 10^{16}\zeta^{1/2}E_{53}^{1/2}
n_0^{-1/2}\Gamma_{2.5}^{-2}T_{30}^{-1/2}\;{\rm cm}\ ,
\end{equation}
where $\zeta=(1+z)/3$, $\Gamma_{2.5}=\Gamma_0/10^{2.5}$, $T_{\rm GRB}=
(1+z)\Delta_0/c=30T_{30}\;$s is the observed duration of the GRB and
\begin{equation}\label{R_gamma}
R_\Gamma=\left[\frac{(3-k)E}{4\pi Ac^2\Gamma_0^2}\right]^{1/(3-k)} =
\frac{l_{\rm S}}{\Gamma_0^{2/(3-k)}} = \left\{\matrix{ 5.4\times
10^{16}E_{53}^{1/3}n_0^{-1/3}\Gamma_{2.5}^{-2/3}\;{\rm cm} & (k=0)\ ,
\cr & \cr 1.8\times 10^{14}E_{53}A_*^{-1}\Gamma_{2.5}^{-2}\;{\rm cm}&
(k=2)\ , }\right.
\end{equation}
is the radius where a rest mass $E/\Gamma_0^2c^2$ of the external
medium is swept up. In this work $T$ denotes the observed time (at
which photons reach the observer), while $t$ denotes the lab frame
time. The observed times corresponding to $R_{N,0}$ and $R_\Gamma$ are
\begin{eqnarray}\label{t_N}
T_{N,0} &=& (1+z)\frac{R_{N,0}}{bc\Gamma_0^2}=13\zeta^{3/2}E_{53}^{1/2}
n_0^{-1/2}\Gamma_{2.5}^{-4}T_{30}^{-1/2}\;{\rm s}\ ,
\\ \label{t_gamma}
T_\Gamma &=& (1+z)\frac{R_\Gamma}{bc\Gamma_0^2}= \left\{\matrix{
27\,\zeta E_{53}^{1/3} n_0^{-1/3} \Gamma_{2.5}^{-8/3}\;{\rm s} 
& (k=0)\ , \cr & \cr
0.089\,\zeta E_{53} A_*^{-1} \Gamma_{2.5}^{-4}\;{\rm s}  & (k=2)\ , }\right.
\end{eqnarray}
where $b\sim 1-2$ and $b \sim 2$ reflect the typical photon arrival
times from regions 3 and 2, respectively, and $b=2$ is used to obtain
the numerical values. Two additional important radii are the spreading
radius $R_s\sim\Gamma_0^2\Delta_0 \sim R_\Gamma^{3-k}R_{N,0}^{k-2}$
mentioned above (where the shell starts spreading radially), and the
radius at which the reverse shock finishes crossing the shell,
$R_\Delta \sim f^{1/2}\Gamma_0\Delta\sim (E\Delta/Ac^2)^{1/(4-k)}\sim
\max(R_\Gamma,R_{\Delta,0})$ where $R_{\Delta,0} \sim
(R_sR_\Gamma^{3-k})^{1/(4-k)} \sim (E\Delta_0/Ac^2)^{1/(4-k)}$.  It is
also convenient to define the parameter
\begin{eqnarray}
\nonumber
\Upsilon_0 &\equiv & \frac{R_\Gamma}{R_s}=\left[\frac{(3-k)E}
{4\pi Ac^2\Gamma_0^{2(4-k)}\Delta_0^{3-k}}\right]^{1/(3-k)}=
\frac{l_{\rm S}}{\Delta_0}\,\Gamma_0^{-2(4-k)/(3-k)}
\\ \nonumber \\ \label{upsilon}
&=& \left\{\matrix{
1.8\,\zeta E_{53}^{1/3}n_0^{-1/3}\Gamma_{2.5}^{-8/3}T_{30}^{-1}
& (k=0)\ , \cr & \cr
5.9\times 10^{-3}\zeta E_{53}A_*^{-1}\Gamma_{2.5}^{-4}T_{30}^{-1} 
& (k=2)\ , }\right.
\end{eqnarray}
and\footnote{Note that $\Upsilon = \xi^{2-k}$, where $\xi$ is
essentially the same parameter that was defined in \citet{SP95}.}
$\Upsilon = \Upsilon_0(\Delta_0/\Delta) = (l_{\rm
S}/\Delta)\Gamma_0^{-2(4-k)/(3-k)}$. Note that $R_\Gamma=R_{\Gamma,0}$
and $R_s=R_{s,0}$ since $R_\Gamma$ does not depend on $\Delta$ and
$R_s$ depends on $\Delta_0$ rather than on $\Delta$. Thus, we have
\begin{equation}\label{R_ord}
\Upsilon^{-1/(2-k)}R_{N} \sim R_\Gamma \sim 
\Upsilon^{1/(4-k)}R_{\Delta} \sim \Upsilon_0 R_s\ ,
\end{equation}
so that the initial relative ordering of the different radii is
determined by the value of $\Upsilon_0$, while the evolution of this
ordering is determined by that of $\Upsilon$.

The condition $\Upsilon_0>1$ can be written as
$\Delta_0<\Delta_{\rm cr}$ or $\Gamma_0<\Gamma_{\rm cr}$ where
\begin{eqnarray}\label{delta_cr}
\Delta_{\rm cr} &=& \left[\frac{(3-k)E}{4\pi
    Ac^2\Gamma_0^{2(4-k)}}\right]^{1/(3-k)} =
\frac{l_{\rm S}}{\Gamma_0^\frac{2(4-k)}{(3-k)}}
=\left\{\matrix{
5.4\times 10^{11}\,E_{53}^{1/3}n_0^{-1/3}\Gamma_{2.5}^{-8/3}\;{\rm cm}
& (k=0)\ , \cr & \cr
1.8\times 10^{9}E_{53}A_*^{-1}\Gamma_{2.5}^{-4}\;{\rm cm}
& (k=2)\ , }\right.
\\ \nonumber
\\ \label{eta_cr}
\Gamma_{\rm cr} &=& \left[\frac{(3-k)E}{4\pi Ac^2\Delta_0^{3-k}}
\right]^{1/2(4-k)}=\left(\frac{l_{\rm S}}{\Delta_0}\right)^\frac{(3-k)}{2(4-k)}
=\left\{\matrix{
395\,\zeta^{3/8}E_{53}^{1/8}n_0^{-1/8}T_{30}^{-3/8}
& (k=0)\ , \cr & \cr
88\,\zeta^{1/4}E_{53}^{1/4}A_*^{-1/4}T_{30}^{-1/4}
& (k=2)\ , }\right.
\end{eqnarray}
so that this case is often referred to as a ``thin'' or ``slow''
shell.  Similarly, the case $\Upsilon_0<1$ corresponds to
$\Delta_0>\Delta_{\rm cr}$ or $\Gamma_0>\Gamma_{\rm cr}$ and is
referred to as a ``thick'' or ``fast'' shell. Note that
\begin{equation}\label{xi2}
\Upsilon_0=\fracb{\Delta_0}{\Delta_{\rm cr}}^{-1} =
\fracb{\Gamma_0}{\Gamma_{\rm cr}}^{-2(4-k)/(3-k)}\ .
\end{equation}

\subsection{Thin Shells}
\label{subsec:thin_hydro}

For $\Upsilon_0 > 1$ (a thin or slow shell) and $k < 2$, the initial
ordering of the critical radii is $R_s < R_{\Delta,0} < R_\Gamma <
R_{N,0}$ and the shell starts spreading early on\footnote{If there is
no significant spreading of the shell (i.e.  $\delta\Gamma_0 \ll
\Gamma_0$) then the reverse shock will cross the shell while it is
still Newtonian, and the energy extraction would proceed via a semi
steady state of Newtonian shocks and refraction waves traveling back
and forth in the shell \citep{SP95}.  In this case we do not expect to
have significant radiation from the original shell of ejecta during
its deceleration.}  so that at $R > R_s$ we have $\Delta\sim
R/\Gamma_0^2 \sim (R/R_s)\Delta_0$ and
$\Upsilon\sim\Upsilon_0(R/R_s)^{-1}$ starts decreasing, which leads to
a triple coincidence, $R_\Delta\sim R_\Gamma \sim R_N$ with
$\Upsilon\sim 1$ at that radius (see Eq.~[\ref{R_ord}]). In this case
the reverse shock is mildly relativistic during the period when most
of the energy is extracted from the shell, near the radius
$R_\Delta\sim R_\Gamma \sim R_N$ or the corresponding time $T_\Gamma$
when the reverse shock finishes crossing the shell. At larger times or
radii, most of the energy has already been transferred to the shocked
external medium and the flow approaches the adiabatic (i.e. with a
constant energy $E$) self-similar
\citet[][hereafter BM76]{BM76} solution.
 
For $k=2$, $f$ is initially (at $R < R_s$) independent of radius and
$f/\Gamma_0^2 = l_{\rm S}/\Gamma_0^4\Delta = \Upsilon$ ($=\Upsilon_0$
as long as the shell does not spread significantly). Therefore, for
thin shells the reverse shock is Newtonian with a constant shock
velocity at $R<R_s$. However, for thin shells $R_s$ is smaller than
all other critical radii, so that the shell begins to spread early
on. Therefore, again at $R > R_s$ we have $\Delta\sim R/\Gamma_0^2\sim
(R/R_s)\Delta_0$ and $\Upsilon\sim\Upsilon_0(R/R_s)^{-1}$ starts
decreasing with radius, leading to $R_\Delta\sim R_\Gamma \sim R_N$
with $\Upsilon\sim 1$ at that radius, so that the reverse shock is
mildly relativistic by the time it finishes crossing the shell, at
$T_\Gamma$.

For $2<k<3$ and $\Upsilon_0>1$, the initial ordering of the critical
radii is $R_{N,0} < R_s < R_{\Delta,0} < R_\Gamma$ and $f$ initially
(at $R<R_s$) increases with radius (and time). Hence the reverse shock
is initially relativistic until $R_{N,0}$ ($T_{N,0}$) and then becomes
Newtonian. At $R>R_s$ the shell begins to spread and from this point
on $\Delta \sim (R/R_s)\Delta_0$ and therefore $f$ and $\Upsilon$
begin to decrease with radius (as $R^{k-3}$ and $R^{-1}$,
respectively).  This again leads to $R_\Delta\sim R_\Gamma\sim R_N$
with $\Upsilon\sim 1$ at that radius, where the reverse shock finishes
crossing the shell. Here $R_N$ is the radius where the reverse shock
becomes relativistic again, i.e. it becomes mildly relativistic when
it finishes crossing the shell, at $T_\Gamma$.

\subsection{Thick Shells}
\label{subsec:thick_hydro}

For $\Upsilon_0<1$ (a thick or fast shell) and $k < 2$, the initial
ordering of the critical radii is $R_{N,0} < R_\Gamma < R_{\Delta,0} <
R_s$. Since $R_s$ is the largest of the
critical radii, spreading is unimportant, and therefore $\Delta \approx
\Delta_0$, $R_N \approx R_{N,0}$, and $R_\Delta \approx
R_{\Delta,0}$. The reverse shock becomes relativistic before it
crosses most of the shell, and therefore in this case most of the
kinetic energy is converted to internal energy (of the shocked shell
and the shocked external medium) at $R_{\Delta,0}$ corresponding to an
observed time $T_E \sim (1+z)R_{\Delta,0}/c\Gamma^2_{\rm BM}(R_{\Delta,0})
\sim (1+z)\Delta_0/c \sim T_{\rm GRB}$, where $\Gamma_{\rm BM}(R) \sim
(E/Ac^2)^{1/2}R^{(k-3)/2}$ is the Lorentz factor of the adiabatic BM76
self-similar solution. Here $R_\Gamma$ is no longer relevant since the
relativistic reverse shock implies that $\Gamma\ll\Gamma_0$ so that
the energy in the swept up external medium of rest mass $M$ is now
$\Gamma^2 Mc^2\ll\Gamma_0^2 Mc^2$, and an external medium of rest mass
much larger than $M_0/\Gamma_0 = E/\Gamma_0^2c^2$ (by a factor of
$\Gamma_0^2/\Gamma^2 \approx 2\Gamma_0/f^{1/2} \gg 1$, where $M_0$ is
the original shell's rest mass) needs to be shocked in order for it to
reach an energy comparable to $E$ (and this occurs only at
$R_{\Delta,0}$).

I now generalize the results of \citet{Sari97}, which are for a
uniform external density ($k = 0$), to a more general power law
external density \citep[with $k < 3$; see also][]{GR-R11}. At
$T<T_{N,0}$ ($R < R_{N,0}$) we have $\Gamma
\approx \Gamma_0$, while at $T_{N,0}<T<T_E$ ($R_{N,0} < R <
R_{\Delta,0}$) we have $\Gamma \approx 2^{-1/2}\Gamma_0^{1/2}f^{1/4}$,
which can be expressed as
\begin{equation}\label{gamma}
\Gamma \approx \left(\frac{E}{16\pi Ac^2\Delta_0}\right)^{1/4}R^{(k-2)/4}
\approx \left(\frac{E}{16\pi b^{2-k}Ac^{4-k}\Delta_0}
\right)^{1/2(4-k)}T_z^{-(2-k)/2(4-k)}\ .
\end{equation}
For $k=2$, $\Gamma$ remains constant at this stage while for $k<2$ it
decreases with time (see below). Since $T_z = T/(1+z) \propto
R/\Gamma^2 \propto R^{(4-k)/2}$ we have $dR/dT_z = [2/(4-k)]dR/dT_z$
and the observed rate of production of internal energy in the forward shock is
\begin{equation}\label{L}
L_{\rm int,obs} = \frac{dE}{dT_z} = \frac{dE}{dR}\,\frac{dR}{dT_z} = 
\frac{8\pi b^{3-k}}{(4-k)}Ac^{5-k}\Gamma^{2(4-k)}T_z^{2-k}\ ,
\end{equation}
where $dE/dR \approx 4\pi R^2\rho_1(R)c^2\Gamma^2(R) = 4\pi
Ac^2\Gamma^2 R^{2-k}$.  Substituting Eq.~(\ref{gamma}) into
Eq.~(\ref{L}) we see that regardless of the value of $k$, the
luminosity of the forward shock is constant, $L_{\rm int,obs} = (1+z)E/T_E$ where
$T_E = [2(4-k)/b](1+z)\Delta_0/c = [2(4-k)/b]T_{\rm GRB} \approx
30(4-k)\,T_{30}\;$s is the time when the energy in the shocked
external medium becomes comparable to $E$.  The Lorentz factor at this
time is independent of the initial Lorentz factor $\Gamma_0$,
\begin{equation}\label{gamma_t_E}
\Gamma(T_E) = \fracsb{(4-k)^{k-2}E}{2^{6-k}\pi Ac^2\Delta_0^{3-k}}^{1/2(4-k)}
\sim \Gamma_{\rm cr}\ .
\end{equation}
After the time $T_E$ most of the energy is in the forward shock, which
quickly approaches the BM76 self-similar solution, in which its Lorentz
factor scales as $\Gamma_{\rm BM}\propto R^{-(3-k)/2}\propto
T^{-(3-k)/2(4-k)}$, which implies $L_{\rm int,obs}\propto T^{-1}$.

For $k=2$ and thick shell ($\Upsilon_0 < 1$) we have $R_s >
R_{\Delta,0}$ so that the shell hardly spreads radially
($\Delta\approx\Delta_0$) while it is crossed by the reverse shock.
This implies that $f/\Gamma_0^2 = l_{\rm S}/\Gamma_0^4\Delta =
\Upsilon \approx \Upsilon_0$, i.e. the reverse shock is relativistic and
its strength (or $\Gamma_{43}$) is constant with radius until it
finishes crossing the shell at $R_{\Delta,0}$ (corresponding to an
observed time $T_E$). Therefore, for thick shells $R_{N,0}$ and
$T_{N,0}$ go to zero, and the Lorentz factor of the shocked fluid is
constant in time, $\Gamma(T<T_E) = \Gamma(T_E) \sim \Gamma_{\rm cr}$
(note that this value is $\ll \Gamma_0$). At $T > T_E$ (or
equivalently, $R > R_{\Delta,0}$) the flow approaches the BM76
self-similar solution.

For $2 < k < 3$ and a thick shell ($\Upsilon_0 < 1$), the initial
ordering of the critical radii is $R_\Gamma < R_{\Delta,0} < R_s <
R_{N,0}$ and $f$ increases with radius (and time). Therefore the
reverse shock is relativistic until it finishes crossing the shell at
$R_{\Delta,0}$ (or $T_E$). Again, $\Gamma$ is given by
Eq.~(\ref{gamma}) at $T < T_E$ (or $R < R_{\Delta_0}$), where it
increases with time (and radius) at this stage, while at $T > T_E$ (or
$R > R_{\Delta,0}$) it is given by the BM76 self-similar solution,
$\Gamma_{\rm BM}(R) \sim (E/Ac^2)^{1/2}R^{(k-3)/2}$.

\section{Acceleration of a highly magnetized impulsive flow into vacuum}
\label{sec:high-sigma-acc}

This was addressed in great detail in paper I, and here I summarize
the main results that were derived in there. Paper I has studied,
under ideal MHD, the test case of a cold (with a negligible thermal
pressure) finite shell of initial (at $t=0$) width $l_0$ (occupying
$-l_0<x<0$) and magnetization $\sigma_0 = B_0^2/4\pi\rho_0c^2\gg 1$,
whose back end leans against a conducting wall (at
$x=-l_0$)\footnote{Such a ``wall'' can be the center of a planar shell
surrounded by vacuum on both sides, which splits into two parts going
in opposite directions, with reflection symmetry about its center,
which remains at rest.} and with vacuum in front of it (at $x>0$),
where the magnetic field is perpendicular to the direction of
motion. A correspondence was shown in this case between the dynamical
equations in planar and spherical geometries. A strong rarefaction
wave develops at the vacuum interface and propagates toward the wall
at the initial fast magnetosonic speed of the unperturbed shell,
$c_{\rm ms,0} =\beta_{\rm ms,0}c$, reaching the wall at $t=t_0 =
l_0/c_{\rm ms,0}\approx l_0/c$. For a cold shell the dimensionless
fast-magnetosonic speed is given by $\beta_{\rm
ms,0}=\sqrt{\sigma_0/(1+\sigma_0)}$ and corresponds to a Lorentz
factor of $\Gamma_{\rm ms,0} = (1-\beta_{\rm ms,0}^2)^{-1/2}
=\sqrt{1+\sigma_0}$ and a dimensionless 4-velocity of $u_{\rm ms,0}
=\Gamma_{\rm ms,0}\beta_{\rm ms,0} =\sigma_0^{1/2}$. In our case
$\sigma_0\gg 1$ so that $\Gamma_{\rm ms,0} \approx u_{\rm ms,0} =
\sigma_0^{1/2}\gg 1$ and $\beta_{\rm ms,0}\approx 1$. The rarefaction
wave accelerates the shell to a typical (or weighted mean over the
energy in the lab frame) Lorentz factor of
$\langle\Gamma\rangle(t_0)\sim\sigma_0^{1/3}$ while the typical
magnetization drops to
$\langle\sigma\rangle(t_0)\sim\sigma_0^{2/3}$. This result has a
simple explanation: as long as $\langle\sigma\rangle \gg 1$ and most
of the energy is in electromagnetic form, energy conservation implies
that $\langle\Gamma\rangle\langle\sigma\rangle
\sim\sigma_0$; such very fast acceleration can occur only as long as
the flow pushes against the ``wall'' (or static source), and stops
when the flow looses causal contact with it, i.e. when it becomes
super-fast-magnetosonic, $\langle\Gamma\rangle \sim \Gamma_{\rm ms}
~\sim\langle\sigma\rangle^{1/2} \sim
\sigma_0^{1/2}\langle\Gamma\rangle^{-1/2}$, which corresponds to
$\langle\Gamma\rangle \sim \sigma_0^{1/3}$ and $\langle\sigma\rangle
\sim \sigma_0^{2/3}$. Such a shell is broadly similar to a uniform
(quasi-) spherical outflow from a static source that lasts a finite
time, $t_0$, during which it reaches a radius $R_0\approx ct_0$,
Lorentz factor $\langle\Gamma\rangle(t_0)\sim\sigma_0^{1/3}$ and
magnetization $\langle\sigma\rangle(t_0) \sim\sigma_0^{2/3}$, being
quickly accelerated from $\Gamma \sim 1$ and $\sigma = \sigma_0$ near
the source.

In a spherical steady-state flow the acceleration becomes inefficient
once the flow loses causal contact with the static source (or
``wall'') and there is no significant subsequent acceleration so that
$\mean{\Gamma}\sim\sigma_0^{1/3}$ also asymptotically, at very large
distances from the source~\citep{GJ70}. For a non-spherical flow
collimation can result in further acceleration up to
$\mean{\Gamma}\sim\sigma_0^{1/3}\theta_j^{-2/3}$
\citep[e.g.,][]{Lyub09}, where $\theta_j$ is the asymptotic
half-opening angle of the jet (at which point lateral causal contact
across the jet is lost, so the center of the jet cannot push against
the ambient material; for simplicity, factors of order unity are
discarded here and until the end of this subsection).  However, for an
impulsive source, which corresponds to a shell of finite width $l_0$
or a outflow lasting for a finite time $t_0 \approx l_0/c$, efficient
subsequent acceleration (at $t > t_0$) does occur. This happens since
the shell pushes against itself and significantly expands in its own
rest frame, under its own magnetic pressure (while its width in the
lab frame remains constant, $\Delta =\Delta'/\Gamma
\sim l_0$, since its comoving width $\Delta'$ increases linearly with
its Lorentz factor $\Gamma$ as it accelerates). While in the comoving
frame the expansion is roughly symmetric between the back and front
parts of the shell, in the lab frame most of the energy remains in the
front part of the the shell, resulting in a constant effective width
($\Delta \sim l_0$, where most of the energy resides).

The radial expansion of the shell in its own rest frame as its
accelerates results in a dispersion $\delta\Gamma \sim \mean{\Gamma}$
in its Lorentz factor.  This causes the shell width in the lab frame
to increase as $\Delta\sim R_0 + R/\mean{\Gamma}^2$.  Ideal MHD
implies that the shell's electromagnetic energy scales as $E_{\rm EM}
\propto 1/\Delta$. Therefore, at the radius $R_c$ where the shell
doubles its initial width, half of the initial magnetic energy is
converted into kinetic form, so that $\mean{\sigma} = E_{\rm
EM}/E_{\rm kin} \sim 1$ and $\mean{\Gamma}\sim \sigma_0$ at this
radius. Therefore, $R_c$ must correspond to the coasting radius where
the acceleration saturates and after which the shell becomes
kinetically dominated and starts coasting at $\mean{\Gamma} \sim
\sigma_0$. This, in turn, implies that $R_c \sim R_0\sigma_0^2$, which
provides the scaling of $\mean{\Gamma}$ with $R$ during the
acceleration phase: $d\log\mean{\Gamma}/d\log R =
\log[\mean{\Gamma}(R_c)/\mean{\Gamma}(R_0)]/\log(R_c/R_0) =
\log(\sigma_0^{2/3})/\log(\sigma_0^2) = 1/3$, so that $\mean{\Gamma}\sim
(\sigma_0 R/R_0)^{1/3}$ during this phase, which ends at the coasting
time, $t_c \sim t_0\sigma_0^2$, distance $l_c \approx ct_c \approx
l_0\sigma_0^2$ or radius $R_c \approx ct_c
\approx R_0\sigma_0^2$. 

At $t > t_c$ the flow becomes essentially unmagnetized (i.e. with a
low magnetization, $\sigma < 1$), its internal (magnetic) pressure
becomes unimportant dynamically, and each fluid element within the
shell coasts at a constant speed (ballistic motion). As we have seen
above, the shell starts spreading radially significantly in the lab
frame at $R_c$, and subsequently its width grows linearly with $R$,
$t$ or $x$,
\begin{equation}
\frac{\Delta}{l_0} \sim \left\{\matrix{1
\quad & \zeta_c < 1\ ,\cr\cr 
\zeta_c \quad & \zeta_c > 1\ ,}\right. 
\end{equation}
where $\zeta_c = t/t_c \approx x/l_c = R/R_c$ (where $R_c \approx
l_c$). Moreover, the growth in the width of the shell causes a
significant drop in its magnetization: $\sigma(t>t_c) \sim t_c/t$. One
can summarize this result in terms of $\zeta_0 = t/t_0 \approx x/l_0 =
R/R_0$ (where $R_0 \approx l_0$) or $\zeta_c$,
\begin{equation}
\langle\Gamma\rangle \sim \left\{\matrix{(\sigma_0\zeta_0)^{1/3} 
\quad & 1 < \zeta_0 < \sigma_0^2\ ,\cr\cr 
\sigma_0 \quad & \zeta_0 > \sigma_0^2\ ,}\right. 
\quad\quad\quad\quad
\langle\sigma\rangle \sim \left\{\matrix{\sigma_0^{2/3}\zeta_0^{-1/3} 
\quad & 1 < \zeta_0 < \sigma_0^2\ ,\cr\cr 
\sigma_0^2\,\zeta_0^{-1} \quad & \zeta_0 > \sigma_0^2\ ,}\right.
\end{equation}

\begin{equation}
\langle\Gamma\rangle \sim \left\{\matrix{\sigma_0\zeta_c^{1/3} 
\quad & \sigma_0^{-2} < \zeta_c < 1\ ,\cr\cr 
\sigma_0 \quad & \zeta_c > 1\ ,}\right. 
\quad\quad\quad\quad
\langle\sigma\rangle \sim \left\{\matrix{\zeta_c^{-1/3} 
\quad & \sigma_0^{-2} < \zeta_c < 1\ ,\cr\cr 
\zeta_c^{-1} \quad & \zeta_c > 1\ .}\right.
\end{equation}

\section{Acceleration and Deceleration of an Impulsive High-$\sigma$ 
Relativistic Outflow}
\label{sec:high-sigma-acc-dec}

\subsection{The general framework, and a spherical self-similar solution for $k = 2$}

For concreteness, let us specify to a spherically symmetric flow
expanding into a power-law external density profile, $\rho_1 =
Ar^{-k}$, where $r$ is the spherical radial coordinate. The outflow is
taken to be cold (with no thermal pressure), and with a high initial
magnetization, $\sigma_0\gg 1$. The original outflow remains cold as
long as it is not shocked by a reverse shock. The shocked swept-up
external medium, however, is typically heated to relativistic
temperatures.  The motion is in the radial direction ($\hat{\beta} =
\hat{r}$) and the magnetic field is tangential ($\hat{r}\cdot\vec{B} =
0$).

It has been shown in paper~I that the relevant cold (no thermal
pressure) MHD equations for spherical and planar geometries are
identical when written in terms of the normalized, barred variables,
which can apply to both a planar and a spherical geometry,
\begin{equation}\label{eq:sph=plane}
(\bar{r},\,\bar{b},\,\bar{\rho}) = \left\{\matrix{
(x,\,b,\,\rho) & \quad{\rm (planar)}\ ,\cr\cr 
(r,\,rb,\,r^2\rho) & \quad{\rm (spherical)}\ ,}\right.
\end{equation}
where $b = B/(\sqrt{4\pi}\,\Gamma)$ is the normalized comoving
magnetic field. When there is thermal pressure then it violates this
rescaling\footnote{This occurs since in the momentum equation there is
a term $\partial_r p$ or $\partial_x p$, while in spherical geometry
this rescaling requires $\bar{p} = r^2p$, which would instead give
$r^{-2}\partial_r r^2p = \partial_r p+2p/r$, i.e. a spurious extra
term.}. There is a convenient analytic solution for the relevant
planar the Riemann problem with a uniform unmagnetized external medium
\citep[paper I;][]{Lyut10}, which has a corresponding spherical solution
according to the above rescaling, for $k=2$.  This solution would be
valid within the original cold magnetized shell, i.e. at $r<R_{\rm
CD}(t)$, where $R_{\rm CD}$ is the radius of the contact discontinuity
(CD) that forms.\footnote{In this region, for $k=2$, I derive the
expressions for the density $\rho$ for the planar case, $\rho_{\rm
pl}$, and those for the spherical case are given by $\rho_{\rm
sph}(r,t) = (r/R_0)^{-2}\rho_{\rm pl}(x=r,t)$, where $\rho_{\rm
sph}(r,t=0) = \rho_0(r/R_0)^{-2}$ is the initial density profile of
the spherical shell.} A shock propagates into the cold unmagnetized
external medium, with a shock radius $R_{\rm sh}(t)$, which heats the
material passing through it to a relativistically hot
temperature. Therefore, in the region between the shock front and the
CD, $R_{\rm CD}(t)<r<R_{\rm sh}(t)$, the simple self-similar solution
for the planar case where this region is uniform with the same
pressure and velocity as the CD itself, is no longer valid in the
spherical case. However, for the spherical case with $k=2$ and a
constant velocity of the CD ($\Gamma_{\rm CD} = {\rm const}$) there is
a different self-similar solution shown in Figs.~4-6 of BM76,
corresponding to $k=2$, $q=0$ (energy injection by a constant power
source), and $m=0$ (the Lorentz factor has no explicit time
dependence, and instead depends only on the value of the self-similar
variable, $\chi = (1+2\Gamma_{\rm sh}^2)(1-\xi)$ where $\xi=r/ct$, so
that the Lorentz factor of the shock front, $\Gamma_{\rm sh}$, or the
CD, $\Gamma_{\rm CD}$, are constant). In our case, this unmagnetized
($\sigma = 0$) solution would apply in the region between the CD and
the shock front, $R_{\rm CD}(t)<r<R_{\rm sh}(t)$, while the inner part
($r< \xi_u ct < R_{\rm CD}(t)$, where $\xi_u$ is introduced below) of
the global solution is given by the self-similar solution mentioned
above for the cold magnetized shell, which can be simply scaled from
planar to spherical geometry. Please note that the shock location,
$\chi=1$, corresponds to $\xi_{\rm sh} =\beta_{\rm sh}\approx
1-1/2\Gamma_{\rm sh}^2$ and that the CD location is (from Table I of
BM76) $\chi_{\rm CD} \approx 1.77$, corresponding to $\xi_{\rm CD} =
\beta_{\rm CD} = 1-\chi_{\rm CD}/2\Gamma_{\rm sh}^2$ and therefore
\begin{equation}\label{chi_CD}
\frac{\Gamma_{\rm sh}^2}{\Gamma_{\rm CD}^2} = \chi_{\rm CD} \approx 1.77\ .
\end{equation}
However, the Lorentz factor of the material just behind the shock
front is $\Gamma(\chi=1)\approx \Gamma_{\rm sh}/\sqrt{2}$, and
therefore $\Gamma_{\rm CD}/\Gamma(\chi=1) = (2/\chi_{\rm CD})^{1/2}
\approx 1.06$.  This shows that the Lorentz factor of the shocked
external medium increases only by about $6\%$ from just behind the
shock front to the CD (and its square increases by 13\%, as can be
seen in Fig. 5 of BM76). Therefore, a uniform Lorentz factor is a
reasonable approximation for this region (and I shall occasionally
use this approximation). Moreover, the normalized width of this region
is
\begin{equation}\label{eq:width}
\frac{R_{\rm sh}-R_{\rm CD}}{R_{\rm CD}} \approx 
\frac{R_{\rm sh}-R_{\rm CD}}{R_{\rm sh}} \approx 
\frac{\chi_{\rm CD}-1}{2\Gamma_{\rm sh}^2} =
\frac{1-\chi_{\rm CD}^{-1}}{2\Gamma_{\rm CD}^2}
\approx \frac{0.435}{2\Gamma_{\rm CD}^2}\ .
\end{equation}

\begin{figure}
\centerline{\includegraphics[width=15cm]{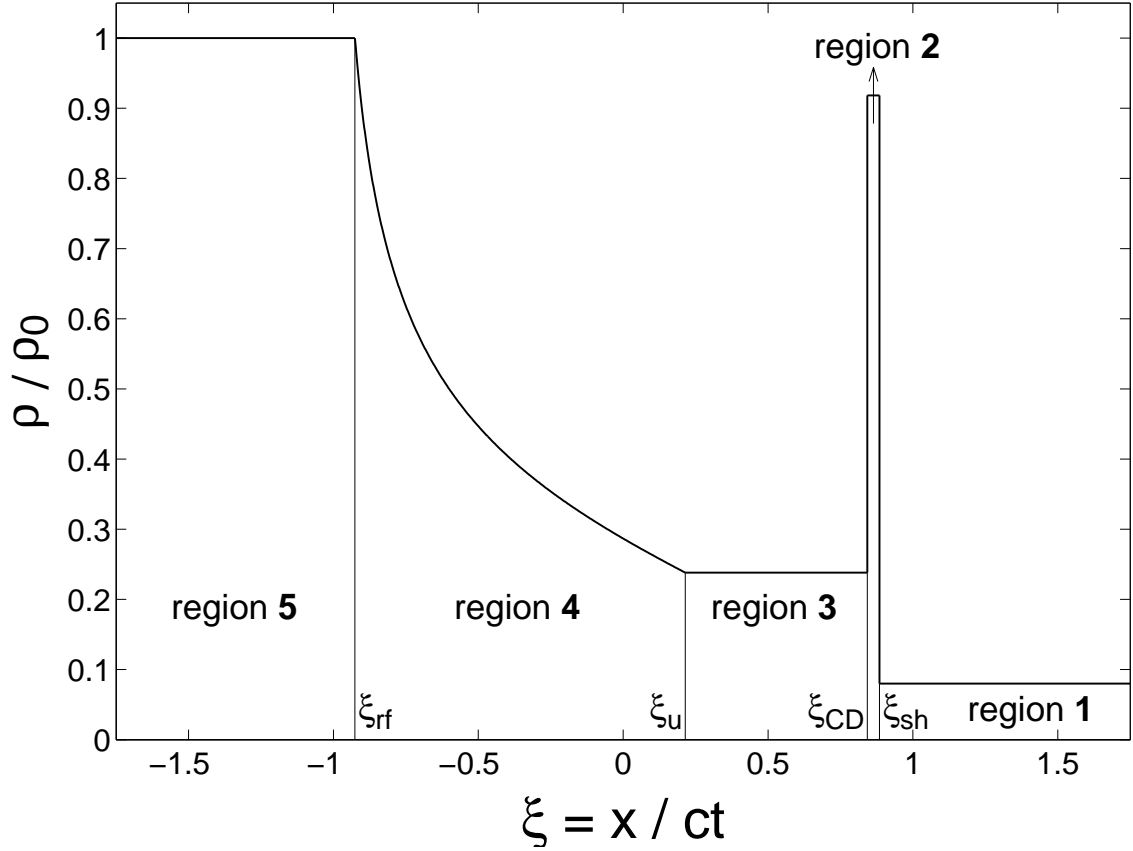}}
\caption{
The self-similar structure when a cold magnetized shell initially at
rest (occupying $x<0$ at $t=0$) accelerates into an unmagnetized
external medium (initially at rest and occupying $x>0$ at $t=0$).  For
concreteness, I show the proper density normalized by its initial
value in the magnetized shell ($\rho_0$ -- the density in region 5),
for $\sigma_0=6$ and $\rho_1/\rho_0 = 0.08$. Such a self-similar
solution in planar symmetry, where $\xi = x/ct$, also corresponds to a
solution in spherical symmetry, where $\xi = r/ct$ and $(x,b,\rho) \to
(r,rb,r^2\rho)$ (see Eq.~[\ref{eq:sph=plane}]).}
\label{fig:RW-diagram}
\end{figure}

This spherical self-similar solution for $k=2$ is a very useful
starting point for the current discussion. It will be described it in
terms of the corresponding planar solution, where a uniform region
would correspond to an $r^{-2}$ dependence of the density or magnetic
pressure in the spherical solution. We are interested in an initially
highly magnetized flow ($\sigma_0\gg 1$), and for all cases of
interest (except regime IV, which is described separately in
\S~\ref{sec:regIV}) the shock that is driven into the external medium is (at
least initially) highly relativistic, the shock front moving with
$\Gamma_{\rm sh} = (1-\beta_{\rm sh}^2)^{-1/2}\gg 1$. The planar
Riemann problem contains 5 regions (see Fig.~\ref{fig:RW-diagram}):
(1) at $\xi >\xi_{\rm sh}$, where $\xi\equiv x/ct$ and $\xi = \xi_{\rm
sh} = x_{\rm sh}(t)/ct =\beta_{\rm sh}$ at the location of the shock
front, there is cold, unmagnetized, unperturbed uniform external
medium at rest with rest mass density $\rho_1$, (2) at $\xi_{\rm CD}
<\xi <\xi_{\rm sh}$ there is a uniform\footnote{As discussed above, in
this region there is a deviation from the simple scaling between the
planar and spherical cases, and the BM76 solution with $m = q = 0$ and
$k=2$ holds there in the spherical case.} region of shocked external
medium, moving at $\Gamma_2 = (1-\beta_2^2)^{-1/2} ~\approx
\Gamma_{\rm sh}/\sqrt{2}$ with $e_2 = 3p_2 = 4\Gamma_2^2\rho_1c^2$,
where at $\xi_{\rm CD} =\beta_2$ there is a contact discontinuity
(CD), (3) at $\xi_u < \xi <\xi_{\rm CD}$ there is a uniform region
(moving at $\Gamma_3 =\Gamma_2$ with $e_3 = p_3 =
(B_3/\Gamma_3)^2/8\pi = p_2$) occupied by magnetized material
(originating from region 5, or from the original magnetized outflow in
an astrophysical context) that has passed through a rarefaction wave
(region 4) and is accumulating between the front end of the
rarefaction wave, at\footnote{Here $\beta_{\rm ms}(\beta)$ is the
dimensionless fast magnetosonic speed within the rarefaction wave
(region 4), at the point where the flow velocity is $v = \beta c$.}
$\xi_u = [\beta_2-\beta_{\rm ms}(\beta_2)]/[1-\beta_2\beta_{\rm
ms}(\beta_2)]$, and the CD, (4) at $\xi_{\rm rf} < \xi < \xi_u$ is a
region with a rarefaction wave described by the self-similar solution
in Appendix~A of paper I, where $\xi_{\rm rf} = -\beta_{\rm ms,0} =
-[\sigma_0/(1+\sigma_0)]^{1/2}$ is its tail, and (5) at $\xi <
\xi_{\rm rf}$ is the original unperturbed uniform, cold magnetized shell at
rest with rest mass density $\rho_0$, magnetic field $B_0$, and
magnetization $\sigma_0 = B_0^2/4\pi\rho_0c^2 \gg 1$.

Now, let us consider such an initial shell of finite initial width
$l_0$, whose back end is leaning against a conducting ``wall'' (at $x
= -l_0$).  At $t_0 = l_0/c_{\rm ms,0}$ (where $t_0\approx l_0/c$ for
$\sigma_0\gg 1$) the tail of the leftward moving rarefaction wave
reaches the wall and a secondary right-going rarefaction wave forms
that decelerates the material at the back of the flow. The
head\footnote{Note that I refer to the rightmost point in the
rarefaction wave as its head. In the original rarefaction wave this
was at the vacuum interface while for the secondary rarefaction wave
this is at the interface with the original rarefaction wave.} of the
secondary rarefaction wave is located at $\xi_*(t)=x_*(t)/ct$ and
moves to the right with a dimensionless speed
\begin{equation}
  \beta_* \equiv \frac{1}{c}\frac{dx_*}{dt} = \frac{\beta(\xi_*)+\beta_{\rm
  ms}(\xi_*)}{1+\beta(\xi_*)\beta_{\rm ms}(\xi_*)}\ ,
\label{xs*}
\end{equation}   
where $\beta(\xi\geq\xi_*)$ and $\beta_{\rm ms}(\xi\geq\xi_*)$ are
given by the self-similar solution for the original expansion
(describing a leftward moving rarefaction), since the part of the flow
ahead of the secondary (or ``reflected'') rarefaction wave
($\xi>\xi_*$) does not ``know'' about the existence of the
``wall''. At this stage region 5 described above no longer exists, and
a new region is formed behind the head of the secondary (right-going)
rarefaction wave. This new region carries a very small fraction of the
total energy as long as the magnetization at its head is large,
$\sigma(\xi_*) = \sigma_0\tilde{\rho}_* \gg 1$ where $\tilde{\rho}_*
=\bar{\rho}(\xi_*)/\bar{\rho}_0$, which implies that this rarefaction
is strong and significantly decelerates the fluid that passes through
it (see paper I for details). Therefore, as long as this condition
holds, most of the energy and momentum in the flow, as well as most of
the original rest mass of the magnetized shell, remain in a shell of
constant width $\approx 2l_0$ between $\xi_*$ and $\xi_{\rm sh}$.

The value of $\xi_u$ is determined by pressure balance at the
CD. Since both the normalized pressure, $\bar{p} = \bar{b}^2/2 =
r^2b^2/2$, and the fluid velocity are constant in the range
$\xi_u\leq\xi < \xi_{\rm sh} = \beta_{\rm sh}$ (corresponding to
regions 2 and 3; see Fig.~\ref{fig:RW-diagram}), and $R_{\rm
sh}(t)\approx R_{\rm CD}(t) \equiv R(t) \approx ct$ so that the
external density can be evaluated at either of these radii,
$\rho_1[R_{\rm sh}(t)] \approx\rho_1[R_{\rm CD}(t)]$, we have
\begin{eqnarray}\label{p_condition}
a\,\frac{4}{3}\Gamma_{\rm CD}^2(R)\rho_1(R)c^2 =
p_2(R) = p_3(R) = \fracb{\xi_u}{\xi_{\rm CD}}^2 p_4(\xi_u)
= \fracb{R_0}{R}^2\frac{\sigma_0\rho_0c^2}{2}\tilde{\rho}_u^2(R)\ ,
\\ \label{eq:a}
a = \left\{\matrix{
1 & {\rm uniform\ approximation}\ , \cr & \cr 
0.571& {\rm BM76\ solution}\ , }\right.
\quad\quad\quad\quad\quad\quad\quad\quad\quad\quad
\end{eqnarray}
where $\tilde{\rho} \equiv \bar{\rho}/\bar{\rho}_0$ is the normalized
density (i.e. $\rho/\rho_0$ in the planar case and $\rho
r^2/\rho_0R_0^2$ in the spherical case) and $\tilde{\rho}_u
=\bar{\rho}(\xi_u)/\bar{\rho}_0$ is its value at $\xi_u$, while
Eq.~(\ref{eq:a}) holds for $\Gamma_{\rm CD}\gg 1$.

Although the self-similar solution at $r<R_{\rm CD}(t)$ is strictly
valid only for $k = 2$, for which $\Gamma_{\rm CD}$ is constant, we
shall make the approximation that it still provides a reasonable
description of the flow for $k\neq 2$, in which case $\Gamma_{\rm
CD}$, $\sigma_{\rm CD}$, etc., gradually evolve with time.

Denoting the initial shell to external density ratio by $f_0\equiv
\rho_0/\rho_1(R_0)$, Eq.~(\ref{p_condition}) implies
\begin{equation}\label{eq:rho_sigma_u}
\tilde{\rho}_u(R) = \frac{\sigma_u}{\sigma_0} \cong 
\fracb{8a}{3f_0\sigma_0}^{1/2}\fracb{R}{R_0}^{(2-k)/2}\Gamma_{\rm CD}\ ,\quad
\sigma_u  \cong 
\fracb{8a\sigma_0}{3f_0}^{1/2}\fracb{R}{R_0}^{(2-k)/2}\Gamma_{\rm CD}\ ,
\end{equation}
where $\sigma_u = \sigma(\xi_u)$. For the self-similar rarefaction
wave solution in region 4 (see paper I),
\begin{equation}
\left(\frac{1+\beta}{1-\beta}\right)\left(\sqrt{\sigma}
+\sqrt{\sigma+1}\right)^4 = \mathcal{J}_+ = \left(\sqrt{\sigma_0}
+\sqrt{\sigma_0+1}\right)^4 \approx 16\sigma_0^2\ ,
\end{equation}
where $\sigma = \sigma(\xi) = \sigma_0\tilde{\rho} =
\sigma_0\bar{\rho}(\xi)/\bar{\rho}_0$ is the local value of the magnetization
parameter, and the Riemann invariant $\mathcal{J}_+$ approaches a
value of $16\sigma_0^2$ for $\sigma_0\gg 1$. We are interested
primarily in the relativistic part of region 4, for which $\Gamma_4
\gg 1$ is given by\footnote{The result for the bulk of the rarefaction
wave (where $\Gamma \gg 1$ and $\sigma
\gg 1$) can be understood considering a finite shell of initial width
$l_0$ and energy (per unit area) $E_0 = l_0(B_0^2/8\pi)(1+2/\sigma_0)
\approx l_0B_0^2/8\pi$. After the passage of the rarefaction wave, the
shell width becomes $\approx 2l_0$, and since it is relativistic there
is an electric field in the lab frame that is almost equal to the
magnetic field so that the shell energy is $E \approx
2l_0(B^2/4\pi)$. Now, $E = E_0$ requires $B \approx B_0/2$, and $B =
B_0\Gamma\tilde{\rho}$ since $B/\Gamma\rho = {\rm const}$, implying
$\Gamma \approx 1/2\tilde{\rho}$. More generally, $\Gamma_4 =
(\delta_\beta+\delta_\beta^{-1})/2$, where $\delta_\beta =
[(1+\beta)/(1-\beta)]^{1/2} = 
[(\sqrt{1+\sigma_0}+\sqrt{\sigma_0})/(\sqrt{1+\sigma}+\sqrt{\sigma})]^2$.}
\begin{equation}\label{eq:Gamma_4}
\Gamma_4 \approx \frac{2\sigma_0+1}{\left(\sqrt{\sigma}
+\sqrt{\sigma+1}\right)^2} \approx \left\{\matrix{
(2\sigma_0+1)/(1+2\sqrt{\sigma})\sim 2\sigma_0 & \quad \sigma \ll 1\ ,\cr\cr 
1/2\tilde{\rho} & \quad \sigma \gg 1 \ .}\right.
\end{equation}
At $1\ll \sigma \ll \sigma_0$ the Lorentz factor varies significantly
with $\sigma = \sigma_0\tilde{\rho}$ as $\Gamma_4 \approx
\sigma_0/2\sigma$, while for $\sigma\ll 1$ it approaches a constant
value of $\Gamma_4\approx 2\sigma_0$.  The transition between these
two regimes occurs at $\sigma \sim 1$ for which $\Gamma_4\sim\sigma_0$
[though $\Gamma_4(\sigma = 1) \approx 0.343\sigma_0$, and
$\Gamma_4(\sigma = 1/8) = \sigma_0$]. We are particularly interested
in when this also corresponds to the transition between regions 4 and
3, i.e. $\Gamma_4(\xi_u) =\Gamma_2\sim\sigma_0$ and $\sigma_u\sim 1$,
which according to Eq.~(\ref{eq:rho_sigma_u}) corresponds to
$f_0\sim\sigma_0^{3}(R/R_0)^{2-k}$, or to a radius $R_1$ that can be
defined by $\sigma_u(R_1) = 1$ and is given by
\begin{equation}\label{eq:R1}
R_1 \sim R_0\fracb{f_0}{\sigma_0^3}^{1/(2-k)}\ .
\end{equation}
For $k = 2$ both $\Gamma_{\rm CD} = \Gamma_u$ and $\sigma_{\rm CD} =
\sigma_u$ do not change with radius, so that generally $\sigma_u$ is
either always below 1 or always above 1, corresponding, respectively,
to regimes I and II that are discussed below, so that in this case
there is no radius $R_1$ where $\sigma_u(R_1) = 1$.

If the magnetization in region 3 or just behind the CD is low,
$\sigma_u \approx \sigma_{\rm CD} \ll 1$, then $\Gamma_{\rm CD}
\approx 2\sigma_0$ according to Eq.~(\ref{eq:Gamma_4}), so that
Eq.~(\ref{eq:rho_sigma_u}) implies
\begin{equation}\label{eq:sigma_uI}
\tilde{\rho}_u \approx \left(\frac{32a\sigma_0}{3f_0}\right)^{1/2}
\fracb{R}{R_0}^{\frac{2-k}{2}}\ ,\quad\quad
\sigma_u \approx \sigma_{\rm CD} \approx 
\left(\frac{32a\sigma_0^3}{3f_0}\right)^{1/2}
\fracb{R}{R_0}^{\frac{2-k}{2}} \ll 1\ .
\end{equation}
If, on the other hand, the magnetization in region 3 or just behind
the CD is high, $\sigma_u\approx \sigma_{\rm CD}\gg 1$, then
Eq.~(\ref{eq:Gamma_4}) implies $\Gamma_4\approx 1/2\tilde{\rho}$
(since $\sigma\geq\sigma_u$ in all of the region behind the CD), and
in particular $\Gamma_{\rm CD} =\Gamma_4(\xi_u) \approx
1/2\tilde{\rho}_u$, so that Eq.~(\ref{eq:rho_sigma_u}) gives
\begin{eqnarray}\nonumber
\tilde{\rho}_u \approx \fracb{2a}{3f_0\sigma_0}^{1/4}
\fracb{R}{R_0}^{(2-k)/4}\ ,\quad\quad\quad
\sigma_u \approx \sigma_{\rm CD}\approx \fracb{2a\sigma_0^3}{3f_0}^{1/4}
\fracb{R}{R_0}^{(2-k)/4}\gg 1\ ,
\\ \label{eq:regII}
\Gamma_{\rm CD} \approx \fracb{3f_0\sigma_0}{32a}^{1/4}
\fracb{R}{R_0}^{(k-2)/4} \sim \Gamma_{\rm cr}\fracb{R}{R_{\rm cr}}^{(k-2)/4} 
\ ,\quad\quad\quad\quad\quad
\end{eqnarray}
where $R_{\rm cr} \sim R_0\Gamma_{\rm cr}^2$ is the radius at which
$\Gamma_{\rm CD}$ reaches the value $\Gamma_{\rm cr}$ when $\sigma_u
\approx \sigma_{\rm CD}\gg 1$, and an expression for this radius is
provided in Eq.~(\ref{R_cr_acc}) below.

\subsection{Regime I}
\label{sec:regI}

From the derivation above it becomes clear that for $f_0 \gg
\sigma_0^{7-2k}$ the external medium would hardly affect the
acceleration phase, and the magnetized shell would accelerate
essentially as if it were expanding into vacuum (as described in paper
I, and summarized in \S~\ref{sec:high-sigma-acc}). This can be seen
from the fact that this condition corresponds to $\sigma_u(R_c)\ll 1$,
i.e. that even by the coasting radius $R_c \sim R_0\sigma_0^2$ the
region of the original shell that had been affected by the external
medium (region 3) occupies only a small part of the flow near its head
that carries a small fraction of its energy. The transition, where
$f_0 \sim \sigma_0^{7-2k}$, corresponds to the equality of the
coasting radius (or distance), $R_c \sim R_0\sigma_0^2$, and the
deceleration radius\footnote{The deceleration radius $R_{\rm dec}$ is
the radius at which most of the energy is transferred to the shocked
swept-up external medium. Here its post-shock Lorentz factor is
$\Gamma_2 \sim
\sigma_0$ and therefore the energy given to a swept-up external rest
mass $M$ is $(\Gamma_2^2-1)Mc^2\sim \sigma_0^2Mc^2$, and $R_{\rm dec}$
is given by $E \sim \sigma_0^2M(r<R_{\rm dec})c^2$.}. In planar
symmetry with a constant external density $\rho_1$ (which corresponds
to $k = 2$ in spherical symmetry), conservation of energy implies
$E\sim l_0\sigma_0\rho_0c^2 \sim\sigma_0^2l_{\rm dec}\rho_1c^2$ and
thus the deceleration distance is given by $l_{\rm dec} \sim
l_0\rho_0/\sigma_0\rho_1 = l_0f_0/\sigma_0$ (where for simplicity we
discard factors of order unity) so that indeed $l_{\rm dec} \sim l_c
\sim l_0\sigma_0^2$ corresponds to $f_0\sim \sigma_0^3$, as it
should. For spherical symmetry, energy conservation reads $E \sim
R_0^3\sigma_0\rho_0c^2 \sim\sigma_0^2 A R_{\rm dec}^{3-k}c^2$, implying
a deceleration radius $R_{\rm dec}\sim (E/\sigma_0^2Ac^2)^{1/(3-k)}
~\sim R_0(f_0/\sigma_0)^{1/(3-k)}$, so that indeed $R_{\rm dec}\sim
R_c \sim R_0\sigma_0^2$ corresponds to $f_0\sim\sigma_0^{7-2k}$ or $A
\sim R_0^k\rho_0\sigma_0^{2k-7}$.  Note that in this regime $R_{\rm
dec}$ essentially corresponds to $R_\Gamma$ that is given in
Eq.~(\ref{R_gamma}) where $\Gamma_0$ is replaced by $\Gamma(R_c) \sim
\sigma_0$.

For $k < 2$, $\sigma_u$ increases with radius (see
Eq.~[\ref{eq:sigma_uI}]) and since we have seen that regime I
corresponds to $\sigma_u(R_c)\ll 1$ this implies that $\sigma_u\ll 1$
all along. For $2 < k < 10/3$, on the other hand, $\sigma_u$ decreases
with radius passing through the value of 1 at a radius $R_1$ given by
$R_1/R_c\sim (f_0/\sigma_0^{7-2k})^{1/(2-k)}\ll 1$. This would be
physically interesting only if $R_1> R_0$, which corresponds to
$\sigma_0^{7-2k}\ll f_0\ll\sigma_0^3$. In this parameter regime $R_0 <
R_1 < R_c$, so that Eqs.~(\ref{eq:sigma_uI}) and (\ref{eq:regII})
imply that $\sigma_u \approx \sigma_{\rm CD} \sim (R/R_1)^{(2-k)/4} >
1$ at $R_0 < R < R_1$ while $\sigma_u \approx\sigma_{\rm CD} \sim
(R/R_1)^{(2-k)/2} < 1$ at $R > R_1$ (or at $R_1 < R < R_{\rm cr}$, as
we shall see later). For both $k < 2$ and $2<k<10/3$ we have $\sigma_u
\approx \sigma_{\rm CD}\ll 1$ at $R \gtrsim R_c$.

One can find the time when the reflected rarefaction wave reaches
region 3, $\xi_* = \xi_u$, or the corresponding radius $R_u$. Relying
on the derivations in paper I, one obtains $R_u/R_c \approx
2\sigma_u^{-3/4}$, which upon substitution of $\sigma_u(R_u)$ from
Eq.~(\ref{eq:sigma_uI}) and solving for $R_u$ gives
\begin{equation}\label{eq:Ru_Rc}
\frac{R_u}{R_c} \approx \frac{2}{\sigma_u^{3/4}(R_u)} \approx
\left(\frac{3f_0}{2^{7/3}a\sigma_0^{7-2k}}\right)^{\frac{3}{14-3k}}
\sim\fracb{R_{\rm RS}}{R_c}^{\frac{12-3k}{14-3k}} \gg 1\ ,
\end{equation}
where $R_{\rm RS}$ (discussed below) is the radius where a strong
reverse shock develops. Therefore, clearly $R_u < R_{\rm RS}$, and the
reflected rarefaction reaches region 3 well before a strong reverse
shock develops. The rarefaction wave also reaches the CD ($\xi_*
=\xi_{\rm CD}$ at a radius $R_{\rm *,CD}$) within a single dynamical
time from reaching $\xi_u$ (i.e. $R_{\rm *,CD}\sim R_u$),
\begin{equation}\label{eq:DR*}
\frac{\Delta R}{R_u} \approx \frac{\Delta t}{t_u} = 
\frac{\xi_{\rm CD}-\xi_u}{\beta_*(\beta_{\rm CD})-\beta_{\rm CD}} =
\frac{\beta_{\rm CD}-\frac{\beta_{\rm CD}-\beta_{\rm ms}(\beta_{\rm CD})}
{1-\beta_{\rm CD}\beta_{\rm ms}(\beta_{\rm CD})}}
{\frac{\beta_{\rm CD}+\beta_{\rm ms}(\beta_{\rm CD})}
{1+\beta_{\rm CD}\beta_{\rm ms}(\beta_{\rm CD})}-\beta_{\rm CD}} =
\frac{1+\beta_{\rm CD}\beta_{\rm ms}(\beta_{\rm CD})}
{1-\beta_{\rm CD}\beta_{\rm ms}(\beta_{\rm CD})} \sim 1\ ,
\end{equation}
where $\Delta R = R_{\rm *,CD}-R_u$ and the last approximate equality
is valid since $\beta_{\rm ms}(\beta_{\rm CD})\approx
\sigma_u^{1/2}\ll 1$. Once the right-going rarefaction wave reaches
the CD, this triggers a gradual deceleration of the CD, which is
initially weak as the rarefaction is weak at this stage since $u_{\rm
ms} =\sqrt{\sigma_u}\ll 1$.

\begin{figure}
\centerline{\includegraphics[width=15cm]{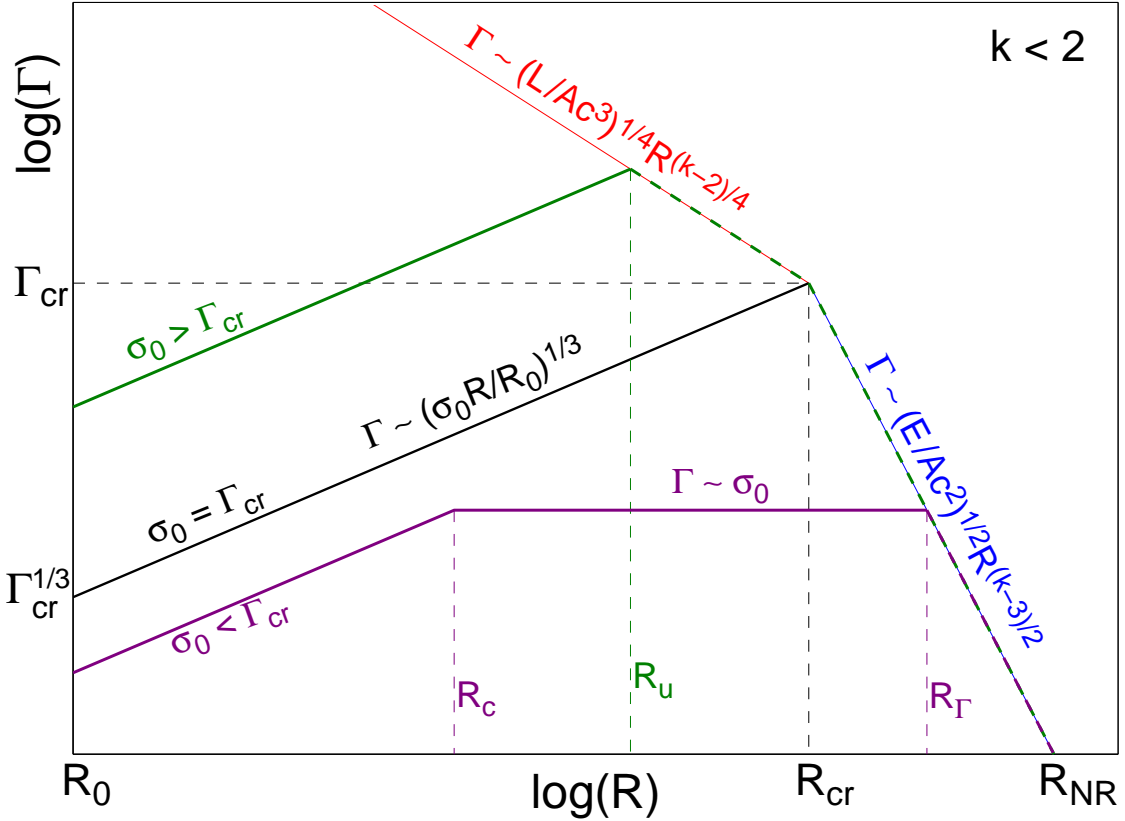}}
\caption{
Evolution of the typical Lorentz factor of the flow (where most of the
energy resides), $\mean{\Gamma}$, as a function of radius $R$ for $k <
2$ and for different values of the initial magnetization $\sigma_0$
(and $\rho_0\propto 1/\sigma_0$) and fixed values of the initial time
or length scale ($t_0 \approx R_0/c$ or $R_0$), energy ($E \sim Lt_0
\approx LR_0/c$ or $L$), and external density ($k$ and $A$ or
$\rho_1(R_0) = AR_0^{-k}$), which imply fixed $\Gamma_{\rm cr}$ and
$R_{\rm cr}$. In most cases of interest $\Gamma_{\rm cr} \gg 1$, so
this is assumed to be the case here. The green and purple lines
correspond to regimes I ($1<\sigma_0 < \Gamma_{\rm cr}$) and II
($\Gamma_{\rm cr} < \sigma_0 <\Gamma_{\rm cr}^{3(4-k)/2}$),
respectively. In regime III ($\sigma_0 >\Gamma_{\rm cr}^{3(4-k)/2}$),
$\mean{\Gamma}(R\geq R_0)$ becomes independent of $\sigma_0$ and
follows the thin solid red and blue lines. (The particular slopes in
this figure are plotted for $k=0$, but the general scalings are
clearly indicated).}
\label{fig:Gamma_R}
\end{figure}

\begin{figure}
\centerline{\includegraphics[width=15cm]{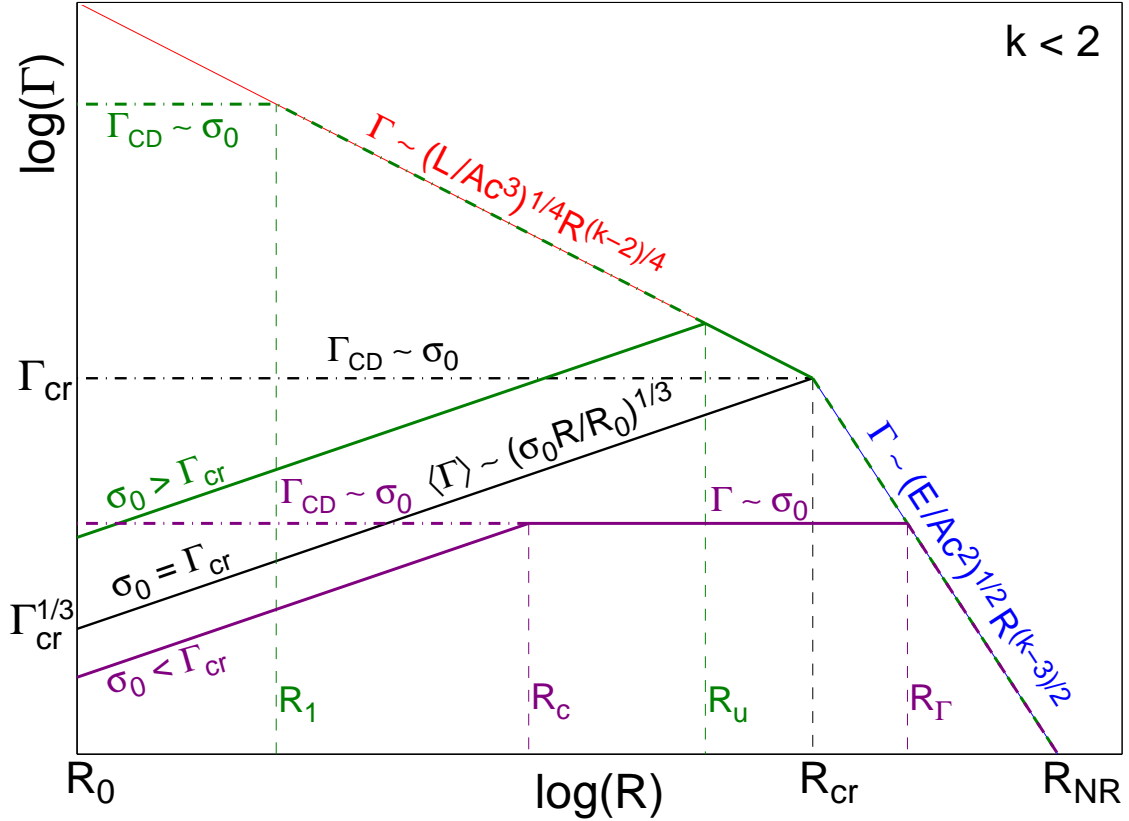}}
\caption{
The same as Fig.~\ref{fig:Gamma_R} but with the addition of the
Lorentz factor of the contact discontinuity, $\Gamma_{\rm CD}$ ({\it
dashed-dotted lines}), until it becomes similar to the typical Lorentz
factor, $\langle\Gamma\rangle$ ({\it solid lines}). The two remain
similar up to the deceleration radius, after which $\Gamma_{\rm CD}$
starts falling behind $\langle\Gamma\rangle$ (at which stage only
$\langle\Gamma\rangle$ is shown in the figure for clarity; {\it dashed
lines}).  (The particular slopes in this plot are drawn for $k=0$, but
the general scalings are clearly indicated).}
\label{fig:Gamma_R2}
\end{figure}

\begin{figure}
\centerline{\includegraphics[width=15cm]{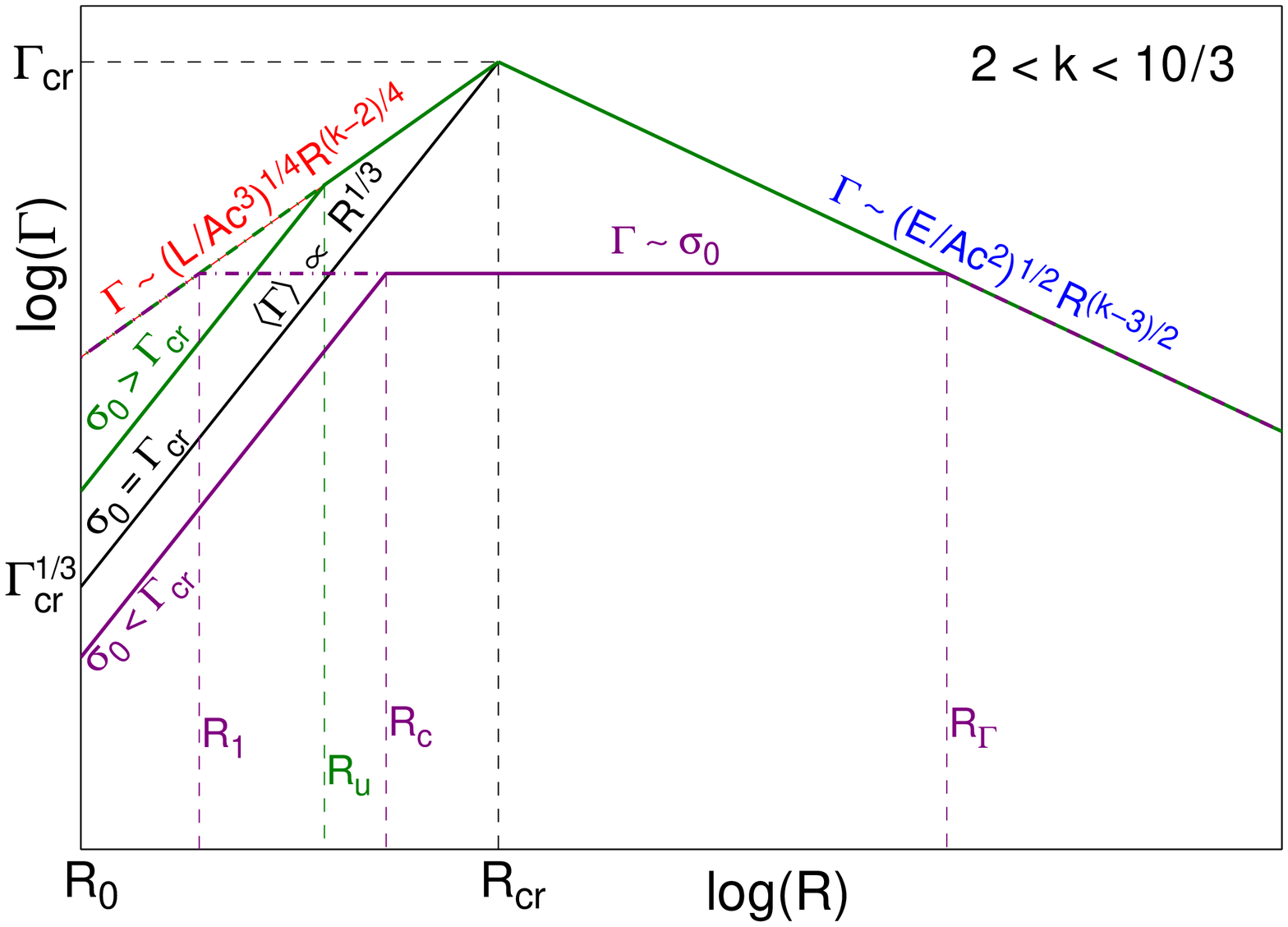}}
\caption{
The same as Fig.~\ref{fig:Gamma_R2} but for $2 < k < 10/3$.}
\label{fig:Gamma_R2_k2p}
\end{figure}

\begin{figure}
\centerline{\includegraphics[width=15cm]{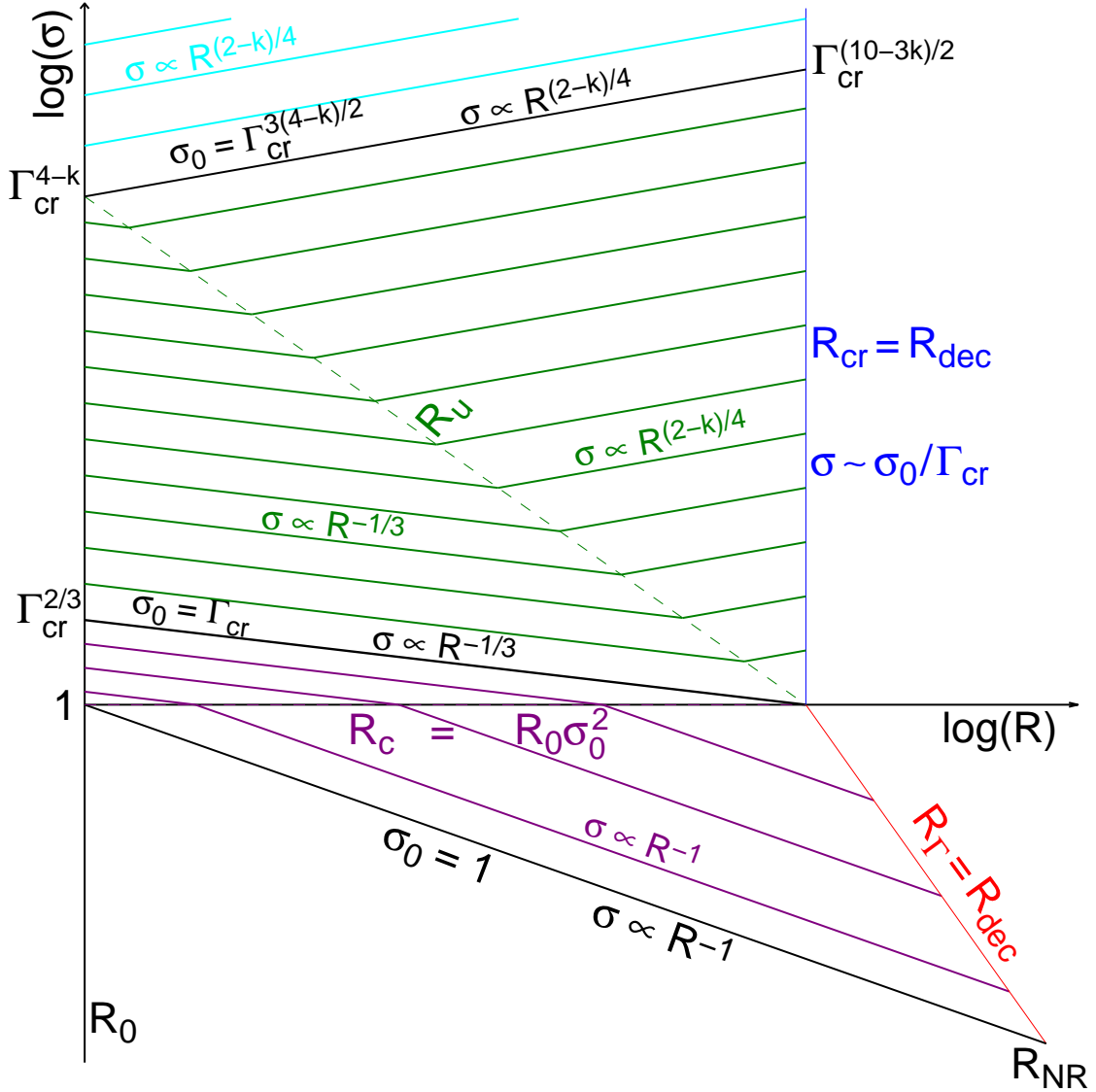}}
\caption{
Evolution of the typical magnetization $\mean{\sigma}$ of the outflow
as a function of radius $R$, corresponding to Fig.~\ref{fig:Gamma_R}
(i.e. for $k < 2$, where each of the solid lines originating at $R =
R_0$ corresponds to a different value of $\sigma_0$). The different
regimes identified in the text are plotted using lines of different
colors: regime I ($1<
\sigma_0 < \Gamma_{\rm cr}$) in green, regime II ($\Gamma_{\rm cr} <
\sigma_0 < \Gamma_{\rm cr}^{3(4-k)/2}$) in purple, and regime III
($\sigma_0 > \Gamma_{\rm cr}^{3(4-k)/2}$) in cyan. The lines
corresponding to relevant critical radii (some of which depend on
$\sigma_0$) are also shown.}
\label{fig:sigma_R} 
\end{figure}

\begin{figure}
\centerline{\includegraphics[width=15cm]{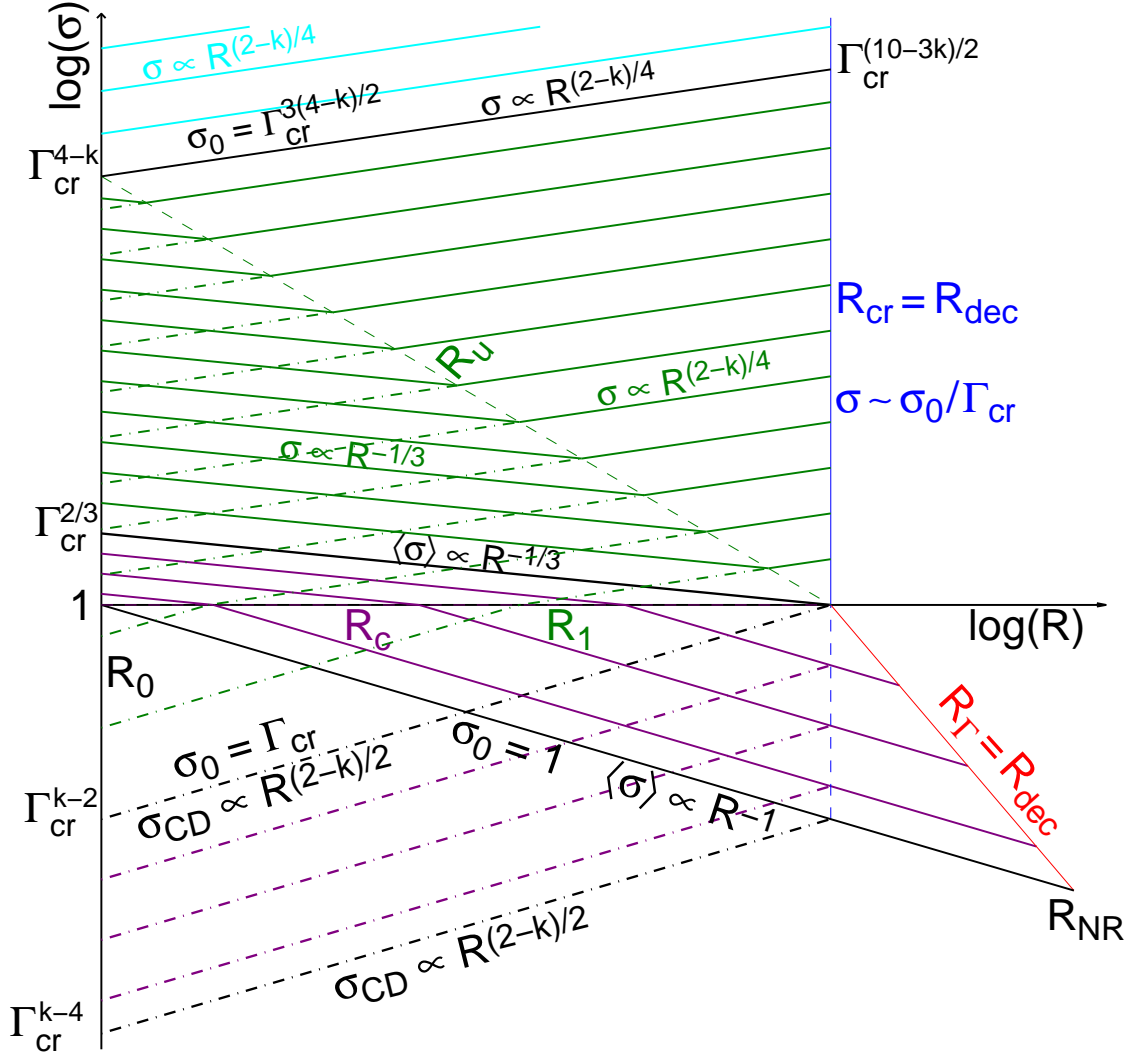}}
\caption{
The same as Fig.~\ref{fig:sigma_R} but with the addition of the
magnetization just behind the contact discontinuity, $\sigma_{\rm CD}$
({\it dashed-dotted lines}), until it becomes similar to the typical
magnetization, $\langle\sigma\rangle$.}
\label{fig:sigma_R2} 
\end{figure}

\begin{figure}
\centerline{\includegraphics[width=15cm]{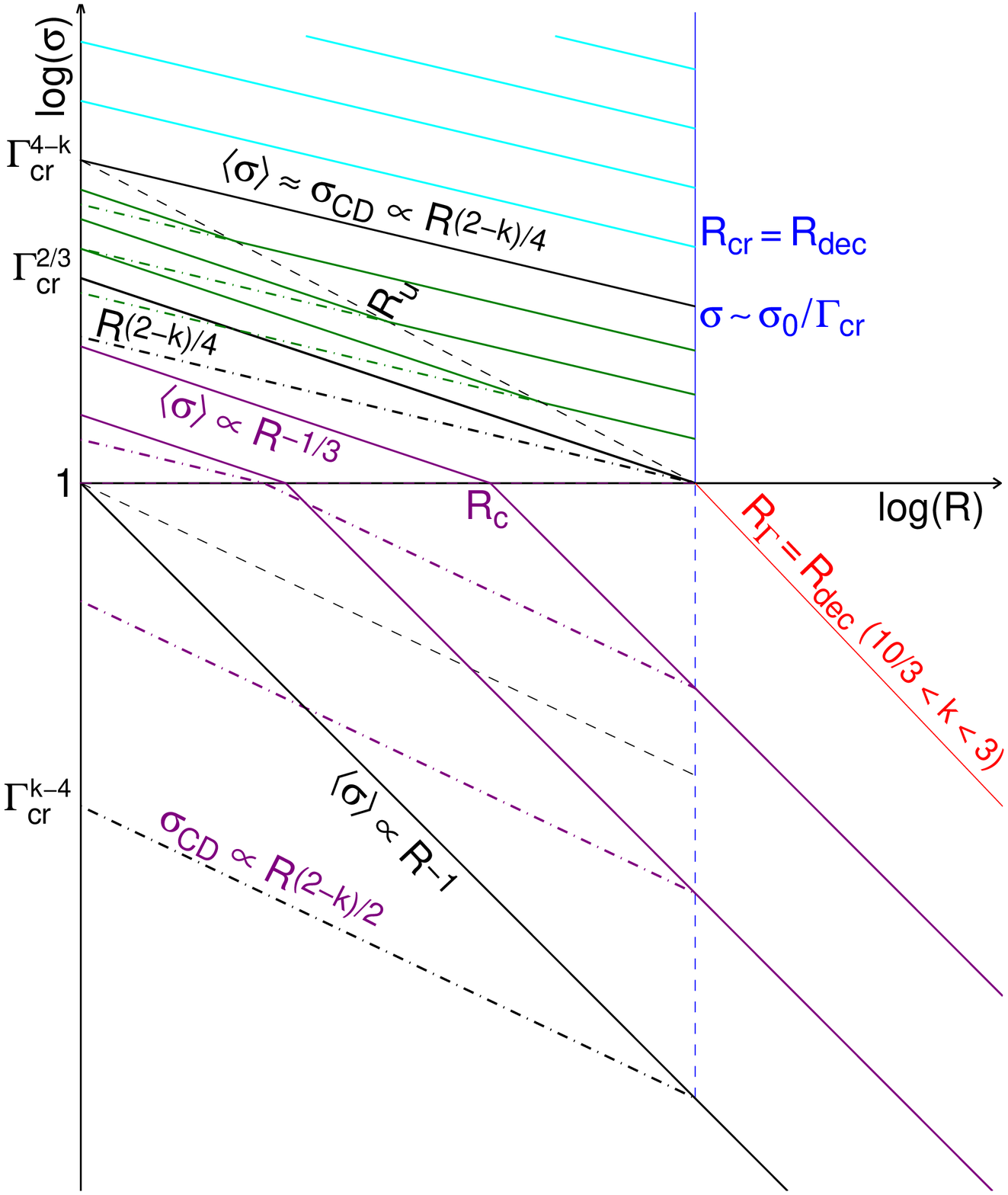}}
\caption{
The same as Fig.~\ref{fig:sigma_R2} but for $2 < k < 10/3$.}
\label{fig:sigma_R2_k2p} 
\end{figure}

In regime I, which corresponds to $f_0
\gg\sigma_0^{7-2k}\Longleftrightarrow\sigma_0 \ll \Gamma_{\rm cr}\sim
(f_0\sigma_0)^{1/(8-2k)}$ or $R_c\ll R_{\rm dec}\sim R_\Gamma$, there
are three main stages in the dynamics of the shell (see
Figs.~\ref{fig:Gamma_R} through
\ref{fig:sigma_R2_k2p}): (i) initially (at $R_0<R<R_c$) the shell accelerates, its
typical Lorentz factor increasing as $\langle\Gamma\rangle \sim
(\sigma_0R/R_0)^{1/3}$ while its typical magnetization decreases as
$\langle\sigma\rangle \sim\sigma_0^{2/3}(R/R_0)^{-1/3}$ (since
magnetic energy is converted into kinetic energy while the total
energy is conserved, $\langle\Gamma\rangle\langle\sigma\rangle
\sim\sigma_0$), (ii) at the coasting radius, $R_c \sim R_0\sigma_0^2$,
the kinetic energy becomes comparable to the magnetic energy,
$\langle\sigma\rangle \sim 1$, so that at $R_c < R < R_{\rm dec}$ most
of the energy is already in kinetic form and the shell coasts at
$\langle\Gamma\rangle \sim
\sigma_0$ while its magnetization decreases as $\langle\sigma\rangle
\sim R_c/R$, (iii) at $R_{\rm dec}\sim (E/\sigma_0^2Ac^2)^{1/(3-k)}
\sim R_\Gamma(\Gamma_0\to\sigma_0)$ most of the energy is transfered
to the shocked external medium\footnote{At $R<R_{\rm dec}\sim
R_\Gamma$ the shocked external medium holds only a small fraction of
the total energy, $E_{\rm ext}/E \sim (R/R_\Gamma)^{3-k} < 1$ (for $k
< 3$ for which the forward shock decelerates).}, and at $R>R_{\rm
dec}$ the flow approaches the BM76 self-similar solution where
$\langle\Gamma\rangle\sim (E/Ac^2)^{1/2}R^{(k-3)/2}$. This is
summarized in the following equation:
\begin{equation}
\langle\Gamma\rangle \sim \left\{\matrix{
\sigma_0(R/R_c)^{1/3} & \quad R_0 < R < R_c \ ,\cr\cr 
\sigma_0 & \quad R_c < R < R_{\rm dec} \ ,\cr\cr 
\sigma_0(R/R_{\rm dec})^{\frac{k-3}{2}} & \quad R > R_{\rm dec}\ ,}\right.
\quad\quad
\langle\sigma\rangle \sim \left\{\matrix{
(R/R_c)^{-1/3} & \quad R_0 < R < R_c \ ,\cr\cr 
(R/R_c)^{-1} & \quad R_c < R < R_{\rm dec} \ .}\right.
\end{equation}
Please note that in regime I, $\Gamma_{\rm CD} \sim \sigma_0$ at $R
>\max(R_0,R_1) \ll R_c \ll R_{\rm dec}$. However, at $R>R_{\rm dec}$
as the original magnetized shell becomes part of the BM76 self-similar
solution its Lorentz factor is $\sim\Gamma_{\rm CD}$ and it decreases
with time along with its magnetization and total energy. As long as it
is relativistically hot and thus part of the BM76 solution, its
Lorentz factor scales as $\Gamma_{\rm CD}\propto R^{(2k-7)/2}\propto
T^{(2k-7)/[4(4-k)]}$ while its magnetization decreases as
$\sigma\propto R^{(2k-9)/2}\propto T^{(2k-9)/[4(4-k)]}$, where $T \sim
R/c\Gamma_{\rm CD}^2$ is the time when radiation from the original
magnetized shell reaches the observer. However, since the reverse
shock is only mildly relativistic the shell's temperature quickly
becomes sub-relativistic and it deviates from the BM76 solution (and
the corresponding scalings above), decelerating more slowly
\citep{KS00}.

In regime I, the typical magnetic pressure in the ejecta shell at $R_c$
is $p_m(R_c) \sim \rho(R_c)c^2 \sim\rho_0c^2\sigma_0^{-5}$ (where
$\rho$ is its typical or average proper density), while the pressure
of the shocked external medium is $p_2(R_c)
\sim\rho_1(R_c)c^2\sigma_0^2=\rho_1(R_0)c^2\sigma_0^{2-2k}$, so that
the typical or average magnetic pressure in the shell is much larger,
$p_m(R_c)/p_2(R_c) \sim f_0/\sigma_0^{7-2k} \gg 1$. However, at larger
radii the two pressure scale as $p_m \propto R^{-4}$ and $p_2
\propto R^{-k}$ so that their ratio drops with radius as $p_m/p_2
\propto R^{k-4}$ and the two pressures become comparable at $R_{\rm
RS}$, where
\begin{equation}\label{eq:R_RS}
R_{\rm RS} \sim R_c\fracb{f_0}{\sigma_0^{7-2k}}^{1/(4-k)} 
\sim R_{\rm cr}\ ,\quad\quad
R_{\rm dec} \sim R_c\fracb{f_0}{\sigma_0^{7-2k}}^{1/(3-k)}\ .
\end{equation}
A strong reverse shock must form at $R\sim R_{\rm RS}$, since at that
stage the magnetic pressure can no longer balance the thermal pressure
of the shocked external medium at the CD, and a new source of pressure
is needed, which comes in the form of thermal pressure that is
generated by the reverse shock that develops and soon becomes
dominant. While a weak reverse shock might develop earlier, at
$R<R_{\rm RS}$ the thermal pressure it generates would be much smaller
than the magnetic pressure, so that it would not have a significant
effect on the dynamics and would dissipate only a small fraction of
the total energy.  The reverse shock is initially Newtonian, until it
becomes mildly relativistic at $R_{\rm dec}$. This can be seen by
balancing the pressure behind the forward shock,
$p_2\sim\rho_1(R)c^2\sigma_0^2\sim \rho_1(R_0)c^2\sigma_0^{2-2k}
(R/R_c)^{-k}$, with the (predominantly thermal at $R>R_{\rm RS}$)
pressure behind the reverse shock, $p_{\rm RS}\sim\rho(R)c^2u_{\rm
RS}^2\sim\rho_0c^2\sigma_0^{-5} (R/R_c)^{-3}u_{\rm RS}^2$, which
implies a reverse shock upstream to downstream relative 4-velocity of
$u_{\rm RS}\sim (R/R_{\rm dec})^{(3-k)/2}$.

This is the familiar ``thin shell'' case for the deceleration of an
unmagnetized initially coasting shell (described in
\S~\ref{sec:low-sigma}). The shell starts spreading significantly (in
the lab frame) at $R_c \sim R_0\Gamma^2(R_c)\sim R_0\sigma_0^2$,
resulting in the formation of a reverse shock that becomes thermal
pressure dominated around $R_{\rm RS}$, and gradually strengthens
until it becomes mildly relativistic near its shell crossing radius,
which is the deceleration radius, $R_{\rm dec}\sim
(E/\sigma_0^2Ac^2)^{1/(3-k)}$. Near $R_{\rm dec}$, where most of the
energy is given to the shocked external medium, and where the reverse
shock crosses most of the shell, the typical magnetization of the
shell is low, $\langle\sigma\rangle \sim R_c/R_{\rm dec} \sim
(\sigma_0/\Gamma_{\rm cr})^{2(4-k)/(3-k)} \sim
(\sigma_0^{7-2k}/f_0)^{1/(3-k)} \ll 1$ (where I have identified
$\Delta_0$ in Eq.~[\ref{eta_cr}] with $R_0$). Note that this regime
corresponds to $\Gamma(R_c)\sim\sigma_0\ll\Gamma_{\rm cr}$, which can
also be expressed as $\Gamma_{\rm cr}/\sigma_0 \sim
(f_0\sigma_0^{2k-7})^{1/(8-2k)}\sim\Upsilon_0^{(3-k)/(8-2k)}\gg 1$
where in the expression for $\Upsilon_0$ (Eq.~[\ref{upsilon}]) one
substitutes $\Delta_0\to R_0$ and $\Gamma_0\to\sigma_0$, thus clearly
corresponding to the unmagnetized (or low magnetization) thin shell
case. A larger magnetic field downstream (and also somewhat upstream)
of the reverse shock is possible due to magnetic field amplification
in the reverse shock itself, which may allow for a reasonable
radiative efficiency coupled to the rather effective energy
dissipation in the mildly relativistic reverse shock.

Altogether, I find that $R_{\rm *,CD}\sim R_u$, $R_{\rm RS} \sim R_{\rm
cr}$, $R_{\rm dec}\sim R_\Gamma$, and
\begin{equation}
\fracb{R_u}{R_c}^{\frac{14-3k}{12-3k}} \sim \frac{R_{\rm cr}}{R_c} \sim
\fracb{R_{\rm dec}}{R_c}^\frac{3-k}{4-k} \sim \fracb{R_1}{R_c}^\frac{2-k}{4-k}
\sim \fracb{\Gamma_{\rm cr}}{\sigma_0}^2
\sim \fracb{f_0}{\sigma_0^{7-2k}}^{1/(4-k)} \gg 1\ .
\end{equation}
The ordering of the relevant critical radii in the different regimes
are given in Tables~\ref{tab:regimes} and \ref{tab:regimes_k>10/3}.
In regime I with $k < 2$ or with $2 < k < 10/3$ and $f_0 > \sigma_0^3$
we have $R_0 < R_c < R_u\sim R_{\rm *,CD} < R_{\rm RS}\sim R_{\rm cr}
< R_{\rm dec} \sim R_\Gamma$ while for $2 < k < 10/3$ and
$\sigma_0^{7-2k}< f_0 < \sigma_0^3$ we also have the critical radius
$R_1$ so that $R_0 < R_1 < R_c < R_u\sim R_{\rm *,CD} < R_{\rm RS}\sim
R_{\rm cr} < R_{\rm dec} \sim R_\Gamma$.

\subsection{Regime II}
\label{sec:regII}

This regime corresponds to $\sigma_0^{1/3} \ll f_0\ll\sigma_0^{7-2k}$
$\Longleftrightarrow$ $\sigma_0^{2/(12-3k)} \ll \Gamma_{\rm
cr}\ll\sigma_0$ $\Longleftrightarrow$ $\Gamma_{\rm
cr}\ll\sigma_0\ll\Gamma_{\rm cr}^{2/(12-3k)}$, where the condition
$\Gamma_{\rm cr} \ll\sigma_0$ corresponds to $R_{\rm dec}\sim R_{\rm
cr} \sim R_0\Gamma_{\rm cr}^2 \ll R_0\sigma_0^2 \sim R_c$. As we shall
see below, this also implies that $\sigma_0^{1/3}\ll\Gamma_{\rm
CD}(R_u)\ll\sigma_0$ and $1\ll\sigma_u(R_u)\ll\sigma_0^{2/3}$.

For $k < 2$, $\sigma_u$ increases with radius, and since in regime II
it is larger than 1 at $R_u$, it passes through the value of 1 at a
smaller radius $R_1$ that is given by $R_1/R_c \sim
(f_0/\sigma_0^{7-2k})^{1/(2-k)}\ll 1$, and the ordering of the
critical radii is $R_1 < R_u < R_{\rm cr} < R_c$. As in regime I, also
here in regime II, $R_1$ is physically interesting only if $R_1>R_0$,
which now corresponds to $\sigma_0^3 < f_0 < \sigma_0^{7-2k}$. In this
parameter range $\sigma_u \approx \sigma_{\rm CD}$ increases with
radius as $\sigma_u\approx \sigma_{\rm CD} \sim (R/R_1)^{(2-k)/2} < 1$
at $R_0 < R < R_1$ and as $\sigma_u\approx\sigma_{\rm CD}\sim
(R/R_1)^{(2-k)/4} > 1$ at $R_1 < R < R_{\rm cr}$. For $\sigma_0^{1/3}
< f_0 < \sigma_0^3$ we have $R_1 < R_0$, and $\sigma_u
\approx\sigma_{\rm CD}\sim (R/R_1)^{(2-k)/4} \gg 1$ all
along. Altogether, for $k<2$ we have $\sigma_u\approx \sigma_{\rm CD}
\sim (R/R_1)^{(2-k)/2} < 1$ at $R_0 < R <\min(R_0,R_1)$ and as
$\sigma_u\approx\sigma_{\rm CD}\sim (R/R_1)^{(2-k)/4} > 1$ at
$\min(R_0,R_1) < R < R_{\rm cr}$.  For $2 < k < 10/3$, on the other
hand, $R_1/R_c \sim (R_1/R_{\rm cr})^{(4-k)/2}
(f_0/\sigma_0^{7-2k})^{1/(2-k)}\gg 1$ so that $R_1 > R_c > R_{\rm cr}$
and $\sigma_u \approx \sigma_{\rm CD}\sim (R/R_1)^{(2-k)/4} \gg
1$ all along.

For $k = 2$ we have a self-similar solution for the rarefaction wave,
thanks to the equivalence of the cold MHD equations for a spherical
flow to those for a planar flow, which make it easier to
explicitly calculate much of the relevant dynamics. For a general
value of $k$ we do not have this privilege, and I have relied on the
approximation that this self-similar solution still approximately
holds in this case where $\Gamma_{\rm CD} = \Gamma(\xi_u)$ and $\xi_u$
gradually change with time. In order to further justify this, I now
provide an alternative derivation of Eq.~(\ref{eq:regII}).  The
pressure balance at the CD reads $(B_{\rm CD}/\Gamma_{\rm CD})^2/8\pi
\approx a(4/3)\Gamma_{\rm CD}^2\rho_1c^2$, implying
\begin{equation}\label{eq:Gamma_CD2}
\Gamma_{\rm CD} \approx \left(\frac{3 L_{\rm CD}}{32a\pi
Ac^3}\right)^{1/4}R^{(k-2)/4}\ ,
\end{equation}
where $L_{\rm CD} \approx c B_{\rm CD}^2 R_{\rm CD}^2$ is the
instantaneous Poynting flux through a static spherical surface at
$r=R_{\rm CD}$. Note that $L_{\rm CD}$ is close to the mean (isotropic
equivalent) luminosity (or power) of the source, $L \approx Ec/2R_0
\approx\pi\rho_0\sigma_0c^3R_0^2$ [identifying the initial width of
the shell $\Delta_0$ with its initial radius $R_0$, where the shell
initially occupies the region $0<r\leq R_0$, while $E\approx E_{\rm
EM,0}=2\pi\rho_0\sigma_0c^2R_0^3$ and $\rho_0=\rho(t=0,r=R_0)$], only
where the magnetization parameter just before the CD is large,
$\sigma_{\rm CD}\gg 1$, which corresponds to $\Gamma_{\rm
CD}\ll\sigma_0$. In this case $L_{\rm CD}\approx L$ and we have
\begin{eqnarray}\nonumber
\Gamma_{\rm CD} \approx \left(\frac{3 L R^{k-2}}{32a\pi Ac^3}\right)^{1/4} 
\approx \left(\frac{3ER^{k-2}}{64a\pi Ac^2R_0}\right)^{1/4} \approx
\fracb{3f_0\sigma_0}{32a}^{1/4}\fracb{R}{R_0}^{(k-2)/4}
\\ \label{eq:Gamma_CD_II}
\sim \Gamma_{\rm cr}^{(4-k)/2}\left(\frac{R}{R_0}\right)^{(k-2)/4}
\sim \Gamma_{\rm cr}\left(\frac{R}{R_{\rm cr}}\right)^{(k-2)/4}
\sim\sigma_0\fracb{R}{R_1}^{(k-2)/4}\ ,
\end{eqnarray}
\begin{equation}\label{eq:sigma_CD_II}
\sigma_{\rm CD} = \sigma_u = \sigma_0\tilde{\rho}_u \approx 
\frac{\sigma_0}{2\Gamma_{\rm CD}} \approx 
\fracb{2a\sigma_0^3}{3f_0}^{1/4}\fracb{R}{R_0}^{(2-k)/4}
\approx\fracb{R}{R_1}^{(2-k)/4}\gg 1\ .
\end{equation}
This is valid as long as the value of the lab-frame magnetic field $B$
at the CD (i.e. the head of the outflow) is close to its original
value, i.e. for $\sigma_{\rm CD} = \sigma_u\gg 1$, which holds at
$\max(R_0,R_1) < R < R_{\rm dec}$.

The condition that $f_0 \gg \sigma_0^{1/3}$ in regime II implies that
$\Gamma_{\rm CD}(R_0) \sim \Gamma_{\rm cr}^{(4-k)/2} \sim
(f_0\sigma_0)^{1/4} \gg \sigma_0^{1/3}$, and therefore at $t_0$ region
4 (see Fig.~\ref{fig:RW-diagram}) holds most of the volume and energy,
and $\langle\Gamma\rangle \sim
\sigma_0^{1/3}(R/R_0)^{1/3} \approx (\sigma_0t/t_0)^{1/3}$ at 
$t \geq t_0$.  At this stage the typical or mean value (weighted
average over the energy in the lab frame) of the Lorentz factor within
the magnetized shell, $\langle\Gamma\rangle$, increases with time,
while for $k=2$ the Lorentz factor of the uniform region at its front,
$\Gamma(\xi_u) = \Gamma_3 = \Gamma_{\rm CD}$, remains constant. More
generally, $\Gamma_{\rm CD}(R_0\leq R\leq R_{\rm cr})$ is given by the
smaller between the expression in Eq.~(\ref{eq:Gamma_CD_II}) and
$2\sigma_0$. This acceleration (increase in $\langle\Gamma\rangle$)
lasts until the secondary (or reflected) rarefaction wave finishes
crossing region 4, i.e. until $\xi_* \approx 1-2R_0/ct\approx
1-2R_0/R$ equals
\begin{equation}
\xi_u = \frac{\beta_{\rm CD}-\beta_{\rm ms}(\beta_{\rm CD})}
{1-\beta_{\rm CD}\beta_{\rm ms}(\beta_{\rm CD})}
\approx \frac{1-(\Gamma_{\rm ms}/\Gamma_{\rm CD})^2}{1+(\Gamma_{\rm ms}/\Gamma_{\rm CD})^2}
\approx 1-\frac{\sigma_0}{\Gamma_{\rm CD}^3} 
\approx 1 - \fracb{32a\sigma_0^{1/3}}{3f_0}^{3/4}\fracb{R}{R_0}^\frac{3(2-k)}{4}\ ,
\end{equation}
(where since $\Gamma_{\rm ms} \gg 1$, we have $\Gamma_{\rm ms}^2 \approx
u_{\rm ms}^2 = \sigma = \sigma_0\tilde{\rho} \approx \sigma_0/2\Gamma$
and specifically $\Gamma_{\rm ms}^2(\beta_{\rm CD}) \approx
\sigma_0/2\Gamma_{\rm CD}$), at $R = R_u$, which corresponds to
\begin{equation}\label{eq:R_u2}
\frac{R_u}{R_0} \approx 
\fracb{3f_0}{2^{11/3}a\sigma_0^{1/3}}^\frac{3}{10-3k}\ ,\quad
\langle\Gamma\rangle(R_u) \sim \left(\frac{\sigma_0R_u}{2R_0}\right)^{1/3} \approx
\fracb{3f_0\sigma_0^{3-k}}{2^{7-k}a}^\frac{1}{10-3k} \approx \Gamma_{\rm CD}(R_u)\equiv\Gamma_u\ .
\end{equation}
This implies
\begin{equation}
\frac{R_u}{R_c} \approx 
\fracb{3f_0}{2^{11/3}a\sigma_0^{7-2k}}^\frac{3}{10-3k}
\approx \frac{1}{4\sigma_u^3(R_u)}\ll 1\ ,
\end{equation}
which is different from the result for regime I (see
Eq.~[\ref{eq:Ru_Rc}]), where $R_u/R_c
\approx 2\sigma_u^{-3/4}(R_u)\gg 1$.

At this stage ($R=R_u$ or $t=t_u$) most of the energy in the flow is
in\footnote{This can be seen as follows for $k=2$. The pressure is
continuous across the CD, and therefore the energy density of regions
2 and 3 in the lab frame is similar, and their relative energy is
determined by their relative width in the lab frame. For region 2,
using the uniform velocity approximation $\xi_{\rm sh} - \xi_{\rm CD}
=\beta_{\rm sh} - \beta_{\rm CD} \approx 1/2\Gamma_{\rm
CD}^2-1/2\Gamma_{\rm sh}^2 \approx 1/4\Gamma_{\rm CD}^2$ (for the BM76
solution $\xi_{\rm sh} -\xi_{\rm CD} \approx (1-\chi_{\rm
CD}^{-1})/2\Gamma_{\rm CD}^2 \approx 1/4.60\Gamma_{\rm CD}^2$, which
is rather similar), while for region 3, $\xi_{\rm CD}-\xi_u\approx
\sigma_0/\Gamma_{\rm CD}^3-1/2\Gamma_{\rm CD}^2 =
(2\sigma_0/\Gamma_{\rm CD}-1)/2\Gamma_{\rm CD}^2$, and therefore the
width of region 3, $\Delta_3$, is larger than that of region 2,
$\Delta_2$, by a factor of $\Delta_3/\Delta_2 \approx
2(2\sigma_0/\Gamma_{\rm CD}-1) \gg 1$, since $\Gamma_{\rm CD} \ll
\sigma_0$ in this regime.} region 3, which moves with $\Gamma_3 \approx
\Gamma_{\rm CD}$ given by Eq.~(\ref{eq:Gamma_CD_II}), that represents 
$\langle\Gamma\rangle$ at this stage ($R_u<R<R_{\rm *,CD}$). Region 3
is gradually crossed by the right going rarefaction wave, until it
reaches the CD at $R_{\rm *,CD}\sim R_{\rm cr} \sim R_{\rm dec}$ (as
shown in detail below), which marks the end of this stage. At that
point most of the energy is in the shocked external medium\footnote{At
$R_u<R<R_{\rm dec}\sim R_{\rm cr}$ only a small fraction of the total
energy is in the shocked external medium, $E_{\rm ext}/E \sim
(R/R_{\rm cr})^{(4-k)/2}$.}, and the flow approaches the BM76
self-similar solution (first the rarefaction wave crosses region 2,
within a few dynamical times,\footnote{For $k=2$, making the
approximation that the region between the CD and shock front has the
constant velocity of the CD and that $\Gamma_{\rm sh}
=\sqrt{2}\Gamma_{\rm CD}$ and using Eq.~(\ref{eq:width}) one obtains
that during the time the rarefaction wave travels from the CD to the
shock front the radius increases by a factor of $1 + (1-\chi_{\rm
CD}^{-1})2(\sqrt{3}+1)/(3-\sqrt{3})\approx 2.87$. If we
self-consistently use the above assumption to estimate the width of
this region (even though this is not fully self-consistent) this gives
$(R_{\rm sh}-R_{\rm CD})/R_{\rm CD}
\approx 1/4\Gamma_{\rm CD}$ instead of Eq.~(\ref{eq:width}), and a
growth in radius during the rarefaction crossing by a factor of
$1+(\sqrt{3}+1)/(3-\sqrt{3}) \approx 3.15$. In both cases it is close
to a factor of $\sim 3$. This factor is relatively large since the
sound speed in region 2 in ``only'' $c_s\approx c/\sqrt{3}$ (as it is
unmagnetized but relativistically hot, while regions 3 and 4 are cold
but highly magnetized) and the shock front moves somewhat faster than
the fluid in region 2.} and then the adiabatic BM76 self-similar
solution is quickly approached).

The width of region 3 at $t_u$ (when the rarefaction wave reaches
$\xi_u$) in the lab frame is $\Delta_3 = ct_u(\xi_{\rm CD}-\xi_u)
\approx R_u\sigma_0\Gamma_{\rm CD}^{-3}(R_u)
\approx 2R_0$. In region 3,
\begin{equation}
\beta_* = \frac{\beta_{\rm CD}+\beta_{\rm ms}(\beta_{\rm CD})}
{1+\beta_{\rm CD}\beta_{\rm ms}(\beta_{\rm CD})}
\approx 1-\frac{1}{8\Gamma_{\rm CD}^2\Gamma_{\rm ms}^2(\beta_{\rm CD})}
\approx 1-\frac{1}{4\sigma_0\Gamma_{\rm CD}}\ ,
\end{equation}
so that $(1-\beta_*) \ll (1-\beta_{\rm CD}) \approx 1/2\Gamma_{\rm
CD}^2$ and therefore $\Delta v = (\beta_*-\beta_{\rm CD})c \approx c
/2\Gamma_{\rm CD}^2$ and the increase in radius, $\Delta R = R_{\rm
*,CD}-R_u$ (or time, $\Delta t=t_{\rm *,CD}-t_u\approx\Delta R/c$),
during the time it takes the rarefaction wave to cross region 3 is
$\Delta R\approx 2R_0c/\Delta v \approx 4R_0\Gamma_{\rm CD}^2 \sim
R_{\rm dec}$ for $k=2$, while more generally
\begin{eqnarray}\nonumber
2R_0\approx \int_{t_u}^{t_{\rm *,CD}}dt\,\Delta v \approx 
\int_{R_u}^{R_{\rm *,CD}}\frac{dR}{2\Gamma_{\rm CD}^2(R)} =
\frac{R_u}{2\Gamma_{\rm CD}^2(R_u)}\int_1^{R_{\rm *,CD}/R_u}d\tilde{R}\,\tilde{R}^\frac{2-k}{2}\ ,
\quad\quad
\\ \nonumber
\Longrightarrow\quad
\frac{2}{(4-k)}\left[\left(1+\frac{\Delta R}{R_u}\right)^{(4-k)/2}-1\right] \approx
\frac{4R_0\Gamma_{\rm CD}^2(R_u)}{R_u} \approx \frac{2\sigma_0}{\Gamma_{\rm CD}(R_u)}
\approx 4\sigma_{\rm CD}(R_u) \gg 1\ ,
\\ \label{eq:DR}
\Longrightarrow\quad \frac{\Delta R}{R_u} \approx 
\left[2(4-k)\sigma_{\rm CD}(R_u)\right]^{2/(4-k)} \sim
\fracb{R_c}{R_u}^{2/(12-3k)} \sim \frac{R_{\rm cr}}{R_u}\ ,
\quad\quad\quad
\end{eqnarray}
so that the rarefaction reaches the CD at $R_{\rm *,CD} \approx \Delta
R\sim R_{\rm cr} \sim R_{\rm dec}$. 

The deceleration radius in this regime can be obtained by equating the
initial magnetic energy to the energy of the swept-up external medium,
$E\approx 2\pi\rho_0\sigma_0c^2R_0^3 \approx [4\pi/(3-k)]Ac^2R_{\rm
dec}^{3-k}\Gamma_{\rm CD}(R_{\rm dec})^2$, which implies
\begin{equation}
\frac{R_{\rm dec}}{R_0} \approx 
\fracsb{(3-k)^2 8af_0\sigma_0}{3}^{1/(4-k)}
\sim \Gamma_{\rm CD}^2(R_{\rm dec}) \sim \Gamma_{\rm cr}^2 \sim \frac{R_{\rm cr}}{R_0}\ ,
\end{equation}
and therefore $R_{\rm dec} \approx R_{\rm cr}$
where $R_{\rm cr}$ is the radius at which $\Gamma_{\rm
CD} = \Gamma_{\rm BM}$, and
\begin{equation}\label{Gamma_BM}
\Gamma_{\rm BM} \approx 
\left[\frac{(3-k)E}{4\pi Ac^2}\right]^{1/2}R^{-(3-k)/2}\ ,
\end{equation}
is the typical Lorentz factor during the subsequent constant energy
self-similar (BM76) stage. Estimating the value of $R_{\rm cr}$ from
Eqs.~(\ref{eq:Gamma_CD_II}) and (\ref{Gamma_BM}) and identifying $R_0$
with $\Delta_0$ gives,
\begin{equation}\label{R_cr_acc}
R_{\rm cr} \approx 
\left[\frac{4(3-k)^2 a E\Delta_0}{3\pi Ac^2}\right]^{1/(4-k)} 
\approx \left\{\matrix{9.3\times 10^{16}a^{1/4}\zeta^{-1/4}n_0^{-1/4}
E_{53}^{1/4}T_{30}^{1/4}\;{\rm cm} & (k = 0)\ ,\cr\cr
5.3\times 10^{15}a^{1/2}\zeta^{-1/2}A_*^{-1/2}
E_{53}^{1/2}T_{30}^{1/2}\;{\rm cm} & (k = 2)\ ,}\right.
\end{equation}
or
\begin{equation}\label{R_cr2}
\frac{R_{\rm cr}}{R_0} \sim \frac{1}{R_0}\fracb{E R_0}{Ac^2}^{1/(4-k)}
\sim \left(f_0\sigma_0\right)^{1/(4-k)} \sim \Gamma_{\rm cr}^2\ .
\end{equation}

During the initial acceleration (at $R > R_0$), $\langle\Gamma\rangle
\sim (\sigma_0R/R_0)^{1/3}$. This lasts until most of 
the energy is transfered to the part of the magnetized shell with
$\Gamma \sim \Gamma_{\rm CD}$, which occurs at a radius $R_u$, Lorentz
factor $\Gamma_u$, and magnetization $\sigma_u(R_u)$ given by
\begin{equation}
\frac{R_u}{R_0} \sim 
\left(f_0\sigma_0^{-1/3}\right)^{3/(10-3k)}\ ,
\quad\quad
\Gamma_u \sim \frac{\sigma_0}{\sigma_u(R_u)} \sim
\sigma_0^{1/3}\left(f_0\sigma_0^{-1/3}\right)^{1/(10-3k)}\ .
\end{equation}
Moreover,
\begin{equation}\label{eq:Gamma_u}
\Gamma_u\sim \Gamma_{\rm cr}\fracb{\Gamma_{\rm cr}}{\sigma_0}^{(k-2)/(10-3k)}
\sim\sigma_0\fracb{\Gamma_{\rm cr}}{\sigma_0}^{2(4-k)/(10-3k)}\ ,
\end{equation}
so that near the transition to regime I, $\Gamma_u \sim \Gamma_{\rm
cr} \sim \sigma_0$ and $R_u \sim R_{\rm cr} \sim R_c$.

In Regime II we have $1\ll
f_0\sigma_0^{-1/3}\ll\sigma_0^{2(10-3k)/3}$, which corresponds to
$1\ll R_u/R_0 \ll\sigma_0^2$ (i.e. $R_0\ll R_u\ll R_c$),
$\sigma_0^{1/3} \ll \Gamma_u\ll \sigma_0$ and $1 \ll
\sigma_u(R_u)\ll\sigma_0^{2/3}$. The different critical radii are related by
\begin{equation}
\fracb{R_u}{R_{\rm cr}}^\frac{10-3k}{2} \sim\
\fracb{R_1}{R_u}^\frac{(2-k)(10-3k)}{4(4-k)} \sim 
\fracb{R_1}{R_c}^\frac{2-k}{4-k} \sim 
\fracb{R_u}{R_c}^{\frac{10-3k}{12-3k}} \sim 
\frac{R_{\rm cr}}{R_c} \sim \fracb{\Gamma_{\rm cr}}{\sigma_0}^2\ll 1\ ,
\end{equation}
so that for $2<k<10/3$ or for $k<2$ and $\sigma_0^{1/3} < f_0 <
\sigma_0^3$ we have $R_1 < R_0 < R_u < R_{\rm cr} \sim R_{\rm dec}
\sim R_{\rm *,CD} < R_c$ and $R_1$ is irrelevant (as $\sigma_u \gg 1$
all along), while for $k<2$ and $\sigma_0^3 < f_0 < \sigma_0^{7-2k}$
we have $R_0 < R_1 < R_u < R_{\rm cr} \sim R_{\rm dec} \sim R_{\rm
*,CD} < R_c$ and $R_1$ is relevant. In all cases $R_c$ is not
relevant physically (since it looses its meaning as a coasting
radius). 

The typical magnetization of the shell in the
intermediate stage is
\begin{equation}\label{eq:sigma_int}
\langle\sigma\rangle(R_u < R < R_{\rm cr}) \sim \sigma_{\rm CD} \sim
\frac{\sigma_0}{\Gamma_{\rm CD}} \sim
\frac{\sigma_0}{\Gamma_u}\left(\frac{R}{R_u}\right)^{(2-k)/4} \sim 
\frac{\sigma_0}{\Gamma_{\rm cr}}\left(\frac{R}{R_{\rm cr}}\right)^{(2-k)/4}\ ,
\end{equation}
so that at $R_{\rm dec} \sim R_{\rm cr}$ we have $\langle\sigma\rangle
\sim \sigma_0/\Gamma_{\rm cr} \gg 1$. Thus, altogether in regimes I and II we have
\begin{equation}\label{sigma_R_dec}
\langle\sigma\rangle(R_{\rm dec}) \sim 
\left\{\matrix{(\sigma_0/\Gamma_{\rm cr})^{2(4-k)/(3-k)} & 
\sigma_0 < \Gamma_{\rm cr} \ \ {\rm (regime\ I)}\ , \cr\cr
\sigma_0/\Gamma_{\rm cr} & \sigma_0 > \Gamma_{\rm cr} \ \ {\rm (regime\ II)}\ , }\right.
\end{equation}
while in regime II we have
\begin{equation}\label{gamma_regime_II}
\langle\Gamma\rangle(R) \sim 
\left\{\matrix{(\sigma_0R/R_0)^{1/3} \sim \Gamma_u(R/R_u)^{1/3} & R_0 < R < R_u\ , \cr\cr 
\Gamma_{\rm cr}(R/R_{\rm cr})^{(k-2)/4} & R_u < R < R_{\rm cr}\ , \cr\cr
\Gamma_{\rm cr}(R/R_{\rm cr})^{(k-3)/2} & R > R_{\rm cr}\ , }\right.
\end{equation}

\begin{equation}\label{gamma_CD_regime_II}
\Gamma_{\rm CD}(R) \sim 
\left\{\matrix{\sigma_0\ \ & R_0 < R < \max(R_0,R_1)\ , \cr\cr 
\sigma_0(R/R_1)^{(k-2)/4} \sim 
\Gamma_{\rm cr}(R/R_{\rm cr})^{(k-2)/4} & \max(R_0,R_1) < R < R_{\rm cr}\ ,}\right.
\end{equation}

\begin{equation}\label{sigma_regime_II}
\langle\sigma\rangle(R) \sim 
\left\{\matrix{\sigma_0^{2/3}(R/R_0)^{-1/3} \sim (\sigma_0/\Gamma_u)(R/R_u)^{-1/3}
& R_0 < R < R_u\ , \cr\cr
(\sigma_0/\Gamma_u)(R/R_u)^{(2-k)/4} \sim (\sigma_0/\Gamma_{\rm cr})(R/R_{\rm cr})^{(2-k)/4} 
& R_u < R < R_{\rm cr}\ .}\right.
\end{equation}

\begin{equation}\label{sigma_CD_regime_II}
\sigma_{\rm CD}(R) \sim 
\left\{\matrix{(R/R_1)^{(2-k)/2} \sim (\sigma_0^3/f_0)^{1/2}(R/R_0)^{(2-k)/2}
& R_0 < R < \max(R_0,R_1)\ , \cr\cr
(R/R_1)^{(2-k)/4} \sim (\sigma_0/\Gamma_{\rm cr})(R/R_{\rm cr})^{(2-k)/4} 
& \max(R_0,R_1) < R < R_{\rm cr}\ .}\right.
\end{equation}

\subsection{Regime III}
\label{sec:regIII}

In regime II we had $\Gamma_{\rm CD}(R_0) \gg\sigma_0^{1/3}$ so that
the plasma near the CD was super-fast-magnetosonic with respect to the
``wall'' already at $R \sim R_0$, with $\Gamma_{\rm CD}/\Gamma_{\rm
ms} \approx (2\Gamma_{\rm CD}^3/\sigma_0)^{1/2}\gg 1$, and thus not in
causal contact with the source. Here, in regime III, we consider what
happens when $1 \ll \Gamma_{\rm CD}(R_0) \ll\sigma_0^{1/3}$.
In all the regions behind the CD, the fast magnetosonic Lorentz factor
is given by $\Gamma_{\rm ms}^2\approx u_{\rm ms}^2 =\sigma \geq
\sigma_{\rm CD}\approx\sigma_0/2\Gamma_{\rm CD}\gg 1$, so that as long
as $\Gamma_{\rm CD}\ll \sigma_0^{1/3}$ the flow remains in causal
contact with the ``wall'' or central source, $\Gamma/\Gamma_{\rm ms} <
(2\Gamma_{\rm CD}^3/\sigma_0)^{1/2}\ll 1$.  Thus, the flow
remains roughly uniform and the conditions just behind the CD are
representative of the typical values in the shell,
$\langle\Gamma\rangle(R_0)\sim\Gamma_{\rm CD}(R_0)$, implying
\begin{eqnarray}\label{Gamma_R0_III}
\langle\Gamma\rangle(R_0) &\sim& \Gamma_{\rm cr}^{(4-k)/2} \sim 
\sigma_0^{1/3}\left(f_0\sigma_0^{-1/3}\right)^{1/4}
= \left(f_0\sigma_0\right)^{1/4} \ ,\quad
\\ \label{sigma_R0_III}
\langle\sigma\rangle(R_0) &\sim& \sigma_0\Gamma_{\rm cr}^{(k-4)/2}
\sim \sigma_0^{2/3}\left(f_0\sigma_0^{-1/3}\right)^{-1/4}
=\sigma_0\left(f_0\sigma_0\right)^{-1/4} \ , 
\end{eqnarray}
so that the conditions near $R_{\rm cr}$ are very similar to those in
regime II: $R_{\rm dec}\sim R_{\rm cr} \sim R_0\Gamma_{\rm cr}^2$,
$\langle\Gamma\rangle(R_{\rm cr})\sim \Gamma_{\rm cr}$,
$\langle\sigma\rangle(R_{\rm cr})\sim\langle\sigma\rangle(R_0)(R_{\rm
cr}/R_0)^{(2-k)/4}\sim\sigma_0/\Gamma_{\rm cr}$, and $\Gamma_{\rm
CD}(R_{\rm dec})\sim\langle\Gamma\rangle(R_{\rm dec}) \sim\Gamma_{\rm
cr}$.  This implies that regime III, defined above through
the condition $1 \ll \Gamma_{\rm CD}(R_0) \ll\sigma_0^{1/3}$,
corresponds to $\sigma_0^{-1}\ll f_0\ll\sigma_0^{1/3}$,
$1\ll\Gamma_{\rm cr}\ll\sigma_0^{2/(12-3k)}$ or
$\sigma_0\gg\Gamma_{\rm cr}^{(12-3k)/2}\gg 1$.
 
In this regime $\Gamma_{\rm CD}$ is still given by
Eq.~(\ref{eq:Gamma_CD2}) while
\begin{equation}
\xi_u = \frac{\beta_{\rm CD}-\beta_{\rm ms}(\beta_{\rm CD})}
{1-\beta_{\rm CD}\beta_{\rm ms}(\beta_{\rm CD})}
\approx -\frac{1-(\Gamma_{\rm CD}/\Gamma_{\rm ms})^2}{1+(\Gamma_{\rm CD}/\Gamma_{\rm ms})^2}
\approx -\frac{1-\frac{2\Gamma_{\rm CD}^3}{\sigma_0}}{1+\frac{2\Gamma_{\rm CD}^3}{\sigma_0}}
\approx \frac{4\Gamma_{\rm CD}^3}{\sigma_0}-1 \sim -1\ ,
\end{equation}
so that region 3 initially occupies most of the volume,
$\Delta_3(t<t_u)/ct =\xi_{\rm CD}-\xi_u \approx 2 - 1/2\Gamma_{\rm
CD}^2+4\Gamma_{\rm CD}^3/\sigma_0\sim 2$. This demonstrates again that
already at $t=t_0$ the conditions just behind the CD (region 3)
dominate the average values over the original magnetized shell,
so that $\mean{\Gamma}\approx\Gamma_{\rm CD}$ and
$\mean{\sigma}\approx\sigma_{\rm CD}$ are given by
Eqs.~(\ref{eq:Gamma_CD_II}) and (\ref{eq:sigma_CD_II}),
respectively. Region 4 occupies only a small fraction of the total
volume already at $t=t_0$, $\Delta_4(t_0)/ct_0 =\xi_u+\beta_{\rm
ms,0}\approx 4\Gamma_{\rm CD}^3/\sigma_0\ll 1$, and it is very quickly
crossed by the right-going rarefaction wave, which reaches region 3
($\xi_* = \xi_u$) at $t = t_u$ (and $R=R_u$) that corresponds to
\begin{equation}
\frac{t_u}{t_0}-1 \approx \frac{\Delta_4(t_0)}{[\beta_*(t_0)-\xi_u]ct_0}
\approx \frac{\Delta_4(t_0)}{2ct_0} \approx \frac{2\Gamma_{\rm CD}^3}{\sigma_0}\ll 1\ .
\end{equation}
This implies that $R_u \approx R_0$ and the time since $t_0$ when the
right-going rarefaction wave reaches the CD ($t_{\rm CD}-t_0$) is
dominated by its propagation time through region 3. For $k=2$, 
\begin{equation}
\frac{t_{\rm CD}}{t_0}-1 \approx 
\frac{\Delta_3(t_0)}{[\beta_*(\beta_{\rm CD})-\beta_{\rm CD}]ct_0}
\approx  4\Gamma_{\rm CD}^2 \gg 1\ ,
\end{equation}
so that it reaches the CD at a radius $R_{\rm *,CD} \approx ct_{\rm
CD} \approx 4\Gamma_{\rm CD}^2R_0 \sim R_{\rm cr} \sim R_{\rm
dec}$. Similarly, since in this case $R_u\approx R_0$ then for a
general $k$-value Eq.~(\ref{eq:DR}) implies $R_{\rm
*,CD}/R_0\approx\Delta R/R_0\approx [2(4-k)\Gamma_{\rm
CD}^2(R_0)]^{2/(4-k)} \sim \Gamma_{\rm cr}^2 \sim R_{\rm cr}/R_0$, so
that again, $R_{\rm *,CD} \sim R_{\rm cr} \sim R_{\rm dec}$.

The effect of the external medium in this regime is very large, in the
sense that it causes most of the energy to be in the uniform region 3,
with a sub-fast-magnetosonic speed relative to the
``wall''. Nevertheless, since this region is still relativistic, it
takes the rarefaction wave that is reflected from the wall a long time
to cross this region in the lab (or wall) frame, and this occurs at a
large distance from the wall, $R_{\rm *,CD} \sim R_{\rm dec}
\sim R_{\rm cr} \sim R_0\Gamma_{\rm cr}^2 \gg R_0$, near the
deceleration radius where most of the energy is transferred to the
shocked external medium. Altogether, in regime III we have $R_0\sim
R_u < R_{\rm cr}\sim R_{\rm dec}\sim R_{\rm *,CD}<R_c$ (see
Table~\ref{tab:regimes}), and similar regime II, here as well $R_c$ does not
have a physical significance (and the same also holds for $R_1$, since
we always have $\sigma_u\gg 1$ in regime III).

In regime III, $\langle\Gamma\rangle(R \geq R_0)$ becomes independent
of $\sigma_0$ while $\langle\sigma\rangle(R \geq R_0)$ scales linearly
with $\sigma_0$, when fixing $L$, $A$, $k$ and $R_0$ (which fixes
$\Gamma_{\rm cr}$) while letting $\sigma_0$ and $\rho_0 \propto
1/\sigma_0$ vary (since $\sigma_0 \gg 1$ we have $L \sim Ec/R_0 \sim
\sigma_0\rho_0c^3R_0^2 \propto \sigma_0\rho_0$, so that fixing $L$
implies that $\rho_0 \propto 1/\sigma_0$). Such a variation of the
parameters means fixing the overall properties of the flow and
changing only its composition or magnetization (as is done in
Figs.~\ref{fig:Gamma_R} through \ref{fig:sigma_R2_k2p}). In this
regime the global dynamics become insensitive to the exact composition.
This can be thought of as the high magnetization limit, where the
behavior of the outflow approaches that of an electromagnetic wave
that is emitted at the source and reflected by the CD, where the time
when the back end of the finite wave reflects off the CD corresponds
to the time when the right-going rarefaction wave reaches the CD,
$t_{\rm CD} \approx R_{\rm *,CD}/c$.

\begin{figure}
\centerline{\includegraphics[width=15cm]{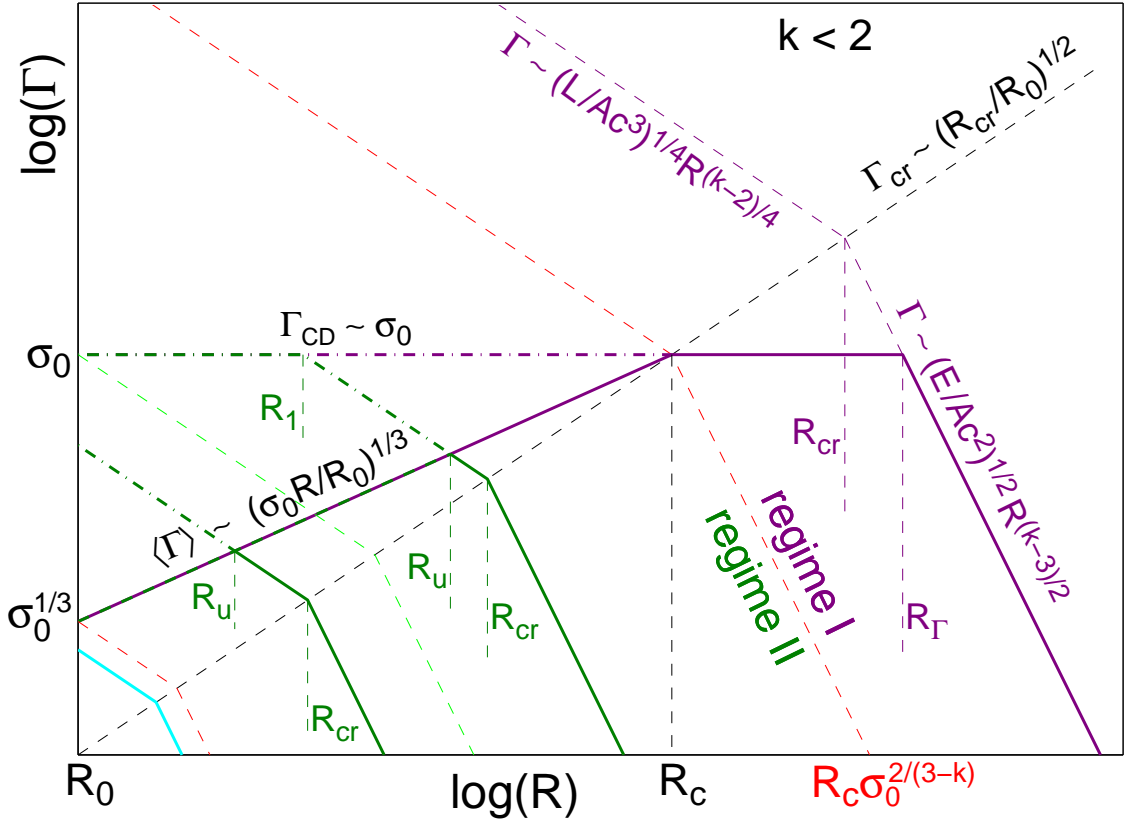}}
\caption{
Evolution of the typical Lorentz factor $\langle\Gamma\rangle$ of the
flow (where most of the energy resides; {\it thick solid lines}) and
the Lorentz factor of the contact discontinuity, $\Gamma_{\rm CD}$
({\it dashed-dotted lines}), as a function of radius $R$, for $k<2$
and for different values of the external density normalization ($f_0 =
\rho_0/\rho_1(R_0)$ or $\rho_1(R_0) = AR_0^{-k}$ or $A$) and fixed
values of all of the other model parameters ($\sigma_0\gg 1$, $k$,
$\rho_0$, $R_0$, and therefore also $E$ and $L$), which imply a
constant $R_c \sim R_0\sigma_0^2$ and varying $\Gamma_{\rm cr} \sim
(f_0\sigma_0)^{1/(8-2k)}$ and $R_{\rm cr}\sim R_0\Gamma_{\rm cr}^2$.
The purple, green and cyan lines correspond, respectively, to regimes
I ($\Gamma_{\rm cr}>\sigma_0>1$ or $R_{\rm cr}>R_c$), II
($\sigma_0^{2/(12-3k)}<\Gamma_{\rm cr} <\sigma_0$ or
$\sigma_0^{-2(3-k)/(4-k)}<R_{\rm cr}/R_c<1$), and III ($1<\Gamma_{\rm
cr}<\sigma_0^{2/(12-3k)}$ or $1<R_{\rm cr}/R_0<\sigma_0^{2/(4-k)}$).
The borderlines between these regimes are indicated by thin dashed red
lines. Within regime II, the thin dashed green line is the border
between the regions with and without a break in $\Gamma_{\rm CD}(R)$
at $R_1>R_0$. (The particular slopes in this plot are drawn for $k=0$,
but the general scalings are clearly indicated).}
\label{fig:vary_A}
\end{figure}

\begin{figure}
\centerline{\includegraphics[width=15cm]{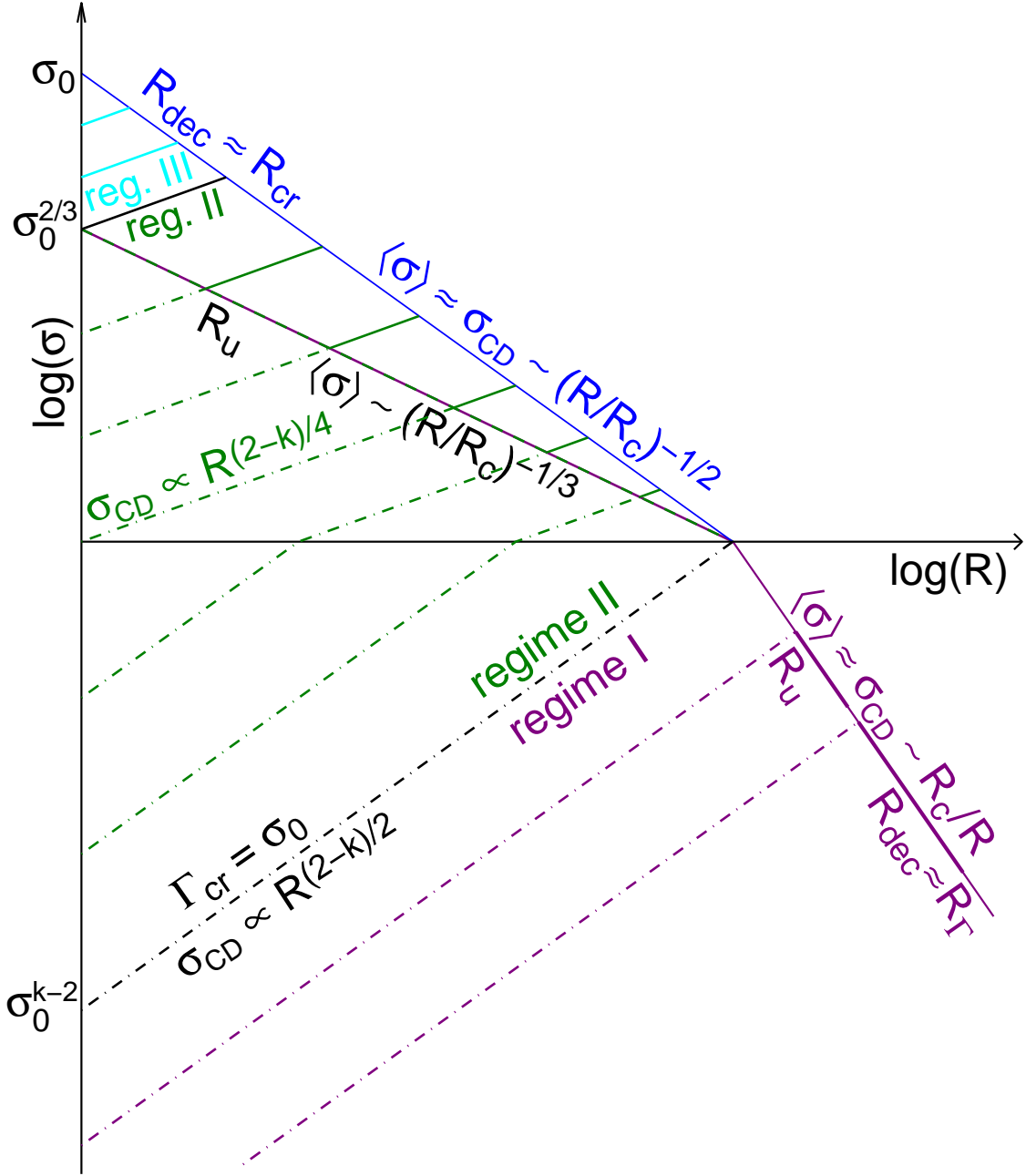}}
\caption{
The evolution of the magnetization $\sigma$ with radius for $k < 2$,
similar to Fig.~\ref{fig:sigma_R2}, but for different values of the
external density normalization ($f_0 =
\rho_0/\rho_1(R_0)$ or $\rho_1(R_0) = AR_0^{-k}$ or $A$) and fixed
values of all of the other model parameters ($\sigma_0\gg 1$, $k$,
$\rho_0$, $R_0$, and therefore also $E$ and $L$), which imply a
constant $R_c \sim R_0\sigma_0^2$ and varying $\Gamma_{\rm cr} \sim
(f_0\sigma_0)^{1/(8-2k)}$ and $R_{\rm cr}\sim R_0\Gamma_{\rm cr}^2$
(similar to Fig.~\ref{fig:vary_A}). (The particular slopes in this
plot are drawn for $k=1$, but the general scalings are clearly
indicated).}
\label{fig:sigma4_R_vary_A}
\end{figure}

Alternatively, as is done in Figs.~\ref{fig:vary_A} and
\ref{fig:sigma4_R_vary_A}, one could fix the properties of the
magnetized flow: $L$, $R_0$, $\sigma_0$, $\rho_0$ (and thus also $R_c
\sim R_0\sigma_0^2$) and vary the normalization of the external
density: $A$ or $\rho_1(R_0) = AR_0^{-k}$ (while fixing its power-law
index, $k$), which effectively varies $\Gamma_{\rm cr}$ and $R_{\rm
cr}$. It can be seen from Figs.~\ref{fig:vary_A} and
\ref{fig:sigma4_R_vary_A} that as the external density goes to zero 
we have $f_0\to\infty$, $\Gamma_{\rm cr}\sim
(f_0\sigma_0)^{1/(8-2k)}\to\infty$ and $R_{\rm dec} \sim R_\Gamma
\to\infty$, and this the solution approaches that of expansion into
vacuum (or the extreme limit of regime I). As the external density
increases $f_0$, $\Gamma_{\rm cr}$ and $R_{\rm dec}$ all decrease,
until when $\Gamma_{\rm cr} \sim
\sigma_0$, $f_0\sim\sigma_0^{7-2k}$ and $R_{\rm dec} \sim R_\Gamma \sim
R_{\rm cr}$ there is a transition to regime II. As the external
density increases even further a transition to regime III occurs when
$\Gamma_{\rm cr} \sim \sigma_0^{2/(12-3k)}$, $f_0\sim \sigma_0^{1/3}$
and $R_u \sim R_0$. Finally, when the external density becomes so large
that $\Gamma_{\rm cr} \sim 1$, $f_0\sim\sigma_0^{-1}$, and $R_{\rm cr}
\sim R_0$, the flow remains Newtonian and there is a transition to
regime IV that is discussed below.

\subsection{Regime IV}
\label{sec:regIV}

For a sufficiently large external density, $f_0 \ll \sigma_0^{-1}$,
the formal expression for $\Gamma_{\rm cr}$ gives $\Gamma_{\rm cr} \ll
1$ and the flow remains Newtonian. If we consider a source that is
active over a time $t_0$ then when the central source finishes
ejecting the highly magnetized outflow, it would be bounded within
$R_{\rm CD}(t_0) \sim \beta_{\rm CD}(t_0)ct_0\approx\beta_{\rm
CD}(t_0)R_0$, where I neglect factors of order unity for simplicity
(here $R_0$ is still defined through the relation $R_0\approx ct_0$,
even though it loses its physical meaning from the relativistic
regime). More generally, at $t \leq t_0$ the radius of the CD
satisfies $R_{\rm CD}(t)\sim \beta_{\rm CD}(t)ct \approx \beta_{\rm
CD}(t)R_0\,t/t_0$. For a tangential magnetic field, which scales as
$B/B_0 \approx R_0/R$, the magnetic pressure at $R_{\rm CD}$, $\sim
B^2[R_{\rm CD}(t)] \sim [R_0/R_{\rm CD}(t)]^2B_0^2
\sim\sigma_0\rho_0c^2(t/t_0)^{-2}\beta_{\rm CD}^{-2}(t)$, would be
balanced by the ram pressure of the shocked external medium at the
frame of the CD, $\sim \rho_1[R_{\rm CD}(t)]\beta_{\rm CD}^2(t)c^2
\sim\rho_1(R_0)c^2(t/t_0)^{-k}\beta_{\rm CD}^{2-k}(t)$, leading to
\begin{equation}
\beta_{\rm CD}(t\leq t_0) \sim 
\left(f_0\sigma_0\right)^{1/(4-k)}
\left(\frac{t}{t_0}\right)^{(k-2)/(4-k)}\ . 
\end{equation}
This implies $\beta_{\rm CD}(t_0) \sim
(f_0\sigma_0)^{1/(4-k)}\sim\Gamma_{\rm cr}^2 \ll 1$, which
demonstrates self-consistency by showing that the flow is indeed
Newtonian in this regime. The magnetic energy in the original outflow,
at $R<R_{\rm CD}(t)$, is given by $E_B[R<R_{\rm CD}(t)]\sim R_{\rm
CD}^3(t)B^2[R_{\rm CD}(t)] \sim (t/t_0)\beta_{\rm CD}(t)R_0^3B_0^2\sim
(t/t_0)\beta_{\rm CD}(t)E_0 \approx \beta_{\rm CD}(t)L_0t$, where $L_0
\approx B_0^2R_0^2c$ and $E_0 \approx L_0t_0$ are the injected
luminosity and corresponding energy over a time $t_0$ for a
relativistic outflow. This would violate conservation of energy, if
the outflow emanating from the central source was indeed relativistic
(this is basically the well-known $\sigma$-problem).\footnote{For a
steady central source that ejects a magnetized outflow over a finite
time $t_0$ this regime resorts back to the well known $\sigma$
problem, where in ideal MHD the stored magnetic energy grows
quadratically with the injection time of the central source, while the
actual injected energy grows only linearly with this time, implying a
breakdown of one or more of the underlying assumptions. This may be
relevant, e.g., for a millisecond magnetar born inside a collapsing
massive star, as a possible progenitor of long duration GRBs.}
However, since in this regime the outflow is sub-sonic (or
sub-fast-magnetosonic) and Newtonian, the information about the
existence of the external medium must propagate back to the source
producing a back reaction that results in a Newtonian outflow with a
speed $\sim\beta_{\rm CD}(t)$. For such a Newtonian magnetized outflow
the electromagnetic luminosity is $L = 4\pi
R^2(c/4\pi)|\vec{E}\times\vec{B}| = B^2R^2\beta c$, and for
$\beta(t)\sim \beta_{\rm CD}(t)$ this gives $L(t) \sim \beta_{\rm
CD}(t)B^2R^2c \approx \beta_{\rm CD}(t)L_0$ and $E(t)\approx tL(t)\sim
\beta_{\rm CD}(t)L_0t$, which is consistent with the above
estimate. Even during the initial injection phase ($t < t_0$) the
shocked external medium holds a good fraction of the total energy at
any given time. After the injection stops, at $t > t_0$, most of the
energy is quickly transfered to the shocked external medium on the
dynamical time (up to $t \sim 2t_0$ or so), and the flow settles into
an adiabatic Sedov-Taylor solution with velocity $\beta(t>t_0)c\sim
[E/At^{3-k}]^{1/(5-k)}\sim \beta_{\rm
CD}(t_0)c(t/t_0)^{-(3-k)/(5-k)}$, radius $R\sim (Et^2/A)^{1/(5-k)}$
and energy $E = E(t_0) \sim \beta_{\rm CD}(t_0)E_0$.

If we start with a magnetized spherical shell or ``ball'' of radius
$R_0$, initially at rest, then in this case the magnetic pressure at
$R_{\rm CD} \sim R_0$ is $\sim B_0^2 \sim \sigma_0\rho_0c^2$ and
equating it to the ram pressure of the shocked external medium, $\sim
\rho_1(R_0)\beta_{\rm CD}^2c^2$, implies $\beta_{\rm CD} \sim
(f_0\sigma_0)^{1/2} \sim \Gamma_{\rm cr}^{4-k} \ll 1$.  Therefore, it
would significantly increase its radius and transfer most of its
energy to the shocked external medium on its dynamical time, which
corresponds to a timescale of
\begin{equation}
t_{\rm dec} \sim \frac{R_0}{\beta_{\rm CD}c} \approx \frac{t_0}{\beta_{\rm CD}} 
\sim (f_0\sigma_0)^{-1/2}t_0 \sim \Gamma_{\rm cr}^{k-4}t_0 \gg t_0\ .
\end{equation}
At $t >t_{\rm dec}$ most of the energy is in the shocked
external medium and the flow settles into an adiabatic Sedov-Taylor
solution with velocity $\beta(t>t_{\rm dec}) c \sim [E/At^{3-k}]^{1/(5-k)}\sim
\beta_{\rm CD}(t_{\rm dec})c(t/t_0)^{-(3-5)/(5-k)}$, radius $R \sim (Et^2/A)^{1/(5-k)}$
and energy $E = E_0 \sim R_0^3\sigma_0\rho_0c^2$.

\subsection{Regime II$^*$ ($10/3<k<4$):}
\label{sec:regII*}

When the external density drops very sharply with radius, $k>10/3$,
then $\Gamma_{\rm CD}$ initially grows with radius faster than
$R^{1/3}$, which has interesting implications. Regime II that exists
for $k<10/3$ disappears for $k=10/3$ and reappears for $10/3<k<4$ in a
different form that we shall call regime II$^*$, which corresponds to
$\sigma_0^{7-2k}\ll f_0\ll\sigma_0^{1/3}$. For $10/3<k<4$, Regime I
holds for $f_0\gg\sigma_0^{1/3}$ and regime III holds for
$\sigma_0^{-1}\ll f_0\ll\sigma_0^{7-2k}$.

In regime I, for $f_0 > \sigma_0^3$ $\Longleftrightarrow$ $\sigma_0 <
\Gamma_{\rm cr}^{(4-k)/2}$ there in no $R_1$ (as $\sigma_{\rm CD}\ll 1$ 
and $\Gamma_{\rm CD} \sim\sigma_0$ all along) and the shocked external
medium decouples from the magnetized shell at a decoupling radius of
$R_{\rm dcp}\sim R_0 < R_c$, as the forward shock accelerates down the
steep external density gradient, and the shocked external medium
carries only a very small fraction of the total energy, $E_{\rm
ext}/E\sim\sigma_0/f_0 <\sigma_0^{-2}\ll 1$. In the parameter range
$\sigma_0^{1/3} < f_0 < \sigma_0^3$ $\Longleftrightarrow$ $\Gamma_{\rm
cr}^{(4-k)/2} <\sigma_0 <\Gamma_{\rm cr}^{(12-3k)/2}$, on the other
hand, there is a radius $R_1$ where $\sigma_{\rm CD}\approx\sigma_u =
1$ and the ordering of the critical radii is $R_0 < R_1 \sim R_{\rm
dcp} < R_c$, so that $\Gamma_{\rm CD}~\sim \sigma_0/\sigma_{\rm
CD}\propto R^{(k-2)/4}$ following Eq.~(\ref{eq:sigma_int}) until
$\Gamma_{\rm CD}\sim\sigma_0$ and $\sigma_{\rm CD}\sim 1$ at $R\sim
R_1\sim(f_0\sigma_0^{-3})^{1/(2-k)}$, and the decoupling of the
forward shock from the magnetized shell occurs at $R_{\rm dcp}
\sim R_1$. In both cases at $R > R_{\rm dcp}$ the shell accelerates 
in the wake of the accelerating forward shock, almost as if into
vacuum.  At $R > R_c$ it starts coasting and spreading radially, where
it can in principle keep coasting indefinitely (or more realistically
until the assumption of a very steep external density profile breaks
down, and enough external mass is swept-up that could decelerate the
forward shock and, in turn, also the shell).

In regime II$^*$ the ordering of the critical radii is $R_0<
R_u<R_1\sim R_{\rm dcp}<R_c<R_{\rm cr}$ (see
Table~\ref{tab:regimes_k>10/3}). Initially, at $R_0<R<R_u$, the
typical Lorentz factor and magnetization are similar to those at the
CD and are determined by the pressure balance at the CD,
$\langle\Gamma\rangle\sim\Gamma_{\rm
CD}\sim\sigma_0/\langle\sigma\rangle ~\sim \sigma_0/\sigma_{\rm
CD}\propto R^{(k-2)/4}$ following Eqs.~(\ref{eq:Gamma_CD_II}) and
(\ref{eq:sigma_int}). At $R_u<R<R_{\rm cr}$ the bulk of the shell
decouples from the CD and accelerates as $\langle\Gamma\rangle\sim
\sigma_0/\langle\sigma\rangle \sim (\sigma_0 R/R_0)^{1/3}$ until
reaching the coasting radius $R_c\sim R_0\sigma_0^2$ (where it starts
to coast and spread, as in regime I), while $\Gamma_{\rm CD}~\sim
\sigma_0/\sigma_{\rm CD}\propto R^{(k-2)/4}$ keeps following
Eq.~(\ref{eq:sigma_int}) until $\Gamma_{\rm CD}\sim\sigma_0$ and
$\sigma_{\rm CD}\sim 1$ at $R\sim
R_1\sim(f_0\sigma_0^{-3})^{1/(2-k)}$. At $R>R_1 \sim R_{\rm dcp}$ we
have $\sigma_{\rm CD}<1$ and $\Gamma_{\rm CD}\sim\sigma_0$, while the
shocked external medium decouples from the shell, carrying with it
only a small fraction of the total energy, $E_{\rm ext}/E \sim
R_1/R_c\sim (f_0\sigma_0^{2k-7})^{1/(2-k)} \ll 1$, as it keeps
accelerating down the steep external density gradient, with
$\Gamma_{\rm BM}\propto R^{(k-3)/2}$ (where the same energy is given
to a decreasing amount of newly swept-up external rest mass).  The
original shell keeps coasting and spreading radially in the evacuated
region in the wake of the accelerating forward shock, essentially as
if into vacuum. Near the transition to regime III we have
$f_0\sim\sigma_0^{7-2k}$ and therefore $E_{\rm ext}/E \sim 1$, so that
the energy in the shocked external medium becomes comparable to the
total energy.

In regime III the ordering of the critical radii is $R_0 < R_{\rm cr}
\sim R_{\rm dcp} \sim R_{\rm dec}$, and there are no $R_u$ or
$R_1$. At $R_0 < R < R_{\rm cr}$ the shell accelerates as
$\mean{\Gamma} \sim \Gamma_{\rm CD}~\sim\sigma_0/\sigma_{\rm CD}\sim
\sigma_0/\mean{\sigma} \propto R^{(k-2)/4}$ following
Eq.~(\ref{eq:sigma_int}), where the typical values of $\Gamma$ and
$\sigma$ are close to those at the CD. The shocked external medium
decouples from the original magnetized shell at $R_{\rm dcp}
\sim R_{\rm cr}$, but since in this case $\sigma_{\rm CD}(R_{\rm
cr})\sim\sigma_0/\Gamma_{\rm cr}\gg 1$ the rarefaction wave is still
strong when it reaches $R_{\rm cr}$ so that it effectively decelerates
the shell and very little energy remains in the original shell at
later times (or larger radii), while most of the energy is transfered
to the shocked external medium, which approaches the BM76 self-similar
solution.

\subsection{Summary}
\label{sec:summary}

\begin{figure}
\centerline{\includegraphics[width=12cm]{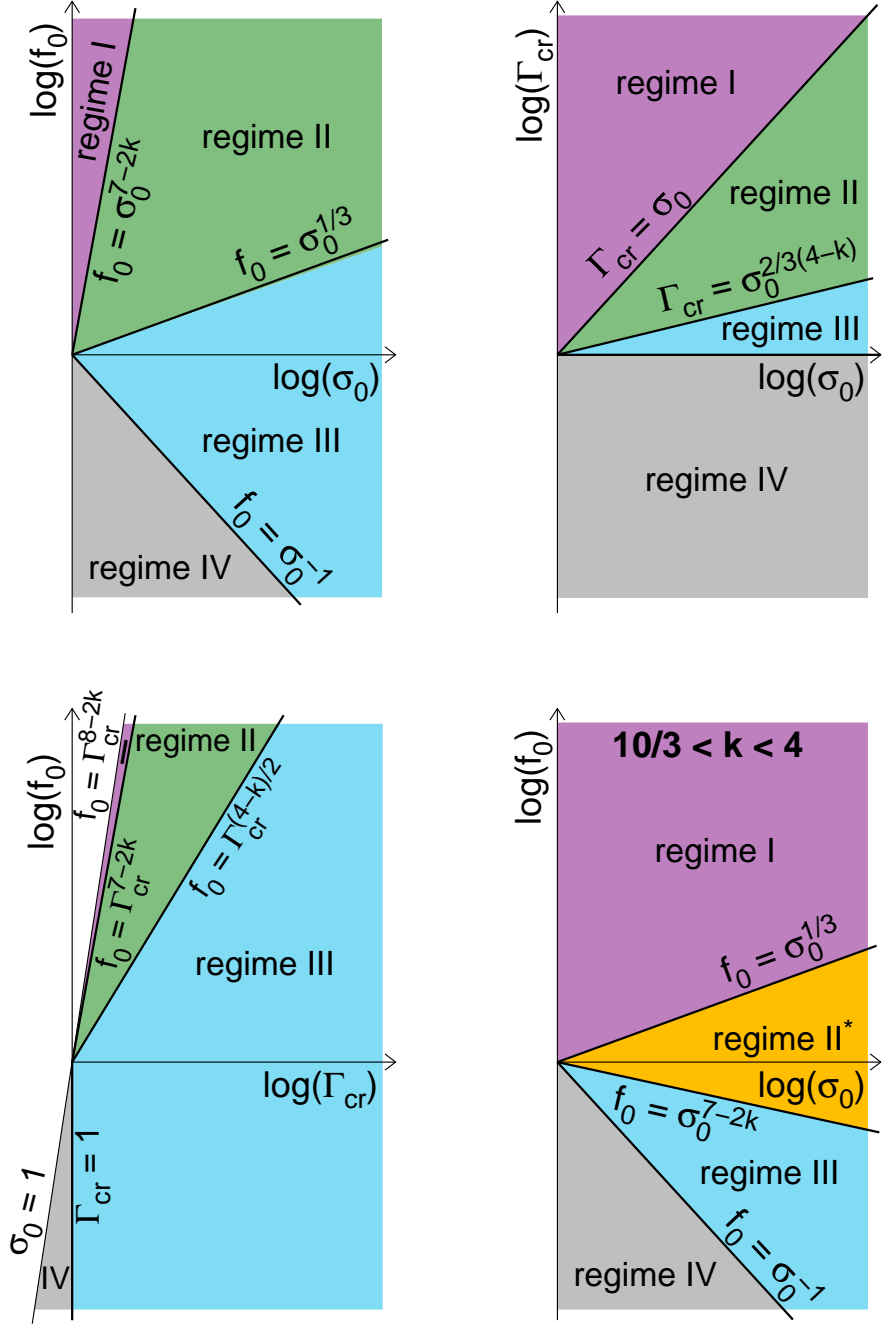}}
\caption{
Phase space diagrams of the different dynamical regimes: in the
$f_0\,$--$\,\sigma_0$ plane for $k<10/3$ ({\it top left panel}), $\Gamma_{\rm
cr}\,$--$\,\sigma_0$ plane for $k<10/3$ ({\it top right panel}), $f_0\,$--$\,\Gamma_{\rm cr}$
plane for $k<10/3$ ({\it bottom left panel}), and in the $f_0\,$--$\,\sigma_0$ plane for
$10/3<k<4$ ({\it bottom right panel}). Each regime is
labeled and denoted by a different color, and the borders between the
different regimes are indicated (by labeled thick black lines).}
\label{fig:regimes}
\end{figure}

The different regimes are summarized in Tables~\ref{tab:regimes} and
\ref{tab:regimes_k>10/3} as well as in Figure~\ref{fig:regimes}. 
Tables~\ref{tab:regimes} and \ref{tab:regimes_k>10/3} provide the
ordering of the various critical radii in the different regimes, along
with the parameter range occupied by each regime, in terms of the
initial shell to external medium density ratio $f_0 =
\rho_0/\rho_1(R_0)$, initial magnetization $\sigma_0$, and
$\Gamma_{\rm cr} \sim(f_0\sigma_0)^{1/(8-2k)}$.
Figure~\ref{fig:regimes} shows the regions of parameter space occupied
by each regime within the two dimensional planes spanned by
$f_0\,$--$\,\sigma_0$ ({\it top left panel} for $k<10/3$ and {\it
bottom right panel} for $10/3<k<4$), $\Gamma_{\rm cr}\,$--$\,\sigma_0$
({\it top right panel} for $k<10/3$), $f_0\,$--$\,\Gamma_{\rm cr}$
plane ({\it bottom left panel} for $k<10/3$).

In order to gain some intuition for these results and better
understand them, it is useful to follow the behaviour of the system
when varying one key physical parameter and leaving the others fixed.
First, I vary the initial magnetization $\sigma_0$ (and $\rho_0
\propto 1/\sigma_0$ for consistency) while keeping fixed the energy 
($E \sim Lt_0 \approx LR_0/c$ or $L$), initial time or length scale
($t_0 \approx R_0/c$ or $R_0$) and external density ($k$ and $A$ or
$\rho_1(R_0) = AR_0^{-k}$), and thus also $\Gamma_{\rm cr}
\sim (f_0\sigma_0)^{1/(8-2k)}$ and
$R_{\rm cr} \sim R_0\Gamma_{\rm cr}^2$. This corresponds to a constant
$\Gamma_{\rm cr}$ and varying $\sigma_0 \propto 1/f_0$, or in
Figure~\ref{fig:regimes} to a horizontal line in the {\it top right
panel}, a vertical line in the {\it bottom left panel}, and a diagonal
line parallel to the $f_0=\sigma_0^{-1}$ line separating regimes III
and IV in the remaining two panels (showing the $f_0\,$--$\,\sigma_0$
plane). The behaviour in this case is summarized in
Figures.~\ref{fig:Gamma_R} through \ref{fig:sigma_R2_k2p}.

In regime I ($1 < \sigma_0 < \Gamma_{\rm cr}$) the acceleration is
almost as if into vacuum: $\mean{\Gamma} \propto R^{1/3}$ and
$\mean{\sigma} \propto R^{-1/3}$ until most of the energy is converted
to kinetic form at the coasting radius $R_c\sim R_0\sigma_0^2$, where
$\mean{\Gamma} \sim \sigma_0$ and $\mean{\sigma} \sim 1$. Then the
shell starts coasting (at $\mean{\Gamma} \sim \sigma_0$) and its width
in the lab frame starts growing linearly with radius resulting in a
fast drop in its magnetization, $\mean{\sigma} \sim R_c/R$. At $R >
R_c$ this regime reverts back to the well-studied unmagnetized ``thin
shell'' case, with a reasonable spread in its Lorentz factor,
$\delta\Gamma\sim\mean{\Gamma}$. A reverse shock develops and is
initially Newtonian, but strengthens as the shell widens, until it
becomes mildly relativistic when it finishes crossing the shell at the
deceleration radius, $R_{\rm dec}\sim R_\Gamma\sim (E/\sigma_0^2
Ac^2)^{1/(3-k)}$, where the magnetization is low,
$\mean{\sigma}(R_{\rm dec})\sim R_c/R_{\rm dec} \sim
(\sigma_0/\Gamma_{\rm cr})^{2(4-k)/(3-k)}\ll 1$.

In regime II ($1 < \Gamma_{\rm cr} < \sigma_0 < \Gamma_{\rm
cr}^{(12-3k)/2}$) the initial acceleration of $\mean{\Gamma} \propto
R^{1/3}$ is limited by the external medium at $R_u$, where most of the
energy is still in magnetic form ($\mean{\sigma}(R_u) \sim \sigma_{\rm
CD}(R_u)\gg 1$), and thus there is no coasting phase. Between $R_u$
and $R_{\rm cr}$ the typical Lorentz and magnetization of the shell
are similar to those just behind the CD and determined by the pressure
balance at the CD, $\mean{\Gamma}\sim \sigma_0/\mean{\sigma} \propto
R^{(k-2)/4}$. A rarefaction wave gradually crosses the shell from its
back to its front, until reaching the CD at $R_{\rm *,CD} \sim R_{\rm
cr} \sim R_{\rm dec}$. At that point the shocked external medium
starts dominating the total energy and the flow approaches the BM76
self-similar solution.

In regime III ($\sigma_0 > \Gamma_{\rm cr}^{(12-3k)/2} > 1$) the
external density is large enough that there is no impulsive
acceleration stage with $\mean{\Gamma}\propto R^{1/3}$. Instead, the
pressure balance at the CD determines the shell's typical Lorentz
factor and magnetization from the very start, $\mean{\Gamma}\sim
\sigma_0/\mean{\sigma} \propto R^{(k-2)/4}$ at $R_0 < R < R_{\rm cr}
\sim R_{\rm dec}$, and the dynamics become insensitive to the exact 
composition [i.e. to the value of $\sigma_0$, when fixing the external
density ($A$ and $k$) and the shell luminosity ($L$) and initial width
($R_0$)]. This is the high-$\sigma$ limit where the flow behaves
like an electromagnetic wave that is emitted at the source and
reflected at the CD.

There are also more ``exotic'' regimes, such as regime IV where the
external density is so high that flow remains Newtonian, or regime
II$^*$ that exists only for a highly stratified external density
($10/3<k<4$) where the external shock accelerates down the steep
external density gradient and decouples from the original shell,
carrying a small fraction of the total energy, while the original
shell travels in its wake essentially as if into vacuum, similar to
regime I. Note that regime IV corresponds to $\Gamma_{\rm cr} < 1$ and
thus cannot be reached when fixing $\Gamma_{\rm cr}$ to a value larger
than 1 and varying $\sigma_0\propto 1/\rho_0$.

A slightly different way to gain perspective about these results is by
varying the external density normalization ($A$ or $\rho_1(R_0) =
AR_0^{-k}$ or $f_0 = \rho_0/\rho_1(R_0)$) while keeping the other
parameters fixed ($k$, $\sigma_0$, $\rho_0$, $R_0$, and therefore also
$E$ and $L$). In this case $R_c \sim R_0\sigma_0^2$ remains constant
while $\Gamma_{\rm cr} \sim (f_0\sigma_0)^{1/(8-2k)}$ and $R_{\rm
cr}\sim R_0\Gamma_{\rm cr}^2$ vary, where both $f_0$ and $\Gamma_{\rm
cr}\sim (f_0\sigma_0)^{1/(8-2k)} \propto f_0^{1/(8-2k)}$ decrease when
the external density increases. In Figure~\ref{fig:regimes} this
corresponds to a vertical line in all but the {\it bottom left panel},
where it corresponds to a slightly diagonal line parallel to the
$f_0=\Gamma_{\rm cr}^{8-2k}$ line (the left boundary of the colored
regions).  The behavior in this case is illustrated in
Figures~\ref{fig:vary_A} and \ref{fig:sigma4_R_vary_A}. For a
sufficiently low external density, corresponding to
$f_0>\sigma_0^{7-2k}$ or $\Gamma_{\rm cr}>\sigma_0$ we are in regime
I, where the expansion is initially essentially as if into vacuum,
reaching the coasting radius at $R_c$ that is independent of $f_0$ and
decelerating significantly only at $R_{\rm dec} \sim R_\Gamma \sim
R_c(f_0\sigma_0^{2k-7})^{1/(3-k)} \propto f_0^{1/(3-k)}$. As the
external density increases, $\Gamma_{\rm CD}(R_0)$ decreases, bringing
about first regime II ($\sigma_0^{1/3}<\Gamma_{\rm CD}(R_0)<\sigma_0$
or $\sigma_0^{-2(3-k)/(4-k)}<R_{\rm cr}/R_c<1$), and at even larger
external densities regime III ($1<\Gamma_{\rm
cr}<\sigma_0^{2/(12-3k)}$ or $1<R_{\rm
cr}/R_0<\sigma_0^{2/(4-k)}$). For the highest external densities
($R_{\rm cr}<R_0$, $\Gamma_{\rm cr}<1$ or $f_0<\sigma_0^{-1}$) the
flow remains Newtonian all along (regime IV).

\section{Comparison with Previous Works}
\label{sec:comp}

The unmagnetized case for the deceleration of a finite uniform
relativistic shell by the external medium has been studied in the
context of GRBs ~\citep{SP95,Sari97,KS00,KZ03,NP04}. The main results
have been summarized in \S~\ref{sec:low-sigma} and extended to a
general power-law of the external density profile, and are consistent
with the previous results. The deceleration of a magnetized
relativistic shell by an unmagnetized external medium has also been
studied ~\citep{ZK05,GMA08,MGA09,Mizuno09,Levinson10,Lyut11}.

\citet[][hereafter ZK05]{ZK05} have both considered arbitrary
``initial'' values for the shell Lorentz factor and magnetization, and
have attached too much importance to the crossing of the shell by the
reverse shock, while for $\sigma\gg 1$ even if a reverse shock exists
its effect on the global dynamics of the system is very small (it
dissipates only a small fraction of the total energy, of the order of
$\sim 1/\sigma$, and by its shell crossing time only a similarly
small fraction of the total energy is transfered to the shocked
external medium). Therefore, the conclusions of that paper are very
different from my results.

\citet{GMA08} have considered a similar initial setting and argued for
a different condition for the formation of a reverse
shock\footnote{Their argument that the shell can be crossed by a
fast-magnetosonic wave (and thus come into causal contact) faster than
by a fast-magnetosonic shock (both starting at the CD) appears to
contradict the basic notion that a shock must always travel faster
than the relevant corresponding wave. It arises since they use the
formula for the radius at which the reverse shock crosses the shell
from Eq.~(38) of ZK05 that is valid only for a strong reverse shock
(with a relativistic upstream to downstream 4-velocity, $u_{\rm RS}\gg
1$, or $\gamma_{34}\gg 1$ in the notation of ZK05) also outside its
range of applicability, while the result for a fast-magnetosonic wave
is approached in the opposite limit of a weak reverse shock ($u_{\rm
RS}\ll 1$ or $\gamma_{34}-1\ll 1$).}. While the condition for the
formation of a reverse shock in the ideal Riemann problem addressed in
ZK05 is correct, such initial conditions are not realistic and the
formation of a reverse shock and its properties can be sensitive to
the exact initial conditions or to fluctuations in the external
density, etc. Moreover, in the high-$\sigma$ limit even if such a
shock exists it has a very small effect on the global dynamics, which
are the main focus of the present work, and therefore this is not
addressed here in detail.  In ~\citet{MGA09} the problem is addressed
with a similar initial setup but using high resolution 1D RMHD
numerical simulations. There, the regime that is argued to have no
reverse shock in ~\citet{GMA08} is correctly found to have either a
weak or no reverse shock. They also demonstrate numerically that the
flow quickly approaches the BM76 self-similar solution after the
deceleration radius.

~\citet{Mizuno09} point out that for the Riemann problem of a
magnetized shell moving relativistically relative to an unmagnetized
region (or ``external medium'') at rest, above some critical value of
magnetization parameter $\sigma$ there is a rarefaction wave that
propagates into the magnetized shell and accelerates it, and only
below that critical value there is a (reverse) shock that decelerates
the shell.  While this observation is correct, this Riemann problem is
not a realistic setup for the deceleration of magnetized GRB ejecta,
since it uses arbitrary ``initial'' conditions near the deceleration
radius.  ~\cite{Lyut11} has analyzed the similar problem of the
deceleration of a shell with arbitrary initial Lorentz factor and
magnetization, concluding that the differences between the magnetized
and unmagnetized cases are rather small, and involve mainly the
existence or strength of the reverse shock at early times (which may
be non-existent or weak for high magnetizations), rather than the
global gross properties of the flow. I find that this is a right
answer for the wrong question, in the sense that the initial setup is
too arbitrary to realistically apply to GRB outflows. The impulsive
acceleration process determines the conditions near the deceleration
radius, which are therefore not arbitrary, and some regions of
parameter space and their corresponding dynamical regimes cannot be
realized under realistic circumstances.

Paper I has addressed mainly the impulsive acceleration into vacuum of
a highly magnetized shell, starting at rest. However, at the end of
its subsection 5.2 it also briefly addressed the expansion of such a
shell into an unmagnetized external medium. There it has outlined the
two main dynamical regimes, which in the current work are referred to
as regimes I and II. ~\citet{Levinson10} has also considered the
acceleration and of an impulsive magnetized shell and its deceleration
due to the interaction with the external medium, following paper I and
treating the latter part in more detail. ~\citet{Levinson10} also
identified regimes I and II. His expressions for the maximal Lorentz
factor of the shell in regime II are only a factor of 1.09 lower than
Eq.~(\ref{eta_cr}) of the current work for $k = 0$, and a factor of
1.57 lower for $k = 2$ (the latter difference might appear larger
since he used $z = 0$, $E_{53} = 0.1$ and $A_* \approx 33$ for his
fiducial values while the current work uses $z = 2$, $E_{53} = 1$ and
$A_* = 1$). The current work finds that for $k < 2$ the maximal value
of $\langle\Gamma\rangle$ is obtained at $R_u$, and is a factor of
$\sim (\sigma_0/\Gamma_{\rm cr})^{(2-k)/(10-3k)} > 1$ (see
Eq.~[\ref{eq:Gamma_u}]) larger than $\Gamma_{\rm cr}$. However, for
the values of $\sigma_0 \lesssim 10^3$ and $\Gamma_{\rm cr} = 180$
considered by \citet{Levinson10} this factor if $\lesssim 1.4$ for $k
= 0$, and thus consistent with his results (see his Fig.~5).

\citet{Levinson10} has argued, however, that multiple sub-shells 
with an initial separation comparable to their initial width would
collide and merge while still highly magnetized, which is incorrect
and in contradiction with paper I. An accompanying
paper~\citep{Granot11} focuses on the possible role of multiple
sub-shells, which can alleviate some of the requirements on the
Lorentz factor of the outflow and may help accommodate GRB
observations much better.

\section{Discussion and Conclusions}
\label{sec:dis}

This work has presented a detailed and unified treatment of the
magnetic acceleration of an impulsive, initially highly magnetized
($\sigma_0\gg 1$) shell and its deceleration by an unmagnetized
external medium (with a power-law density profile). The dynamics
divide into three main regimes (I, II, and III) and two more
``exotic'' regimes (relevant for an external density that either
sharply drops with radius [II$^*$], or is very large [IV], leading
to a Newtonian flow).

In regime I the external density is low enough that the shell
accelerates almost as if into vacuum. At the coasting radius, $R_c\sim
R_0\sigma_0^2$, it reaches its maximal Lorentz factor of
$\langle\Gamma\rangle\sim \sigma_0 < \Gamma_{\rm cr}$ (where
$\Gamma_{\rm cr}$ is given in Eq.~[\ref{eta_cr}]) and becomes kinetic
energy dominated. At $R > R_c$ this regime reverts back to the
well-studied unmagnetized ``thin shell'' case~\citep{SP95}, where the
shell coasts and spreads radially, $\Delta(R>R_c) \sim (R/R_c)R_0$, as
its magnetization rapidly decreases well below unity,
$\sigma(R>R_c)\sim R_c/R$. In this regime the reverse shock is
initially Newtonian, starts dominating the pressure behind the CD at
$R_{\rm RS}\sim R_{\rm cr}$, and becomes mildly relativistic when it
finishes crossing the shell, at $R_\Gamma\sim R_{\rm dec}$. The
deceleration radius, $R_{\rm dec}$, which corresponds to an observed
deceleration time $T_{\rm dec}$, is where most of the energy
dissipation in the shell takes place and most of the energy is
transfered to the shocked external medium. Thus, both the reverse
shock emission and the afterglow emission are expected to peak on the
timescale of $T \sim T_{\rm dec}$. At $R>R_{\rm dec}$ (or $T>T_{\rm
dec}$) the flow quickly approaches the BM76 self-similar solution,
which for GRBs signals the start of the usual long-lived decaying
afterglow emission. The magnetization at the deceleration radius is
low, $\langle\sigma\rangle(R_{\rm dec}) \sim R_c/R_{\rm
dec}\sim(\sigma_0/\Gamma_{\rm cr})^{2(4-k)/(3-k)} \ll 1$. If it is
very low then magnetic field amplification in the mildly relativistic
collisionless (reverse) shock that develops could bring the downstream
magnetic field to within a few percent of equipartition, thus allowing
a good radiative efficiency for synchrotron emission, resulting in a
bright reverse shock emission. In this regime the reverse shock
emission and the afterglow emission both peak on a timescale $T_{\rm
dec}$ that is larger than the duration $T_{\rm GRB}$ of the prompt GRB
emission, $T_{\rm dec}/T_{\rm GRB} \sim R_{\rm dec}/R_c\gg 1$.
Moreover, the degree of magnetization behind the reverse shock,
$\sim\langle\sigma\rangle(R_{\rm dec}) \sim T_{\rm GRB}/T_{\rm dec}\ll
1$, can be directly inferred from the ratio of these two observable
times.

In regimes II or III the shell remains highly magnetized near the
deceleration radius, $\langle\sigma\rangle(R_{\rm dec})\sim
\sigma_0/\Gamma_{\rm cr} \gg 1$, which strongly suppresses the reverse
shock (which either becomes very weak or non-existent) and its
associated emission. The energy in the flow is transfered to the
shocked external medium (mostly near $R_{\rm dec}\sim R_{\rm cr}$)
with very little dissipation within the original shell as long as
ideal MHD holds. This is a highly magnetized ``thick shell'' case, and
the afterglow onset time is similar to the initial shell light
crossing time, $T_{\rm dec}\sim (1+z)R_0/c$. For a single shell
ejected from the source the prompt emission in this case might either
arise from the onset of the forward shock emission (for an external
shock origin, which makes it difficult to account for significant
variability, and in which case $T_{\rm GRB}\sim T_{\rm dec} \sim
(1+z)R_0/c$) or alternatively due to magnetic reconnection events
within the highly magnetized shell. The latter might be induced by the
deceleration of the shell due to the external medium, in which case
they might peak near $R_{\rm dec} \sim R_{\rm cr}$, since the angular
size of causally connected regions ($\sim\Gamma^{-1}$) grows as the
shell decelerates ($\Gamma_{\rm CD} \propto R^{(k-2)/4}$ decreases
with radius for $k<2$) and at $R\lesssim R_{\rm dec} \sim R_{\rm cr}$
most of the energy is still in the original magnetized shell (this
would again lead to $T_{\rm GRB} \sim T_{\rm dec} \sim (1+z)R_0/c$).

For the single shell case that was analyzed in this work there is
either the low magnetization ``thin shell'' (regime I) or the high
magnetization ``thick shell'' (regimes II or III). There is no low
magnetization ``thick shell'' case where a strong highly relativistic
reverse shock develops, which can result in a bright reverse shock
emission on a timescale comparable to that of the prompt gamma-ray
emission in GRBs ($T_{\rm dec} \sim T_{\rm GRB}$). Similarly, there is
no high magnetization ``thin shell'' case where the reverse shock is
severely suppressed by a high magnetization in the shell near the
deceleration radius and the afterglow onset occurs well after the
prompt GRB emission ($T_{\rm dec} \gg T_{\rm GRB}$). Therefore, a
bright reverse shock emission is possible only in the low
magnetization ``thin shell'' case -- regime I, in which case this
reverse shock emission (as well as the afterglow emission) would peak
on a timescale larger than the duration of the prompt GRB emission,
$T_{\rm dec}\gg T_{\rm GRB}\sim (1+z)R_0/c$. An accompanying
paper~\citep{Granot11}, however, shows that if the flow consists of
many distinct sub-shells instead of a single shell, then this may also
allow a low magnetization ``thick shell'' regime.

The Lorentz factor of the emitting region in GRBs must be high enough
to overcome the compactness problem and avoid excessive pair
production within the source~\citep{KP91,Fenimore93,WL95,BH97,LS01}.
It had been recently argued~\citep{Levinson10} that the interaction
with the external medium might not enable an impulsive highly
magnetized outflow in GRBs to accelerate up to sufficiently high
Lorentz factors, and that its maximal achievable Lorentz factor is
largely limited to $\Gamma \lesssim \Gamma_{\rm cr}$. This would pose
a particularly severe problem for a stellar wind-like external medium
($k=2$) for which typically $\Gamma_{\rm cr}\lesssim 10^2$ (see
Eq.~[\ref{eta_cr}]). Recent high-energy observations by the Fermi
Large Area Telescope (LAT) have set a lower limit of $\Gamma\gtrsim
10^3$ for the emitting region in a number of GRBs with a bright
high-energy emission~\citep{GRB080916C,GRB090902B,GRB090510} using a
simplified one-zone model. However, a more detailed and realistic
treatment shows that the limit is lower by a factor of $\sim
3$~\citep{Granot08,GRB090926A,HDMV11}, which would correspond to
$\Gamma\gtrsim 10^{2.5}$ for the brightest Fermi LAT GRBs.
Nevertheless, this might still pose a problem for a single highly
magnetized shell in a stellar-wind environment.

Regime I implies a maximal Lorentz factor of the flow, $\Gamma
\lesssim \sigma_0 \ll \Gamma_{\rm cr}$ (where $\Gamma_{\rm cr}$ is
given in Eq.~[\ref{eta_cr}]). In regime II, a higher maximal Lorentz
factor is possible for $k<2$, and $\langle\Gamma\rangle$ peaks at
$\Gamma_u$ where it exceeds $\Gamma_{\rm cr}$ by a factor of $\sim
(\sigma_0/\Gamma_{\rm cr})^{(2-k)/(10-3k)} > 1$ (see
Eq.~[\ref{eq:Gamma_u}]), while $\Gamma_{\rm CD}$ can reach values as
high as $\sim \sigma_0$ at $R\lesssim R_1$ (however, the material with
such a Lorentz factor would carry only a small fraction of the total
energy, $\sim (\Gamma_{\rm cr}/\sigma_0)^{(8-2k)/(2-k)}\ll 1$ at
$R\sim R_1$). In regime III the typical Lorentz factor is close to
that of the CD, $\langle\Gamma\rangle \sim \Gamma_{\rm CD}$, and for
$k<2$ they both peak at $R_0$ where they exceed $\Gamma_{\rm cr}$ by a
factor of $\sim \Gamma_{\rm cr}^{(2-k)/2} > 1$, while
$\langle\Gamma\rangle\sim\Gamma_{\rm cr}(R/R_{\rm cr})^{(k-2)/4}$ at
$R_0<R<R_{\rm cr}$. All this could help increase $\mean{\Gamma}$ above
$\Gamma_{\rm cr}$ for $k<2$. However, for a stellar wind-like (or
steeper) external medium, $k = 2$ (or $k < 2$), we have $\mean{\Gamma}
\lesssim\Gamma_{\rm cr}$, which makes it very difficult to satisfy the
observational constraints on $\Gamma$ from pair opacity (mentioned
above), and to a lesser extent also those from the onset time of the
afterglow emission~\citep[usually around a few
hundred;][]{SP99,NP05,Molinari07,ZP10,Gruber11}. The accompanying
paper~\citep{Granot11} shows, however, that if instead of a single
shell the flow is initially divided into multiple, well separated
sub-shells, then it can reach $\mean{\Gamma}\gg\Gamma_{\rm cr}$ and
reasonably efficient internal shocks can naturally take place at such
high Lorentz factors. This greatly helps to satisfy the observational
constraints on $\Gamma$.

\acknowledgements
The author thanks A. Spitkovsky, Y.~E. Lyubarsky, T. Piran, A. Levinson and
S.~S. Komissarov for useful comments on the manuscript.  This research
was supported by the ERC advanced research grant ``GRBs''.

\newpage
\begin{table}
\begin{center}
{\scriptsize
\begin{tabular}{lll}
\hline\hline
Notation & Definition & Eq./Sect.\\
\hline 
$\Gamma_0$, $\Delta_0$\dotfill & Initial Lorentz factor and lab-frame width of the unmagnetized shell & 
\S~\ref{sec:low-sigma}\\
$\rho_1 = Ar^{-k}$\dotfill & External medium rest mass density ($r$ is the distance from the origin) & 
\S~\ref{sec:low-sigma}\\
$\rho_4$\dotfill & Proper rest mass density of the unmagnetized shell & 
\S~\ref{sec:low-sigma}\\
$f = \rho_4/\rho_1$\dotfill & Unmagnetized shell to external proper rest mass density ratio & 
Eq.~(\ref{f})\\
$l_{\rm S}$\dotfill & Sedov length (or radius) & Eq.~(\ref{l})\\
$R_N\sim\min(R_\Gamma,R_{N,0})$\dotfill & Radius where the reverse shock becomes Newtonian or relativistic & 
\S~\ref{sec:low-sigma}\\
$R_\Gamma$\dotfill & Radius where an external rest mass of $E/\Gamma_0^2c^2$ is swept up & 
Eq.~(\ref{R_gamma})\\
$R_\Delta \sim \max(R_\Gamma,R_{\Delta,0})$\dotfill & 
Radius where the reverse shock finishes crossing the unmagnetized shell &
\S~\ref{sec:low-sigma}\\
$R_s \sim \Delta_0\Gamma_0^2$\dotfill & 
Radius where the shell starts spreading radially significantly & 
\S~\ref{sec:low-sigma}\\
$R_{N,0}$, $R_{\Delta,0}$\dotfill & Initial values (without radial spreading of the shell) of $R_N$ and $R_\Delta$  & 
Eq.~(\ref{R_N}), \S~\ref{sec:low-sigma}\\
$\Upsilon= \Upsilon_0(\Delta_0/\Delta)$\dotfill & Reverse shock strength parameter 
(Newtonian for $\Upsilon > 1$, rel. for $\Upsilon < 1$) & \S~\ref{sec:low-sigma}\\
$\Upsilon_0$\dotfill & Initial value (without radial spreading of the shell) of $\Upsilon$ & 
Eqs.~(\ref{upsilon}), (\ref{xi2})\\
$\Gamma_{\rm cr}$, $\Delta_{\rm cr}$\dotfill & Critical Lorentz factor and width of the shell, respectively & 
Eqs.~(\ref{delta_cr}), (\ref{eta_cr})\\
$T$, $t$\dotfill & Time when photons reach the observer and lab-frame time, respectively & 
\S~\ref{sec:low-sigma}, \S~\ref{sec:high-sigma-acc}\\
$E$, $E_{\rm ext}$ \dotfill & Total energy and energy in the shocked external medium, respectively &
\S~\ref{sec:low-sigma}, \S~\ref{sec:high-sigma-acc}, \S~\ref{sec:high-sigma-acc-dec}\\
$E_{\rm EM}$, $E_{\rm EM,0}$, $E_{\rm kin}$\dotfill & 
Electromagnetic, initial electromagnetic and kinetic energies, respectively &
\S~\ref{sec:high-sigma-acc}, \S~\ref{sec:high-sigma-acc-dec}\\
$\sigma_0 = B_0^2/4\pi\rho_0c^2\gg 1$\dotfill & Initial value of the magnetization parameter & 
\S~\ref{sec:high-sigma-acc}\\
$B$, $B_0$\dotfill & Lab-frame magnetic field and its initial value (at $R_0$), respectively & 
\S~\ref{sec:high-sigma-acc}, \S~\ref{sec:high-sigma-acc-dec}\\
$\rho_0$\dotfill & Initial proper rest mass density of the magnetized shell (at $R_0$) & 
\S~\ref{sec:high-sigma-acc}, \ref{sec:high-sigma-acc-dec}\\
$\beta_{\rm ms,0}$, $u_{\rm ms,0}$, $\Gamma_{\rm ms,0}$\dotfill & 
Initial fast magnetosonic dimensionless speed, 4-velocity and Lorentz factor & \S~\ref{sec:high-sigma-acc}\\
$\beta_{\rm ms}$, $u_{\rm ms}$, $\Gamma_{\rm ms}$\dotfill & 
Fast magnetosonic dimensionless speed, 4-velocity and Lorentz factor & \S~\ref{sec:high-sigma-acc}\\
$\Gamma$, $\sigma$\dotfill & Lorentz factor and magnetization parameter of the shell, respectively & 
\S~\ref{sec:high-sigma-acc}\\
$\mean{\Gamma}$, $\mean{\sigma}$\dotfill & 
Typical values of $\Gamma$ and $\sigma$ -- weighted means over the lab-frame energy & 
\S~\ref{sec:high-sigma-acc}\\
$R_0\approx ct_0 \sim \Delta_0$\dotfill & Initial radius (or lab-frame width) of the magnetized shell & 
\S~\ref{sec:high-sigma-acc}\\
$R_c\approx ct_c \sim R_0\sigma_0^2$\dotfill & Coasting radius where
the shell becomes kinetically dominated (in vacuum) &
\S~\ref{sec:high-sigma-acc}\\
$R_{\rm CD}$, $\Gamma_{\rm CD}$\dotfill & Radius and Lorentz factor of the contact discontinuity (CD) that & \\
 & separates between the magnetized shell and the shocked external medium & \S~\ref{sec:high-sigma-acc-dec}\\
$R_{\rm sh}$, $\Gamma_{\rm sh}$\dotfill & Radius and Lorentz factor of the shock front for the external shock &
\S~\ref{sec:high-sigma-acc-dec}\\
$\xi = r/ct = x/ct$\dotfill & Similarity variable & \S~\ref{sec:high-sigma-acc-dec}, Fig.~\ref{fig:RW-diagram}\\
$\xi_{\rm CD}$, $\xi_{\rm sh}$\dotfill & 
Values of $\xi$ corresponding, respectively, to $R_{\rm CD}$ and $R_{\rm sh}$ &
\S~\ref{sec:high-sigma-acc-dec}, Fig.~\ref{fig:RW-diagram}\\
$\xi_u$\dotfill & Value of $\xi$ where the uniform region 3 in the Riemann problem starts &
\S~\ref{sec:high-sigma-acc-dec}, Fig.~\ref{fig:RW-diagram}\\
$\xi_{\rm rf} = -\beta_{\rm ms,0}$\dotfill & Value of $\xi$ at the head of the self-similar rarefaction wave &
\S~\ref{sec:high-sigma-acc-dec}, Fig.~\ref{fig:RW-diagram}\\
$\chi$, $\chi_{\rm CD}$\dotfill & Similarity variable of \citet{BM76} and its value at $R_{\rm CD}$ &
\S~\ref{sec:high-sigma-acc-dec}, Eq.~(\ref{chi_CD})\\
$\xi_*$\dotfill & Value of $\xi$ at the head of the secondary (or ``reflected'') rarefaction wave & 
\S~\ref{sec:high-sigma-acc-dec}\\
$\beta_*$\dotfill & Dimensionless speed of the secondary (``reflected'') rarefaction wave head & 
Eq.~(\ref{xs*})\\
$a$\dotfill & Ratio of pressure at the CD for the BM76 solution and a uniform region 2 & 
Eqs.~(\ref{p_condition}), (\ref{eq:a})\\
$R_u\approx ct_u$ & Radius where the secondary rarefaction wave reaches region 3, $\xi_*(t_u) = \xi_u$ & 
Eqs.~(\ref{eq:Ru_Rc}), (\ref{eq:R_u2})\\
$\tilde{\rho} \equiv \bar{\rho}/\bar{\rho}_0 = \sigma/\sigma_0$\dotfill & 
Normalized shell proper rest mass density (or magnetization) &
\S~\ref{sec:high-sigma-acc-dec}\\
$f_0 = \rho_0/\rho_1(R_0)$\dotfill & Initial (at $R_0$) magnetized shell to external proper rest mass density ratio &
\S~\ref{sec:high-sigma-acc-dec}\\
$R_1$\dotfill & Radius where $\sigma = 1$ just behind the contact discontinuity (CD) & Eq.~(\ref{eq:R1})\\
$\sigma_u\approx\sigma_{\rm CD}$, $\tilde{\rho}_u = \frac{\sigma_u}{\sigma_0}$\dotfill & 
Values of $\sigma$ and $\tilde{\rho}$, respectively, at $\xi = \xi_u$ (and also just behind the CD) &
Eqs.~(\ref{eq:rho_sigma_u}), (\ref{eq:sigma_uI}), (\ref{eq:regII})\\
$R_{\rm cr} \sim R_0\Gamma_{\rm cr}^2$\dotfill & 
Critical radius where $\mean{\Gamma}$ reaches $\Gamma_{\rm cr}$ in regimes II and III & 
Eqs.~(\ref{R_cr_acc}), (\ref{R_cr2})\\
$R_{\rm dec}$\dotfill & Deceleration radius where most of the energy is transfered to the shocked & \\
 & external medium & \S~\ref{sec:high-sigma-acc-dec}\\
$R_{\rm RS}$\dotfill & Radius where a strong reverse shock develops in regime I & Eq.~(\ref{eq:R_RS})\\
$R_{\rm *,CD}\approx ct_{\rm CD}$\dotfill & 
Radius where the secondary rarefaction's head reaches the CD ($\xi_*=\xi_{\rm CD}$) &
Eqs.~(\ref{eq:DR*}), (\ref{eq:DR})\\
$u_{\rm RS}$\dotfill & Reverse shock upstream to downstream relative 4-velocity & \S~\ref{sec:regI}\\
$L \approx Ec/2R_0$, $L_{\rm CD}$\dotfill & 
Shell's mean total energy flux through a static sphere \& its value at $R_{\rm CD}$ &
\S~\ref{sec:regII}\\
$\Gamma_u \equiv\Gamma_{\rm CD}(R_u)$\dotfill & 
The CD as well as the typical Lorentz factor at $R_u$, $\Gamma_u \sim \mean{\Gamma}(R_u)$ &
Eqs.~(\ref{eq:R_u2}), (\ref{eq:Gamma_u})\\
$T_{\rm GRB} = (1+z)\Delta_0/c$\dotfill & Observed duration of the prompt GRB emission & 
\S~\ref{sec:low-sigma}, \S~\ref{sec:dis}\\
$T_{\rm dec}\sim\max(${\scriptsize$T_{\rm GRB},T_\Gamma$}$)$\dotfill & 
Duration of peak reverse shock or afterglow emission (deceleration time) & 
Eq.~(\ref{t_gamma}), \S~\ref{sec:dis}\\
\hline
\end{tabular}}
\end{center}
\caption{\footnotesize
Notation and Definition of Some Quantities Used Throughout This Work.
\label{tab:notation}}
\end{table}

\begin{sidewaystable}
\begin{center}
{\scriptsize
\begin{tabular}{|c|c|c|c|c|}
\hline
regime & ordering of critical radii & $f_0 = \rho_0/\rho_1(R_0)$ & 
$\Gamma_{\rm cr}$ & $\sigma_0$ \\ 
\hline
\hline\vspace{-0.3cm}
 & & & & \\
I   & $R_0 < (R_1<)^\dagger R_c < R_u \sim R_{\rm *,CD} < 
       R_{\rm RS}\sim R_{\rm cr}< R_{\rm dec} \sim R_\Gamma$ &
      $f_0\gg\sigma_0^{7-2k}\gg 1$ & 
      $\Gamma_{\rm cr}\gg\sigma_0\gg 1$ & 
      $1\ll\sigma_0\ll\Gamma_{\rm cr}$ \\ 
    & & $\Gamma_{\rm cr}^{8-2k}\gg f_0\gg\Gamma_{\rm cr}^{7-2k}\gg 1$ & 
      $f_0^\frac{1}{7-2k}\gg\Gamma_{\rm cr}\gg f_0^\frac{1}{8-2k}\gg 1$ &
      $1\ll\sigma_0\ll f_0^\frac{1}{7-2k}$ \\
 & & & & \\
II  & $R_0<(R_1<)^\ddagger\,R_u<R_{\rm cr}\sim R_{\rm dec}\sim R_{\rm *,CD}<R_c$ &
      $\sigma_0^{1/3}\ll f_0\ll\sigma_0^{7-2k}$ &
      $\sigma_0^\frac{2}{12-3k}\ll\Gamma_{\rm cr}\ll\sigma_0$ &
      $\Gamma_{\rm cr}\ll\sigma_0\ll\Gamma_{\rm cr}^\frac{12-3k}{2}$ \\ 
     & & $\Gamma_{\rm cr}^\frac{4-k}{2}\ll f_0 \ll \Gamma_{\rm cr}^{7-2k}$ &
      $f_0^\frac{1}{7-2k}\ll\Gamma_{\rm cr}\ll f_0^\frac{2}{4-k}$ &
      $f_0^\frac{1}{7-2k}\ll\sigma_0\ll f_0^3$ \\
 & & & & \\
III & $R_0\sim R_u < R_{\rm cr}\sim R_{\rm dec}\sim R_{\rm *,CD}<R_c$ &
      $\sigma_0^{-1}\ll f_0\ll\sigma_0^{1/3}$ &
      $1\ll\Gamma_{\rm cr}\ll\sigma_0^\frac{2}{12-3k}$ &
      $\sigma_0\gg\Gamma_{\rm cr}^\frac{12-3k}{2}\gg 1$ \\ 
     & & $f_0 \ll \Gamma_{\rm cr}^\frac{4-k}{2}$, $\Gamma_{\rm cr}\gg 1$ &
      $\Gamma_{\rm cr}\gg\max(1,f_0^\frac{2}{4-k})$ &
      $\sigma_0\gg\max(f_0^3,f_0^{-1})$ \\
 & & & & \\
IV  & $R_{\rm dec} \sim R_0$\ ,\quad $t_{\rm dec}/t_0 \sim \Gamma_{\rm cr}^{k-4}\gg 1$ &
      $f_0\ll\sigma_0^{-1}\ll 1$ & 
      $\Gamma_{\rm cr}\ll 1\ll\sigma_0$ &
      $\sigma_0\gg 1\gg\Gamma_{\rm cr}$\\
     & & $f_0\ll\Gamma_{\rm cr}^{8-2k}\ll 1$ &
      $f_0^\frac{1}{8-2k}\ll\Gamma_{\rm cr}\ll 1$ &
      $1\ll\sigma_0 \ll f_0^{-1}$ \\
\hline 
\end{tabular}}
\end{center}
\caption{\footnotesize
The different regimes for $k < 10/3$ expressed in terms of $f_0 =
\rho_0/\rho_1(R_0)$, $\Gamma_{\rm cr} \sim (f_0\sigma_0)^{1/(8-2k)}$
and $\sigma_0$.
\newline
$^\dagger\,$This ordering of $R_1$ is valid only for $2 < k < 10/3$
and $\sigma_0^{7-2k} < f_0 < \sigma_0^3$ $\Longleftrightarrow$
$\Gamma_{\rm cr}^{(4-k)/2} < \sigma_0 <
\Gamma_{\rm cr}$.
\newline
$^\ddagger\,$This ordering of $R_1$ is valid only for $k < 2$ and
$\sigma_0^3 < f_0 < \sigma_0^{7-2k}$ $\Longleftrightarrow$ $\Gamma_{\rm
cr} < \sigma_0 <
\Gamma_{\rm cr}^{(4-k)/2}$.
\label{tab:regimes}}
\end{sidewaystable}

\begin{sidewaystable}
\begin{center}
{\scriptsize
\begin{tabular}{|c|c|c|c|c|}
\hline
regime & ordering of critical radii & $f_0 = \rho_0/\rho_1(R_0)$ & 
$\Gamma_{\rm cr}$ & $\sigma_0$ \\ 
\hline
\hline\vspace{-0.3cm}
 & & & & \\
I   & $^{\dagger}\,R_0 \sim R_{\rm dcp} < R_c < R_{\rm cr}$ &
      $f_0\gg\sigma_0^{1/3}\gg 1$ & 
      $\Gamma_{\rm cr}\gg\sigma_0^\frac{2}{12-3k}\gg 1$ & 
      $1\ll\sigma_0\ll\Gamma_{\rm cr}^\frac{12-3k}{2}$ \\ 
       & $^{\ddagger}\,R_0 < R_1 \sim R_{\rm dcp} < R_c < R_{\rm cr}$ &
      $\Gamma_{\rm cr}^{8-2k}\gg f_0\gg\Gamma_{\rm cr}^\frac{4-k}{2}\gg 1$ & 
      $f_0^\frac{2}{4-k}\gg\Gamma_{\rm cr}\gg f_0^\frac{1}{8-2k}\gg 1$ &
      $1\ll\sigma_0\ll f_0^3$ \\
 & & & & \\
II$^*$  & $R_0< R_u<R_1\sim R_{\rm dcp}<R_c<R_{\rm cr}$ &
      $\sigma_0^{7-2k}\ll f_0\ll\sigma_0^{1/3}$ &
      $1\ll\sigma_0\ll\Gamma_{\rm cr}\ll\sigma_0^\frac{2}{12-3k}$ &
      $1\ll\Gamma_{\rm cr}^\frac{12-3k}{2}\ll\sigma_0\ll\Gamma_{\rm cr}$ \\ 
     & & $\Gamma_{\rm cr}^{7-2k}\ll f_0\ll\Gamma_{\rm cr}^\frac{4-k}{2}$ & & \\
$\frac{10}{3}<k<\frac{7}{2}$ & & & 
      $1\ll f_0^\frac{2}{4-k}\ll \Gamma_{\rm cr}\ll  f_0^\frac{1}{7-2k}$ &
      $1\ll f_0^3\ll\sigma_0\ll f_0^\frac{1}{7-2k}$ \\
$\frac{7}{2}<k<4$ & & & $\Gamma_{\rm cr}>\max(f_0^\frac{2}{4-k},f_0^\frac{1}{7-2k})$ & 
      $\sigma_0>\max(f_0^3,f_0^\frac{1}{7-2k})$ \\ 
 & & & & \\
III & $R_0 < R_{\rm cr}\sim R_{\rm dcp} \sim R_{\rm dec}$ &
      $\sigma_0^{-1}\ll f_0\ll\sigma_0^{7-2k}$ &
      $1\ll\Gamma_{\rm cr}\ll\sigma_0$ &
      $\sigma_0\gg\Gamma_{\rm cr}\gg 1$ \\ 
     & & $f_0\ll\Gamma_{\rm cr}^{7-2k}$, $\Gamma_{\rm cr}\gg 1$ & & \\
$\frac{10}{3}<k<\frac{7}{2}$ & & & 
      $\Gamma_{\rm cr} > \max(1,f_0^{7-2k})$ &
      $\sigma_0>\max(f_0^{-1},f_0^\frac{1}{7-2k})$ \\
$\frac{7}{2}<k<4$ & & & $f_0^{7-2k}\gg \Gamma_{\rm cr}\gg 1$ &
      $1\ll f_0^{-1}\ll\sigma_0\ll f_0^\frac{1}{7-2k}$ \\
 & & & & \\
IV  & $R_{\rm dec} \sim R_0$\ ,\quad $t_{\rm dec}/t_0 \sim \Gamma_{\rm cr}^{k-4}\gg 1$ &
      $f_0\ll\sigma_0^{-1}\ll 1$ & 
      $\Gamma_{\rm cr}\ll 1$ &
      $\sigma_0 \gg 1 \gg \Gamma_{\rm cr}$ \\
    & & $f_0\ll\Gamma_{\rm cr}^{8-2k}\ll 1$ &
      $f_0^\frac{1}{8-2k}\ll\Gamma_{\rm cr}\ll 1$ &
      $1\ll\sigma_0\ll f_0^{-1}$ \\
\hline 
\end{tabular}}
\end{center}
\caption{\footnotesize
The different regimes for $10/3<k<4$ expressed in terms of $f_0 =
\rho_0/\rho_1(R_0)$, $\Gamma_{\rm cr} \sim (f_0\sigma_0)^{1/(8-2k)}$
and $\sigma_0$.
\newline
$^\dagger\,$This ordering holds for $f_0 > \sigma_0^3$
$\Longleftrightarrow$ $\sigma_0 <\Gamma_{\rm cr}^{(4-k)/2}$.
\newline
$^\ddagger\,$This ordering holds for $\sigma_0^{1/3} <
f_0 < \sigma_0^3$ $\Longleftrightarrow$ $\Gamma_{\rm cr}^{(4-k)/2} <
\sigma_0 <\Gamma_{\rm cr}^{(12-3k)/2}$.
\label{tab:regimes_k>10/3}}
\end{sidewaystable}

\end{document}